\documentclass[11pt,letterpaper]{article}
\pdfoutput=1
\usepackage{jcappub}

\usepackage{appendix}
\usepackage{array}
\newcolumntype{x}[1]{>{\centering\arraybackslash}p{#1}}

\usepackage{amsmath}
\usepackage{amsfonts}
\usepackage{amssymb}
\usepackage{eucal}               
\usepackage{mathrsfs}          
\usepackage{slashed}

\usepackage{epsfig}
\usepackage{graphicx}
\usepackage{dcolumn}
\usepackage{enumerate}

\usepackage{color}
\usepackage{ulem}
\hypersetup{colorlinks=false}

\newcommand{\eg}{e.g.~}
\newcommand{\ie}{i.e.~}

\newcommand{\beq}{\begin{equation}}
\newcommand{\eeq}{\end{equation}}
\newcommand{\ud}{\text{d}}

\newcommand{\ER}{E_\text{R}}

\newcommand{\vesc}{v_\text{esc}}
\newcommand{\vmin}{v_\text{min}}
\newcommand{\vmax}{v_\text{max}}

\newcommand{\bsv}
{\vec{v}}
\newcommand{\bsu}
{\vec{u}}

\newcommand{\Lag}{\mathscr{L}}
\newcommand{\Mel}{\mathscr{M}}

\newcommand{\AV}{\text{AV}}
\newcommand{\PS}{\text{PS}}
\newcommand{\rf}{\text{ref}}

\title{Reevaluation of spin-dependent WIMP-proton interactions as an explanation of the DAMA data}

\author{Eugenio Del Nobile,}
\author{Graciela B. Gelmini,}
\author{Andreea Georgescu,}
\author{and Ji-Haeng Huh}

\affiliation{Department of Physics and Astronomy, UCLA,\\
475 Portola Plaza, Los Angeles, CA 90095, USA}

\emailAdd{delnobile@physics.ucla.edu}
\emailAdd{gelmini@physics.ucla.edu}
\emailAdd{a.georgescu@physics.ucla.edu}
\emailAdd{jhhuh@physics.ucla.edu}

\abstract{
We reexamine the interpretation of the annual modulation signal observed by the DAMA experiment as due to WIMPs with a spin-dependent coupling mostly to protons. We consider both axial-vector and pseudo-scalar couplings, and elastic as well as endothermic and exothermic inelastic scattering. We conclude that the DAMA signal is in strong tension with null results of other direct detection experiments, particularly PICASSO and KIMS. }


\notoc

\begin{document}

\maketitle

{\vspace{-17.6cm}\begin{flushright}
NSF-KITP-15-018
\end{flushright}  }

\vspace{16.6cm}

\setcounter{page}{0}

\newpage

\section{Introduction}\label{sec:introduction}

Direct Dark Matter (DM) searches aim at detecting Weakly Interacting Massive Particles (WIMPs) scattering with nuclei in a target material. The most stringent limits on the DM mass and cross section parameter space are currently set by LUX \cite{Akerib:2013tjd} and SuperCDMS \cite{Agnese:2014aze} for WIMPs with spin-independent interactions and spin-dependent interactions with neutrons, and by PICASSO \cite{Archambault:2012pm}, SIMPLE \cite{Felizardo:2011uw}, COUPP \cite{Behnke:2012ys}, and KIMS \cite{Kim:2012rza} for those with protons. While none of these experiments have detected a DM signal so far, three other experiments have signals that can be interpreted as due to WIMP scattering: DAMA and DAMA/LIBRA (called here DAMA from now on) \cite{Bernabei:2013xsa}, CoGeNT \cite{Aalseth:2014eft, Aalseth:2014jpa}, and CDMS-II with silicon detectors \cite{Agnese:2013rvf}. CRESST-II \cite{Angloher:2014myn} has not confirmed a previous DM hint found by the same collaboration \cite{Angloher:2011uu}. Among the potential DM signals, DAMA's observation of an annually modulated rate has the highest statistical significance. However, these potential signals are challenged by the null results of other direct detection experiments which exclude the possibility of WIMP scattering in a large number of particle models. In particular, the scattering cross section fitting the DAMA data for WIMPs with isospin-conserving spin-independent interactions is several orders of magnitude above the $90\%$ CL LUX limit \cite{DelNobile:2014sja}.

Over the years, several particle candidates have been proposed with properties which enhance a potential signal in DAMA and weaken the main limits imposed by direct DM searches with negative results, among them WIMPs with spin-dependent coupling mostly to protons, which we reevaluate in this paper.

Inelastic DM \cite{TuckerSmith:2001hy, TuckerSmith:2004jv, Chang:2008gd, MarchRussell:2008dy, Cui:2009xq, Chang:2010en, Barello:2014uda} scatters to another particle state, either heavier (endothermic scattering) or lighter (exothermic scattering, see \eg \cite{Graham:2010ca,Gelmini:2014psa}), when colliding with a nucleus. Endothermic scattering favors heavier targets, thus enhancing scattering off I in DAMA while reducing scattering off lighter targets such as Ge. Moreover, this type of interaction enhances the annual modulation amplitude, thus pushing the cross section needed to fit the DAMA data to lower values. However, experiments employing Xe as target material, which is heavier than I, rule out endothermic scattering of DM as an explanation to the DAMA data unless there is an additional feature of the interaction that favors a signal in DAMA. Two types of WIMP couplings favor Na and I (DAMA) over Xe (LUX, XENON10) and Ge (CDMS, SuperCDMS): a spin-dependent coupling mostly to protons and a magnetic dipole moment coupling \cite{Chang:2010en,DelNobile:2012tx}. The reason for the first is that the spin of a nucleus is mostly due to an unpaired nucleon and Na and I have an unpaired proton, while Xe and Ge have an unpaired neutron. The reason for the second is the large magnetic moment of both Na and I. We do not consider here inelastic magnetic DM, which Ref.~\cite{Barello:2014uda} recently found still marginally compatible with all negative results of direct searches, for a small value of the I quenching factor (without the aid of inelasticity, instead, this candidate has been shown to be ruled out \cite{DelNobile:2014eta}).

The possibility that a DM candidate with spin-dependent interactions mostly with protons would explain the DAMA signal was, to the best of our knowledge, first studied in Ref.~\cite{Ullio:2000bv}. Compared to spin-independent interactions, spin-dependent couplings reduce the bounds from experiments with heavy targets, most notably LUX, due to the lack of the usual $A_T^2$ enhancement factor proper to the spin-independent interaction ($A_T$ being the mass number of the target nucleus). This interaction might explain why the DAMA signal is not seen by LUX and SuperCDMS. In this case, bounds from PICASSO, SIMPLE, COUPP, and KIMS become relevant, since they contain F, I and Cs, all nuclei with an unpaired proton. This candidate was further studied in Ref.~\cite{Kopp:2009qt}, in the context of both elastic and inelastic endothermic scatterings. The inelastic endothermic kinematics reduces the expected rate in experiments employing F (PICASSO, SIMPLE) because it is light, thus making the COUPP (CF$_3$I) and KIMS (CsI) bounds the most relevant constraints on WIMP scatterings off I in DAMA. Ref.~\cite{Barello:2014uda} found that a small portion of the parameter space favored by DAMA for inelastic spin-dependent couplings with protons can still escape all bounds from null experiments.

Inelastic exothermic scattering \cite{Batell:2009vb,Graham:2010ca} favors lighter targets, so it favors Na in DAMA over heavier nuclei (Ge and Xe). In this case the most important limits are set by experiments containing F (PICASSO and SIMPLE). 

Recently, Ref.~\cite{Arina:2014yna} studied a Dirac WIMP candidate coupled to standard model (SM) fermions through a light pseudo-scalar mediator, and claimed that with a contact interaction and elastic scattering it reconciles the DAMA data with the null results of other experiments at the 99\% credible level. The model produces a non-standard spin-dependent interaction, with the noteworthy feature that, for universal flavor-diagonal quark couplings to the pseudo-scalar mediator, the WIMP couples mainly to protons. The couplings of pseudo-scalar light bosons ($m_\phi<7$ GeV) with quarks are strongly constrained by rare meson decays \cite{Hiller:2004ii,Andreas:2010ms,Dolan:2014ska}, and unless the pseudo-scalar coupling to the DM (called $g_\text{DM}$ below) is very large, $g_\text{DM} \gtrsim 10^3$, the one-particle exchange scattering cross section required in Ref.~\cite{Arina:2014yna} is rejected \cite{Dolan:2014ska}. The flavor physics bounds on pseudo-scalar couplings to quarks proportional to the quark mass are less stringent \cite{Dolan:2014ska}, but in this case Ref.~\cite{Arina:2014yna} found that the resulting proton to neutron coupling ratio is not large enough to reconcile a DM signal in DAMA with the results of other direct detection experiments. Leaving aside the limits from other types of experiments we concentrate here on direct detection.
 
Ref.~\cite{Arina:2014yna} employed a Bayesian analysis, where a number of uncertain parameters like quenching factors and background levels, as well as astrophysical quantities, are marginalized over. While the process of marginalization (\ie integrating over nuisance parameters with assumed prior probability distributions) is the proper treatment of uncertain and uninteresting parameters in the context of Bayesian statistics, it makes it unclear whether there exists at least one set of values of the uncertain parameters, in particular one halo model, that produces the same result of the analysis.

In this paper we reconsider the viability of a signal due to WIMPs with spin-dependent coupling mostly to protons as an explanation of the DAMA data. We study both axial-vector and pseudo-scalar couplings, which lead respectively to $\vec{s}_\chi\cdot\vec{s}_p$ and $(\vec{s}_\chi\cdot\vec{q})(\vec{s}_p\cdot\vec{q})$ couplings in the non-relativistic limit ($\vec{s}_\chi$ and $\vec{s}_p$ are the spins of the WIMP $\chi$ and the proton respectively, and $\vec{q}$ is the momentum transfer). We assume the mediator to be either heavy enough for the contact interaction limit to be valid, or otherwise much lighter than the typical momentum transfer of the scattering process than the typical momentum transfer of the scattering process (we refer to this last case as ``massless''). The possibilities of elastic and inelastic scattering, both endothermic and exothermic, are considered.

In Section~\ref{sec:cross sections} we present the differential cross sections for axial-vector and pseudo-scalar couplings, which can be used in the direct detection rate formula in Section~\ref{sec:direct detection rate}. The analysis methods we adopt for experimental data are described in Section~\ref{sec:data analysis}, and our results assuming a standard model of the dark halo of our galaxy are presented in Section~\ref{sec:data comparison}. In Section~\ref{sec:halo_indep} we describe our halo-independent analysis and present the related results. Our conclusions are given in Section~\ref{sec:conclusion}.

\section{Cross sections}\label{sec:cross sections}

\subsection{Axial-vector (AV) interaction}

An AV coupling leads to the usual spin-dependent interaction. The effective Lagrangian for the elastic scattering of a DM particle $\chi$, either a Dirac or a Majorana fermion, with AV couplings to nucleons, mediated by a vector boson of mass $m_\phi$, is
\beq
\label{L_AV}
\Lag_\text{AV} = \frac{g_{\rm DM}}{2 (m_\phi^2 - q^\mu q_\mu)} \sum_{N = p, n} a_N \, \bar\chi \gamma^\mu \gamma^5 \chi \, \bar N \gamma_\mu \gamma^5 N \ .
\eeq
Here we assumed a one-particle exchange process. $q^\mu$ is the momentum transfer four-vector, and $N$ is a nucleon, $p$ or $n$. $g_{\rm DM}$ and $a_N$ are the mediator coupling constants to $\chi$ and $N$, respectively, and they are real. The scattering amplitude is
\beq
\Mel_\text{AV} = \frac{g_{\rm DM}}{2 (m_\phi^2 - q^\mu q_\mu)} \sum_{N = p, n} a_N \, \bar u^{s'}_\chi \gamma^\mu \gamma^5 u^s_\chi \, \bar u^{r'}_N \gamma_\mu \gamma^5 u^r_N \ .
\eeq
We now follow Ref.~\cite{Fitzpatrick:2012ix} because we will largely use the nuclear form factors given in this reference. We first take the non-relativistic limit of the Dirac spinors, in the chiral representation, for both $\chi$ and $N$:
$u^s(\vec{p})\simeq \sqrt{1/4m}\big(\!\left(2m-\vec{p}\cdot\vec{\sigma}\right)\xi^s, \left(2m+\vec{p}\cdot\vec{\sigma}\right)\xi^s\big)^\text{T}$, where $\vec{\sigma}$ are the Pauli matrices. This limit is justified by the fact that the DM initial speed and the exchanged momentum are small. The matrix element for scattering off a single nucleon then reads
\beq\label{eq:M_AV_nonrel}
\Mel_\text{AV} = - 8 m_N m \frac{g_{\rm DM}}{m_\phi^2 + q^2} \sum_{N = p, n} a_N \, \langle \vec{s}_\chi \rangle \cdot \langle \vec{s}_N \rangle \ ,
\eeq
where $\langle \vec{s}_\chi \rangle = {\xi_\chi^{s'}}^\dagger \frac{\vec{\sigma}}{2} \xi_\chi^s$ and $\langle \vec{s}_N \rangle = {\xi_N^{r'}}^\dagger \frac{\vec{\sigma}}{2} \xi_N^r$
(see e.g.~Eqs.~(44), (47d), and (49) of Ref.~\cite{DelNobile:2013sia}). 
Notice that this matrix element assumes the usual form for the Dirac spinors with normalization $\bar{u}^{s'} \! (\vec{p}) \, u^s(\vec{p}) = 2 m \delta^{ss'}$; Quantum Mechanical amplitudes usually assume a different state normalization, which differs by a factor of $2 \sqrt{m^2 + \vec{p}^{\, 2}}$. With this normalization Eq.~\eqref{eq:M_AV_nonrel} would be replaced by  $\Mel_\text{AV}^\text{QM} = - 2 g_{\rm DM} (q^2 + m_\phi^2)^{-1} \sum_{N = p, n} a_N \, \langle \vec{s}_\chi \rangle \cdot \langle \vec{s}_N \rangle$.

For a model of inelastic DM, one could introduce two Dirac fields, $\chi_1$ and $\chi_2$, with slightly different masses and the DM-nucleon effective Lagrangian
\beq
\label{L_AV inelastic}
\Lag_\text{AV} = \frac{g_{\rm DM}}{2 (m_\phi^2 + q^2)} \sum_{N = p, n} a_N \, \bar\chi_2 \gamma^\mu \gamma^5 \chi_1 \, \bar N \gamma_\mu \gamma^5 N + \text{h.c.} \ . 
\eeq
The $g_\text{DM}$ coupling can now be complex. We use the same symbol, $g_\text{DM}$, for the couplings in Eqs.~\eqref{L_AV} and \eqref{L_AV inelastic}, because then the expression of $\sigma_p^\AV$ in Eq.~\eqref{eq:sigmaAV} is valid both for elastic and inelastic scattering. $\chi_1$ is the DM particle entering the scattering process, with mass $m$, while $\chi_2$ is the DM particle in the final state, with mass $m' = m + \delta$. The sign of the mass splitting $\delta$ determines the different kinematic regimes: $\delta > 0$ implies endothermic scattering, $\delta < 0$ implies exothermic scattering, while $\delta = 0$ implies elastic scattering.

One may attempt to build an inelastic DM model without introducing additional degrees of freedom by assuming the interaction in Eq.~\eqref{L_AV} and adding a small Majorana mass term which produces two almost degenerate Majorana fermions, $\chi_1$ and $\chi_2$, in which case $\chi=\chi_1+i\chi_2$ becomes a quasi-Dirac fermion. However, as noted in Ref.~\cite{Kopp:2009qt}, this interaction produces diagonal terms which result in elastic scattering rather than inelastic. In this case, one can instead write an effective tensor interaction $\bar\chi \sigma^{\mu\nu} \chi \, \bar N \sigma_{\mu\nu} N$, which produces inelastic scattering since the diagonal interaction terms vanish identically. The non-relativistic limit of this operator is also $\vec{s}_\chi\cdot \vec{s}_N$.

The differential cross section for DM-nucleus scattering, for both the elastic and inelastic interactions introduced above in Eqs.~\eqref{L_AV} and \eqref{L_AV inelastic}, is
\beq
\label{eq:AV_diff_cross_section}
\dfrac{\ud \sigma_T^\AV}{\ud \ER} = \sigma_p^\AV \, \dfrac{m_T}{2\mu_p^2} \left(\dfrac{m_\phi^2}{m_\phi^2 + 2 m_T \ER}\right)^2 \dfrac{1}{v^2} F_\AV^2(q^2) \ ,
\eeq
where $\ER = q^2 / 2 m_T$ is the nuclear recoil energy, $v$ is the incoming WIMP speed, and $\mu_p$ is the DM-proton reduced mass. $F_\AV^2(q^2)$ is a nuclear form factor including spin dependence, and will be defined in Section \ref{sec:nuclearFF}.
$\sigma_p^\AV$ is the total DM-proton cross section in the limit of contact interaction $m_\phi \gg q = \sqrt{2m_T \ER}$,
\beq\label{eq:sigmaAV}
\sigma_p^\AV = \dfrac{3 |g_\text{DM}|^2 a_p^2}{4\pi} \dfrac{\mu_p^2}{m_\phi^4} \ .
\eeq
The term in parenthesis in Eq.~\eqref{eq:AV_diff_cross_section} accounts for long-range interactions, when $m_\phi$ is smaller than or comparable to $q=\sqrt{2m_\text{T} \ER}$. For typical target masses of a few tens of GeV and recoil energies around few to tens of keV, the interaction becomes effectively long-range if the mediator mass is smaller than several MeV: $m_\phi \ll q \simeq 20 \text{ MeV} \sqrt{(\ER/10 \text{ keV})(m_T/20 \text{ GeV})}$. 

In order to plot our results for long-range interactions, we express the differential cross section in terms of a reference total cross section $\sigma_p^{\AV,\: \rf} = \sigma_p^{\AV}(m_\phi = m_\phi^\rf)$ corresponding to a reference mediator mass $m_\phi^\rf$, which we set equal to $1$ GeV:
\beq
\label{eq:AV_diff_cross_section_ref}
\dfrac{\ud\sigma_T^{\AV}}{\ud\ER} = \sigma_p^{\AV,\:\rf} \, \dfrac{m_T}{2\mu_p^2} \left(\dfrac{(m_\phi^\rf)^2}{m_\phi^2 + 2 m_T \ER}\right)^2 \dfrac{1}{v^2} F_\AV^2(q^2) \ .
\eeq
The massless mediator limit thus corresponds to setting $m_\phi=0$ in the equation above. In the following we will refer to any scenario with $m_\phi^2 \ll q^2$ as massless mediator limit or long-range limit.

Given that a large value of $a_p / a_n$ is needed to suppress the strong LUX and SuperCDMS constraints, we will assume the maximally isospin-violating coupling $a_n = 0$ for the AV interaction.

\subsection{Pseudo-scalar (PS) interaction}

Here the DM particle is a Dirac fermion $\chi$, coupled to a real PS boson $\phi$ with mass $m_\phi$,
\beq\label{eq:lagrangianDM1}
\Lag_\text{DM} = - i \frac{g_\text{DM}}{\sqrt{2}} \phi \, \bar{\chi} \gamma^5 \chi
\eeq
(as in Refs.~\cite{Arina:2014yna, Boehm:2014hva}), with a real coupling constant $g_\text{DM}$. The PS field couples also to the SM quarks with real coupling $g_q$,
\beq\label{eq:lagrangiana}
\Lag_q = - i \frac{1}{\sqrt{2}} \sum_q g_q  \, \phi \, \bar{q} \gamma^5 q \ .
\eeq
While PS couplings to quarks are usually taken to be proportional to the fermion mass (see \eg \cite{Boehm:2014hva}), we will assume instead a flavor-universal coupling $g_q=g$, which introduces a larger $|a_p/a_n|$ ratio, $a_p/a_n \simeq -16.4$ \cite{Arina:2014yna} (see below). 

To model inelastic scattering we assume a non-diagonal coupling of two Dirac DM fields $\chi_1$ and $\chi_2$ with $\phi$,
\beq\label{eq:lagrangian2}
\Lag_\text{DM} = - i \frac{g_\text{DM}}{\sqrt{2}} \phi \, \bar{\chi}_2 \gamma^5 \chi_1 + \text{h.c.} \ .
\eeq
Again, the $g_\text{DM}$ coupling can now be complex. With this definition of $g_\text{DM}$, Eqs.~\eqref{eq:lagrangianDM1} and \eqref{eq:lagrangian2} yield the same expression for $\sigma_p^\PS$ in Eq.~\eqref{eq:sigmapPS}. Diagonal interaction terms as well as non-diagonal mass terms can be forbidden by assuming a $\mathbb{Z}_2$ symmetry under which both $\phi$, $\chi_1$ (or $\chi_2$), and all the SM electroweak doublets have charge $-1$. 

The DM-nucleon effective Lagrangian, for elastic scattering (for inelastic scattering we should have $\bar\chi_2 \gamma^5 \chi_1$ instead of $\bar\chi \gamma^5 \chi$), and assuming one-particle exchange, is
\beq\label{eq:lagrangianN}
\Lag_\text{PS} = \frac{g_{\rm DM}}{2 (m_\phi^2 - q^\mu q_\mu)} \sum_{N = p, n} a_N \, \bar\chi \gamma^5 \chi \, \bar N \gamma^5 N \ ,
\eeq
and yields in the non-relativistic limit
\beq
\Mel_\text{PS} = 2 \frac{g_{\rm DM}}{m_\phi^2 + q^2} \sum_{N = p, n} a_N \, (\langle \vec{s}_\chi \rangle \cdot \vec{q}) (\langle \vec{s}_N \rangle \cdot \vec{q}) \ ,
\eeq
where $\langle \vec{s}_\chi \rangle$ and $\langle \vec{s}_N \rangle$ were defined after Eq.~\eqref{eq:M_AV_nonrel}. This is a different type of spin-dependent interaction than in Eq.~\eqref{eq:M_AV_nonrel}. Due to the extra factors of $\vec{q}$, the PS cross section receives a large $q^4/m_N^2m^2$ suppression with respect to the AV cross section. Therefore, the normalization of the signal and its spectrum, and also the nuclear form factors are different in the two cases \cite{Fitzpatrick:2012ix} (see Section \ref{sec:nuclearFF}). Given the large momentum suppression, one needs to check the existence of unsuppressed radiative corrections to this tree-level cross section, that would spoil the setup. The PS interaction in Eq.~\eqref{eq:lagrangianN} with a Dirac fermion $\chi$ has been proven not to produce such corrections \cite{Freytsis:2010ne}, while this would not be the case if $\chi$ were a Majorana fermion.

The proton and neutron couplings appearing in Eq.~\eqref{eq:lagrangianN} are given by
\beq\label{eq:sumq}
a_N = g \sum_{f = u, d, s} \frac{m_N}{m_f} \biggl[ 1 - \sum_{f' = u, \dots, t} \frac{\bar{m}}{m_{f'}} \biggr] \Delta_f^{(N)} = g \sum_{f = u, d, s} \frac{m_N}{m_f} \biggl[ \sum_{f' = c, b, t} \frac{\bar{m}}{m_{f'}} \biggr] \Delta_f^{(N)}
\ ,
\eeq
with $\bar{m} \equiv (1 / m_u + 1 / m_d + 1 / m_s)^{-1}$. The subscripts $f$ in \eqref{eq:sumq} indicate quark flavors. The $\Delta_f^{(N)}$ factors parametrize the quark spin content of the nucleon, and are usually determined experimentally or computed with lattice calculations. As in Ref.~\cite{Arina:2014yna}, we adopt the following values from \cite{Cheng:2012qr}:
\begin{align}\label{eq:Delta}
\Delta_u^{(p)} &= \Delta_d^{(n)} = + 0.84 \ ,
&
\Delta_d^{(p)} &= \Delta_u^{(n)} = - 0.44 \ ,
&
\Delta_s^{(p)} &= \Delta_s^{(n)} = - 0.03 \ ,
\end{align} 
with which $a_p\simeq-0.4 g$.
As a natural feature of this model, the proton coupling $a_p$ is larger (in modulus) than the neutron coupling $a_n$, by an amount that depends on the choice of the $\Delta_f^{(N)}$'s. As noted in \cite{Arina:2014yna}, the values in Eq.~\eqref{eq:Delta} are conservative in the sense that they minimize the ratio $a_p / a_n$ with respect to other values encountered in the literature (see \eg Table 4 in~\cite{DelNobile:2013sia}). In this case $a_p/a_n=-16.4$.

The differential cross section for the PS interaction, for both elastic and inelastic scattering, is
\beq
\label{eq:PS_diff_cross_section}
\dfrac{\ud\sigma_T^\PS}{\ud\ER} = \sigma_p^\PS \, \dfrac{3m_T^3\ER^2}{8\mu_p^6} \dfrac{1}{{v^\rf}^4} \left(\dfrac{m_\phi^2}{m_\phi^2 + 2 m_T \ER}\right)^2 \dfrac{1}{v^2} F_\PS^2(q^2) \ .
\eeq
$F_\PS^2(q^2)$, to be defined in Section \ref{sec:nuclearFF}, is the nuclear form factor including spin dependence. $\sigma_p^\PS$ is the total DM-proton cross section in the limit of contact interaction,
\beq\label{eq:sigmapPS}
\sigma_p^\PS = \dfrac{|g_\text{DM}|^2 a_p^2}{12\pi} \dfrac{\mu_p^6}{m_\phi^4} \dfrac{{v^\rf}^4}{m^2m_p^2} \ .
\eeq
In this case, the total DM-proton cross section has a $v^4$ dependence, and, for the purpose of plotting our results in terms of the reference cross section, in Eq.~\eqref{eq:sigmapPS} we evaluate $\sigma_p^\PS$ at a reference speed $v^\rf$. We set $v^\rf$ equal to the rotational speed of our Local Standard of Rest, $220$ km/s, which is representative of the WIMP speeds with respect to Earth. 

For long-range PS interactions we proceed in the same manner as for the long-range AV interactions, by writing the differential cross section in terms of a reference total cross section $\sigma_p^{\PS,\:\rf}=\sigma_p^\PS (m_\phi = m_\phi^\rf)$, with $m_\phi^\rf=1$ GeV:
\beq
\label{eq:PS_diff_cross_section_ref}
\dfrac{\ud\sigma_T^{\PS}}{\ud\ER} = \sigma_p^{\PS,\:\rf} \, \dfrac{3m_T^3\ER^2}{8\mu_p^6} \dfrac{1}{{v^\rf}^4} \left(\dfrac{(m_\phi^\rf)^2}{m_\phi^2 + 2 m_T \ER}\right)^2 \dfrac{1}{v^2} F_\PS^2(q^2) \ .
\eeq

\subsection{Nuclear form factors}\label{sec:nuclearFF}

We adopt the form factors computed in Ref.~\cite{Fitzpatrick:2012ix} using standard shell model techniques, for the nuclides for which they are available, namely the main stable isotopes of Ge, Xe, Na, I, and F. In these cases, we define
\beq
\label{eq:FFO4}
F_\AV^2(q^2) = \dfrac{1}{3 a_p^2} \sum_{N,N'=p,n} a_N a_{N'} \left(F_{\Sigma''}^{(N,N')}(q^2) + F_{\Sigma'}^{(N,N')}(q^2) \right)
\eeq
for the AV interaction, and
\beq
\label{eq:FFO6}
F_\PS^2(q^2) = \dfrac{1}{a_p^2} \sum_{N,N'=p,n} a_N a_{N'} F_{\Sigma''}^{(N,N')}(q^2) 
\eeq
for the PS interaction. The (squared) nuclear form factors $F_{\Sigma'}$ and $F_{\Sigma''}$ are tabulated in \cite{Fitzpatrick:2012ix} for the nuclides mentioned above. These form factors can be employed unmodified also for inelastic scattering \cite{Barello:2014uda}. We include a factor of $1/3$ in the definition of $F_\AV^2$ in order to normalize $F_\AV^2$ and $F_\PS^2$ to be $1$ in the limit of zero momentum transfer when the target is an isolated proton. This factor traces back to $F_{\Sigma'}$ being twice as large as $F_{\Sigma''}$ at $q = 0$, which is consistent with the fact that $F_{\Sigma''}$ corresponds to the component of the nucleon spin along the direction of the momentum transfer, while $F_{\Sigma'}$ corresponds to the transverse component.

$F_\AV^2(q^2)$ can be expressed (see Eqs.~(59), (60) and (77c) in Ref.~\cite{Fitzpatrick:2012ix}) in terms of the usual nuclear spin structure function $S(q^2) = a_0^2 S_{00}(q^2) + a_0a_1 S_{01}(q^2) + a_1^2 S_{11}(q^2)$ \cite{Engel:1992bf} (with $a_0=a_p+a_n$ and $a_1=a_p-a_n$ the isoscalar and isovector parameters):
\beq
F_\AV^2(q^2) = \dfrac{4\pi}{3(2J_T+1)} \dfrac{1}{a_p^2} S(q^2) \ ,
\eeq
with $J_T$ the spin of the target nucleus. At zero momentum transfer
\beq
S(0) = \frac{1}{\pi} \frac{(2J_T+1)(J_T+1)}{J_T} \left(a_p \langle S_p \rangle + a_n \langle S_n \rangle \right)^2 \ ,
\eeq
where $\langle S_p \rangle \equiv \langle J_T, M_T = J_T | S_p^z | J_T, M_T = J_T \rangle$, and where $S_p^z$ is the component of $\vec{S}_p \equiv \sum_{\text{protons}} \vec{s}_p$ along the $z$-axis \cite{Ressell:1993qm} ($\langle S_n \rangle$ is defined analogously). Notice that $\langle S_p \rangle$ and $\langle S_n \rangle$ are often denoted with boldface style in the literature, although they are not vector quantities. $F_\AV^2$ can then be expressed in terms of the usually called spin-dependent form factor $F^2_\text{SD}(q^2)=S(q^2)/S(0)$ as 
\beq
\label{eq:FFGaussianSD}
F_\AV^2(q^2) = \dfrac{4(J_T+1)}{3J_T} \left(\langle S_p \rangle + \dfrac{a_n}{a_p} \langle S_n \rangle \right)^2 F^2_\text{SD}(q^2) \ .
\eeq

For the nuclides for which no form factors have been computed in Ref.~\cite{Fitzpatrick:2012ix} (Cl, C and Cs), we define $F_\AV^2(q^2)$ by means of Eq.~\eqref{eq:FFGaussianSD}, with the spin-dependent form factor in Gaussian form
\beq
F^2_\text{SD}(q^2)= e^{- q^2 R^2 / 4} \ ;
\eeq
here we take $R = \left(0.92 A_T^{1/3} + 2.68 - 0.78 \sqrt{(A_T^{1/3} - 3.8)^2 + 0.2}\right)$ fm, with $A_T$ the mass number of the target nucleus \cite{Belanger:2008sj}. In this case we also assume $F_\PS^2=F_\AV^2$, which we expect to be approximately valid at low $q^2$.
For Cs, a component of KIMS's target material, we take $\langle S_p \rangle = -0.370$, $\langle S_n \rangle = 0.003$ \cite{Iachello:1990ut,Lee.:2007qn}. For SIMPLE, we use $\langle S_p \rangle=-0.051$, $\langle S_n \rangle = -0.0088$ for both ${^{35}}$Cl and ${^{37}}$Cl \cite{Ressell:1993qm}, and $\langle S_p \rangle=-0.026$, $\langle S_n \rangle = -0.155$ for ${^{13}}$C \cite{Engel:1989ix}.
Notice that there are large uncertainties in the hadronic matrix elements $\langle S_p \rangle$ and $\langle S_n \rangle$ and the nuclear form factors, which differ in different nuclear models (see \eg Fig.~1 of Ref.~\cite{Aprile:2013doa} for the Xe nuclear structure functions). Factors of $2$ difference in different calculations are not uncommon \cite{Bednyakov:2004xq}.

\section{Direct detection rate}\label{sec:direct detection rate}

The DM-nucleus scattering rate for a target nuclide $T$ with mass $m_T$ is
\begin{equation}
\label{R-ER}
\frac{\ud R_T}{\ud\ER}(\ER, t) =
\frac{\rho}{m} \int_{v \geqslant v_\text{min}(\ER)} \ud^3 v \, v f(\bsv, t) \frac{\ud \sigma_T}{\ud \ER}(\ER, \bsv) \ ,
\end{equation}
where $m$ is the WIMP mass, $\rho$ is the local DM density, $f(\bsv, t)$ is the DM velocity distribution in Earth's reference frame, and $\ud \sigma_T / \ud \ER$ is the differential cross section for a WIMP scattering off the target $T$. $v_\text{min}(\ER)$ is the minimum WIMP speed needed to impart to the target nucleus a recoil energy $\ER$.

The detectors do not measure directly the recoil energy, but they measure a related energy $E'$ (sometimes expressed in `keV electron equivalent' or keVee, or else in number of photoelectrons). The rate in Eq.~\eqref{R-ER} is related to the event rate measured by an experiment within a detected energy interval $[E'_1,E'_2]$ as
\begin{equation}
\label{R-E'}
R_{[E'_1,E'_2]}(t) = \sum_T \frac{C_T}{m_T} \int_0^\infty \ud \ER \,
\frac{\ud R_T}{\ud \ER}(\ER, t)
 \int_{E'_1}^{E'_2} \ud E' \, \epsilon(E',\ER) G_T(\ER, E') \ ,
\end{equation}
where $C_T$ is the mass fraction of the nuclide $T$ in the detector, $G_T(\ER,E')$ is the target-dependent resolution function of the detector and $\epsilon(E',\ER)$ is the experimental acceptance. The resolution function is defined as the probability distribution for a nuclear recoil energy $\ER$ to be measured as $E'$, and incorporates the mean value $\langle E' \rangle = Q_T(\ER) \ER$, with $Q_T$ the target's quenching factor, and the detector energy resolution. The detector acceptance is a function of both $\ER$ and $E'$ in general, but it is often given to be $\ER$ independent, or the dependencies on $\ER$ and $E'$ are factorizable.

Except for our halo-independent analysis, in this paper we assume the Standard Halo Model (SHM) for the dark halo of our galaxy, where the DM local density is $\rho=0.3$ $\text{GeV}/\text{cm}^3$ and the velocity distribution of WIMPs in the galactic frame is a truncated Maxwell-Boltzmann distribution
\begin{equation}
f_\text{G}({\bsu})=\frac{1}{N_{\rm esc}(v_0\sqrt{\pi})^3}\exp(-u^2/v_0^2) \, \theta(v_{\rm esc}-u).
\end{equation}
Here $v_0$ is the velocity dispersion and $v_{\rm esc}$ is the escape speed from our galaxy. The normalization factor
\begin{equation}
N_{\rm esc} \equiv {\rm erf}(v_{\rm esc}/v_0)-2(v_{\rm esc}/v_0) \exp(-v_{\rm esc}^2/v_0^2)/\sqrt{\pi}
\end{equation}
ensures that $\int d^3 u \, f_\text{G}({\bsu}) = 1$. We consider $v_0$ to be the same as the velocity of the Local Standard of Rest $v_0=220$ km/s, and we take $v_{\rm esc}=533$ km/s, according to recent Radial Velocity Experiment (RAVE) 2013 results \cite{Piffl:2013mla}. The velocity distribution in Earth's frame, $f({\bsv}, t)$ in Eq.~\eqref{R-ER}, can be obtained with the Galilean transformation
\begin{equation}\label{velocity distribution}
f ({\bsv},t)=f_\text{G}({\bsv}_\odot+{\bsv}_\oplus(t)+{\bsv}) \ ,
\end{equation}
where ${\bsv}_\odot$ and ${\bsv}_\oplus(t)$ are the velocity of the Sun with respect to the galaxy and the time dependent velocity of Earth with respect to the Sun, respectively. We take $v_\odot=232$ km/s, and $v_\oplus=30$ km/s in an orbit inclined at $60^\circ$ with respect to the galactic plane \cite{Schoenrich:2009bx}.

\begin{figure}[t!]
\centering
\includegraphics[width=0.49\textwidth]{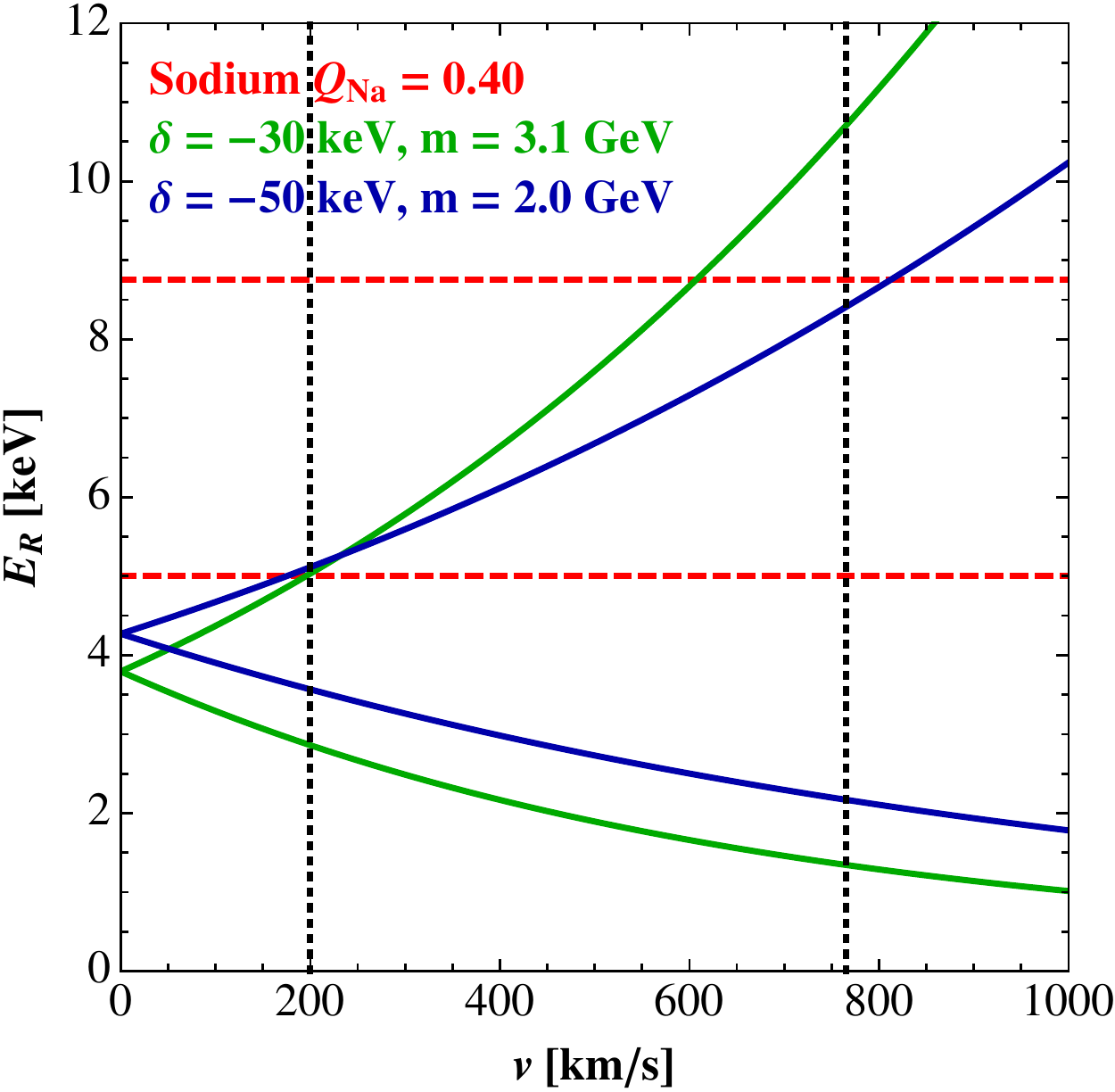}
\includegraphics[width=0.49\textwidth]{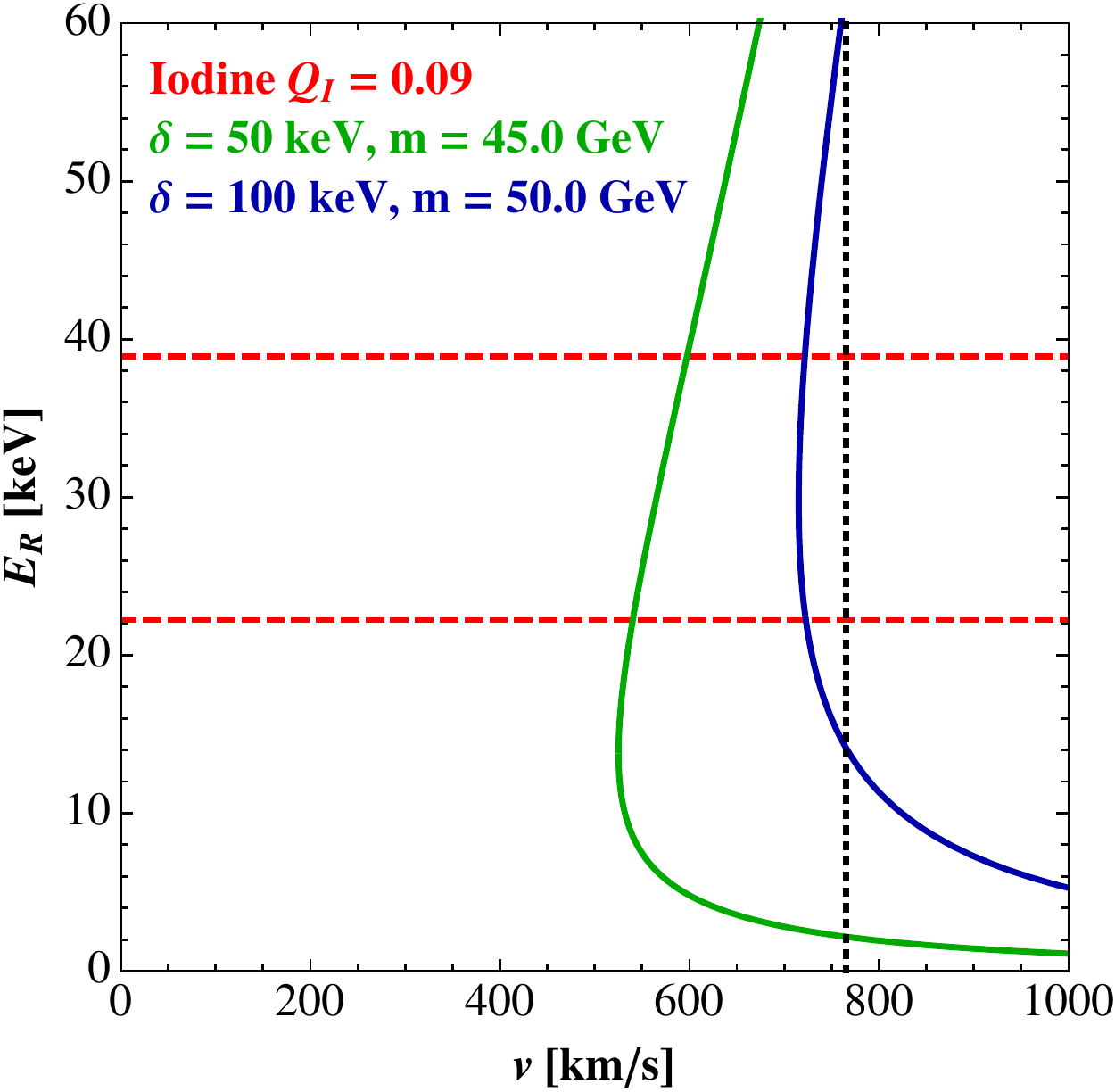}
\caption{\label{fig:ER_vs_vmin}
Recoil energy range for exothermic scattering off Na (left) and endothermic scattering off I (right), as a function of the WIMP speed $v$, for the indicated values of WIMP mass $m$ and mass splitting $\delta$. The two red horizontal lines enclose the $2.0$--$3.5$ keVee energy interval, in which most of the DAMA signal is observed. Here $Q_\text{Na}=0.40$ and $Q_\text{I}=0.09$. The DAMA events can only be between the two vertical lines at $v_\text{min} = 200$ km/s and $v_\text{min}=v_\text{max} = 765$ km/s (see the text).}
\end{figure}

Earth's revolution around the Sun causes the velocity distribution given in Eq.~\eqref{velocity distribution}, and therefore the scattering rate in Eq.~\eqref{R-ER}, to modulate in time. Both the time-average rate and the modulation amplitude of the rate can be measured. In the SHM, the velocity of Earth with respect to the Galaxy is maximum at the end of May or the beginning of June. The rate has a maximum at this moment if $v_\text{min} > 200$ km/s  and it has instead  a minimum at this moment if $v_\text{min}< 200$ km/s. Thus, choosing the modulation phase so that the modulation amplitude is positive for $v_\text{min} > 200$ km/s, the amplitude is negative for $v_\text{min} < 200$ km/s. Only $v_\text{min} > 200$ km/s  are compatible  with the phase of the DAMA modulation data such that the rate has the maximum on the 2nd of June \cite{Bernabei:2013xsa}. The value of $v_\text{min} = 200$ km/s is shown as a vertical line in Fig.~\ref{fig:ER_vs_vmin}.

The minimum speed the DM particle must have in the rest frame of the target nuclide in order to impart a nuclear recoil energy $\ER$ for $\mu_T|\delta|/m^2 \ll 1$ is
\beq
\label{eq:vmin_vs_ER inel}
\vmin = \dfrac{1}{\sqrt{2m_T\ER}}\left|\dfrac{m_T\ER}{\mu_T}+\delta\right| ,
\eeq
where $\mu_T$ is the DM-nucleus reduced mass. The mass splitting $\delta=m'-m$ can be either positive for endothermic scattering~\cite{TuckerSmith:2001hy}, negative for exothermic scattering~\cite{Finkbeiner:2007kk,Batell:2009vb,Graham:2010ca}, or zero for elastic scattering. By inverting this equation one obtains the minimum and maximum recoil energies $\ER^\pm(v)$ that are kinematically allowed for a fixed DM speed $v$,
\beq\label{eq:ER_vs_v}
\ER^\pm(v) =
\frac{\mu_T^2 v^2}{2 m_T} \left( 1 \pm \sqrt{1 - \frac{2 \delta}{\mu_T v^2}} \right)^2 \ .
\eeq
The left panel in Fig.~\ref{fig:ER_vs_vmin} shows the two $\ER^\pm(v)$ branches for the Na component of DAMA, for $\delta=-30$ and $-50\text{ keV}$, for WIMP masses that correspond to the best fit regions we will present later. The right panel shows the $\ER^\pm(v)$ branches for the I component of DAMA, for $\delta=50$ and $100$ keV. 

The minimum possible value of $v$ for the interaction to be kinematically allowed is $v^T_\delta=\sqrt{2\delta/\mu_T}$ for endothermic scattering, and $v^T_\delta = 0$ for exothermic. This speed value corresponds to the point of intersection of the two $\ER^\pm$ branches, which occurs at $E_\delta=\ER^+(v^T_\delta)=\ER^-(v^T_\delta)=\mu_T|\delta|/m_T$. The maximum value of the DM speed allowed for a given halo model is the sum of the escape speed $\vesc$ and the modulus of Earth's velocity in the galactic rest frame, which is equal on average to the Sun's velocity $\vec{v}_\odot$; therefore, for our choice of parameter values, $v_\text{max}=765$ km/s. This is indicated with a vertical line in Fig.~\ref{fig:ER_vs_vmin}. As a result, endothermic scattering with a target nucleus will be kinematically forbidden for $\delta$ larger than $\simeq 3.3 \text{ keV} (\mu_T/\text{GeV})$. In the ranges for $m$ and $\delta$ corresponding to the best fit regions presented later, endothermic scattering with Na in DAMA is kinematically forbidden. On the other hand, scattering off I in DAMA is allowed for endothermic scattering, as can be seen in the right panel of Fig.~\ref{fig:ER_vs_vmin} for $\delta=50$ and $100\text{ keV}$. 

In the following sections we examine the compatibility of the WIMP interpretation of the DAMA annual modulation signal with various null results for the AV and PS models described above. We consider both elastic, endothermic, and exothermic scattering.

\section{Data analysis assuming the SHM}\label{sec:data analysis}

In this section we describe the data analysis we perform assuming the SHM, which follows the procedure already presented in Refs.~\cite{DelNobile:2014eta, DelNobile:2013gba}.

For the DAMA annual modulation signal, we take the data plotted in Fig.~8 of Ref.~\cite{Bernabei:2013xsa}. We determine the DAMA favored regions in the DM parameter space by performing a Maximum Likelihood analysis, assuming the data are Gaussian distributed. Due to the uncertainties residing in the quenching factors of Na and I, which play an important role in the analysis, we choose two values for each target, namely $Q_{\rm Na} = 0.40$ and $0.30$ for Na, and $Q_{\rm I}=0.09$ and $0.06$ \cite{Bernabei:1996vj} for I (see \eg Ref.~\cite{Collar:2013gu} and references therein). In the analysis we adopt the combinations $Q_{\rm Na} = 0.30$ with $Q_{\rm I} = 0.06$, and $Q_{\rm Na} = 0.40$ with $Q_{\rm I} = 0.09$.

We also compute an upper limit on the WIMP cross section using the total rate measured by DAMA, employing the data points plotted in Fig.~1 of Ref.~\cite{Bernabei:2008yi}. We restrict our analysis to energies above the experimental threshold of  $2$ keVee. Given the very large number of observed events in each bin, and the resulting small statistical fluctuations, we compute an upper bound on the cross section by requiring that the predicted rate does not exceed the observed rate in any energy bin. This limit is particularly important for exothermic scattering, which reduces the modulation amplitude with respect to the average rate.

To compute the LUX bound, following Ref.~\cite{DelNobile:2013gba}, we apply the Maximum Gap method \cite{Yellin:2002xd} to the variable $S_1$ in the range $2$--$30$ photoelectrons. We choose several numbers of observed events, i.e.~0, 1, 3, 5 and 24 as described in Ref.~\cite{DelNobile:2013gba}. In our SHM model the maximum WIMP speed is $v_\text{max}=765$ km/s, thus the maximum recoil energy for a WIMP lighter than $11.5$ GeV in an elastic scattering with Xe is $\sim12$ keV. Using the approximated recoil energy contours in Fig.~4 of Ref.~\cite{Akerib:2013tjd}, and dropping all observed events in and above the electron-recoil band (plotted at 1.28$\sigma$) in the same figure, only five observed events remain below $\sim12$ keV. This means that for elastic scattering, choosing $5$ events provides a safe upper limit for WIMP masses $m<11.5$ GeV. However, if $m>11.5$ GeV, only using all the 24 events lying outside the electron-recoil band provides a reliable upper limit. For inelastic scattering with $\delta=-50$, $-30$, $50$ and $100$ keV, the maximum WIMP masses for which using the $5$ events bound is reliable are $8.2$, $9.3$, $19.2$ and $56.2$ GeV, respectively. Since our procedure does not depend on the WIMP distribution in the $S_1$--$\text{log}_{10}(S_2/S_1)$ plane \cite{Akerib:2013tjd}, our Maximum Gap upper limits are conservative and safe to be applied to any WIMP-nucleus interactions.

For the SuperCDMS bounds, we use the data set collected by the seven Ge detectors between October 2012 and June 2013, corresponding to an effective exposure of $577$ kg-days. We use the Maximum Gap method with the eleven observed events listed in Table 1 of Ref.~\cite{Agnese:2014aze}, which passed the selection criteria introduced by the collaboration to discriminate signal from background events. We do not incorporate the uncertainty in recoil energy in our analysis. For the detector acceptance we take the red curve in Fig.~1 of Ref.~\cite{Agnese:2014aze}.

To compute the SIMPLE limits, we consider only the Stage 2 \cite{Felizardo:2011uw}, a C$_2$ClF$_5$ detector with an exposure of $6.71$ kg-day, with one observed event above 8 keV compatible with the expected background of $2.2$ events. We use the Feldman-Cousins method \cite{Feldman:1997qc} to place a $90\%$ CL upper limit of $2.39$ signal events for $2.2$ expected background events and 1 observed event.

For PICASSO we perform a Maximum Likelihood analysis using the data in Fig.~5 of Ref.~\cite{Archambault:2012pm}. The target material in PICASSO is C$_4$F$_{10}$, but the collaboration only considers scattering off F in their analysis \cite{Archambault:2012pm}; we do the same, noting that the contribution of C for DM spin-dependent interactions with protons is anyway negligible. We construct our Gaussian likelihood using the expected rate above each one of the eight energy thresholds adopted by the collaboration ($1.7$, $2.9$, $4.1$, $5.8$, $6.9$, $16.3$, $38.8$, and $54.8$ keV), and the measured rate with its uncertainty, which are already background subtracted.

For KIMS we perform again a Maximum Likelihood analysis using the data points with their $68\%$ CL intervals from Fig.~4 of Ref.~\cite{Kim:2012rza}, assuming Gaussian distributed data. Because Cs and I have similar atomic masses, their quenching factors are not measured separately \cite{KimKimKim}. As for DAMA, we perform our analysis of the KIMS data adopting two values for $Q_{\rm I}=Q_{\rm Cs}$ in CsI: $0.05$ and $0.10$ (see Fig.~2 of \cite{KimKimKim}, and Fig.~5 of \cite{Collar:2014lya} and references therein).

\section{Results assuming the SHM}\label{sec:data comparison}

\begin{figure}[h!]
\centering
\includegraphics[width=0.49\textwidth]{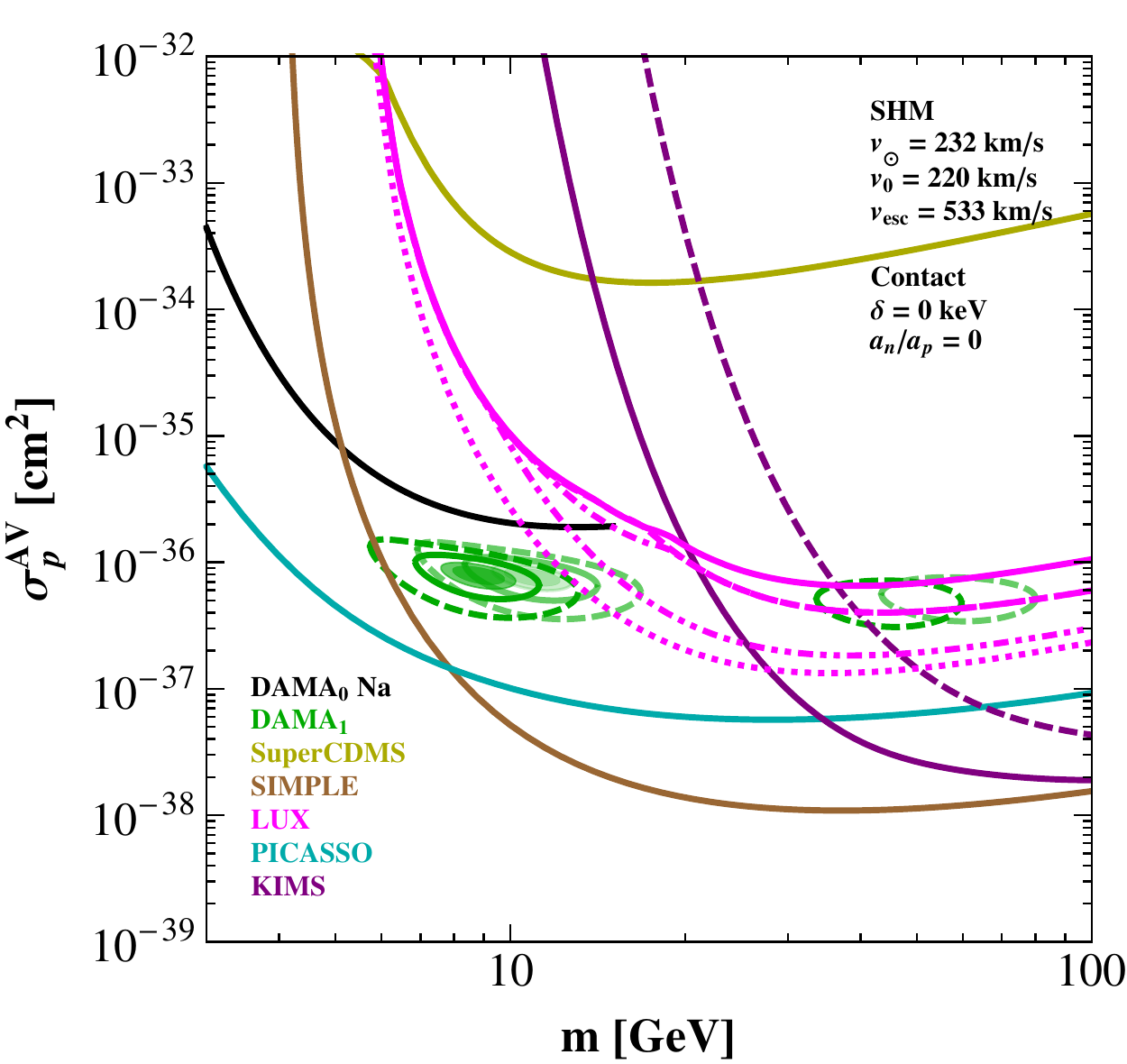} \\
\caption{\label{fig:AVel}
$90\%$ CL bounds and $68\%$ CL, $90\%$ CL, $3 \sigma$, and $5 \sigma$ allowed regions in the WIMP-proton reference cross section $\sigma_p$ vs WIMP mass plane, assuming the SHM, for elastic proton-only contact AV interactions. The unmodulated DAMA rate limit (black) corresponds to a Na quenching factor of $0.40$. Different line styles for the LUX bound correspond, from most to least constraining, to $0$, $1$, $3$, $5$ and $24$ observed events (see text). The KIMS bound is shown for both $Q_{\rm I} = Q_{\rm Cs} = 0.10$ (solid line) and $0.05$ (dashed line).}
\end{figure}

The plots in Figs.~\ref{fig:AVel}--\ref{fig:PSel_otherSHM}, \ref{fig:LongRange_el}--\ref{fig:PSexo}, and \ref{fig:LongRange_exo}--\ref{fig:LongRange_endo} show $90\%$ CL upper bounds and $68\%$ CL (inner and darker shaded region), $90\%$ CL (outer and lighter shaded region), $3 \sigma$ (solid contour) and $5 \sigma$ (dashed contour) allowed regions in the $m$--$\sigma_{p}$ plane. The green shaded regions and green closed contours labeled `$\text{DAMA}_1$' are the allowed regions compatible with the DAMA annual modulation, for quenching factors $Q_\text{Na}=0.40$ and $Q_\text{I}=0.09$ in dark green, and $Q_\text{Na}=0.30$ and $Q_\text{I}=0.06$ in light green. The lower the quenching factor, the higher is the DM mass needed to fit the data. The low and high WIMP mass regions correspond to the interpretation of the DAMA data as the WIMP scattering mostly off Na and I in the detector, respectively. The upper limit due to the DAMA total rate (black, and labeled `$\text{DAMA}_0\:\text{Na}$') is shown for scattering off Na assuming $Q_\text{Na}=0.40$. $90\%$ CL upper limits from LUX data are shown as various magenta curves. As in Ref.~\cite{DelNobile:2013gba} the different dashing styles of the lines indicate different selections of candidate events used in the Maximum Gap analysis: dotted (0 events), double-dot-dashed (1 event), dot-dashed (3 events), dashed (5 events) and solid (24 events) curves. Two purple lines show the 90\% CL upper limits from KIMS data with quenching factors $Q_{\rm I}=Q_{\rm Cs}= 0.10$ (solid) and $0.05$ (dashed). $90\%$ CL upper limit from SIMPLE (brown), PICASSO (cyan), and SuperCDMS (dark yellow) are also drawn. 

\subsection{Elastic contact interactions}\label{sec:elastic_contact}

Fig.~\ref{fig:AVel} shows our results for elastic proton-only contact AV interactions. The most stringent bounds come from two bubble chamber experiments, SIMPLE and PICASSO. Both of these limits exclude all the regions favored by the DAMA modulation signal.

\begin{figure}[t!]
\centering
\includegraphics[width=0.49\textwidth]{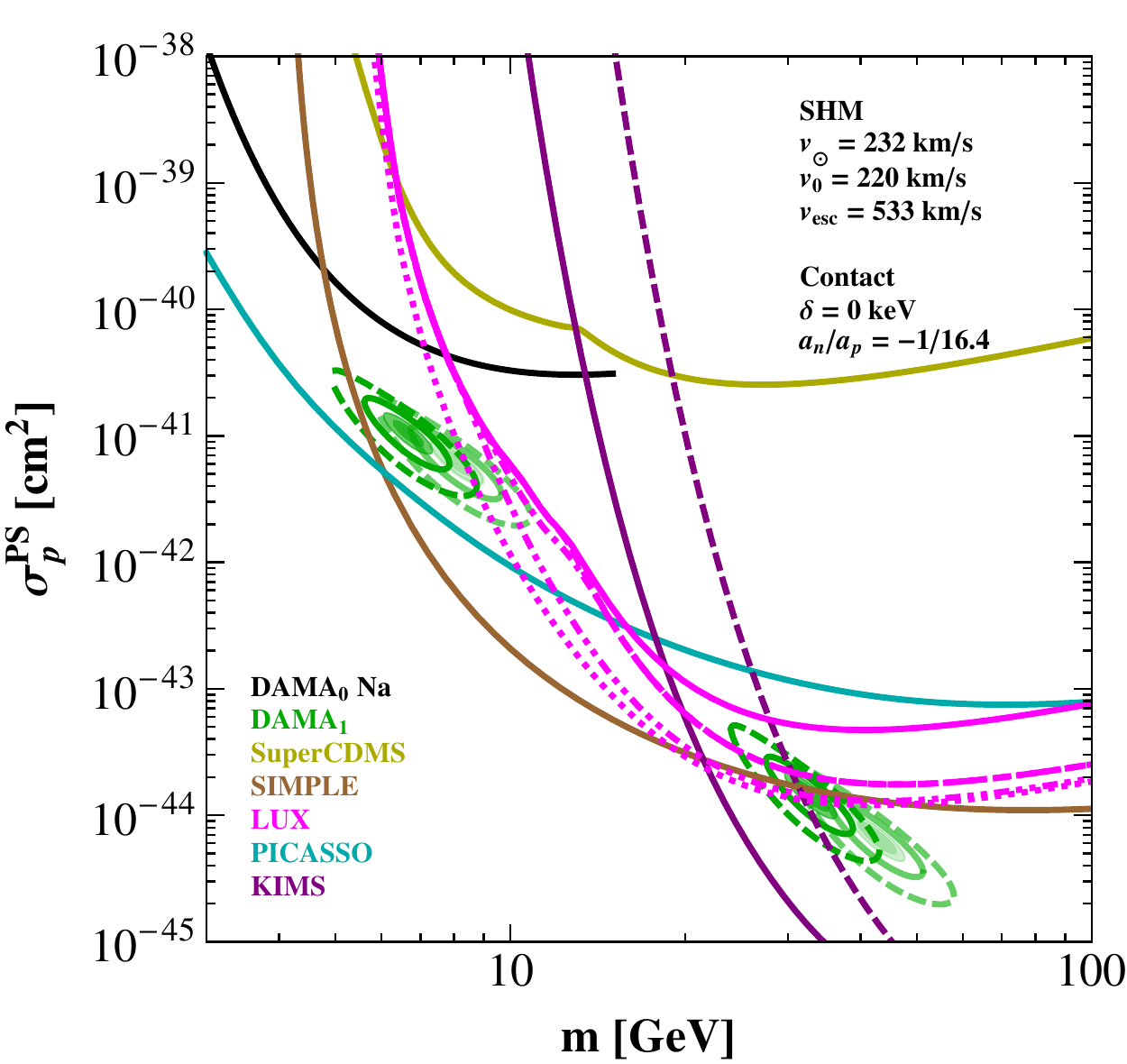}
\includegraphics[width=0.49\textwidth]{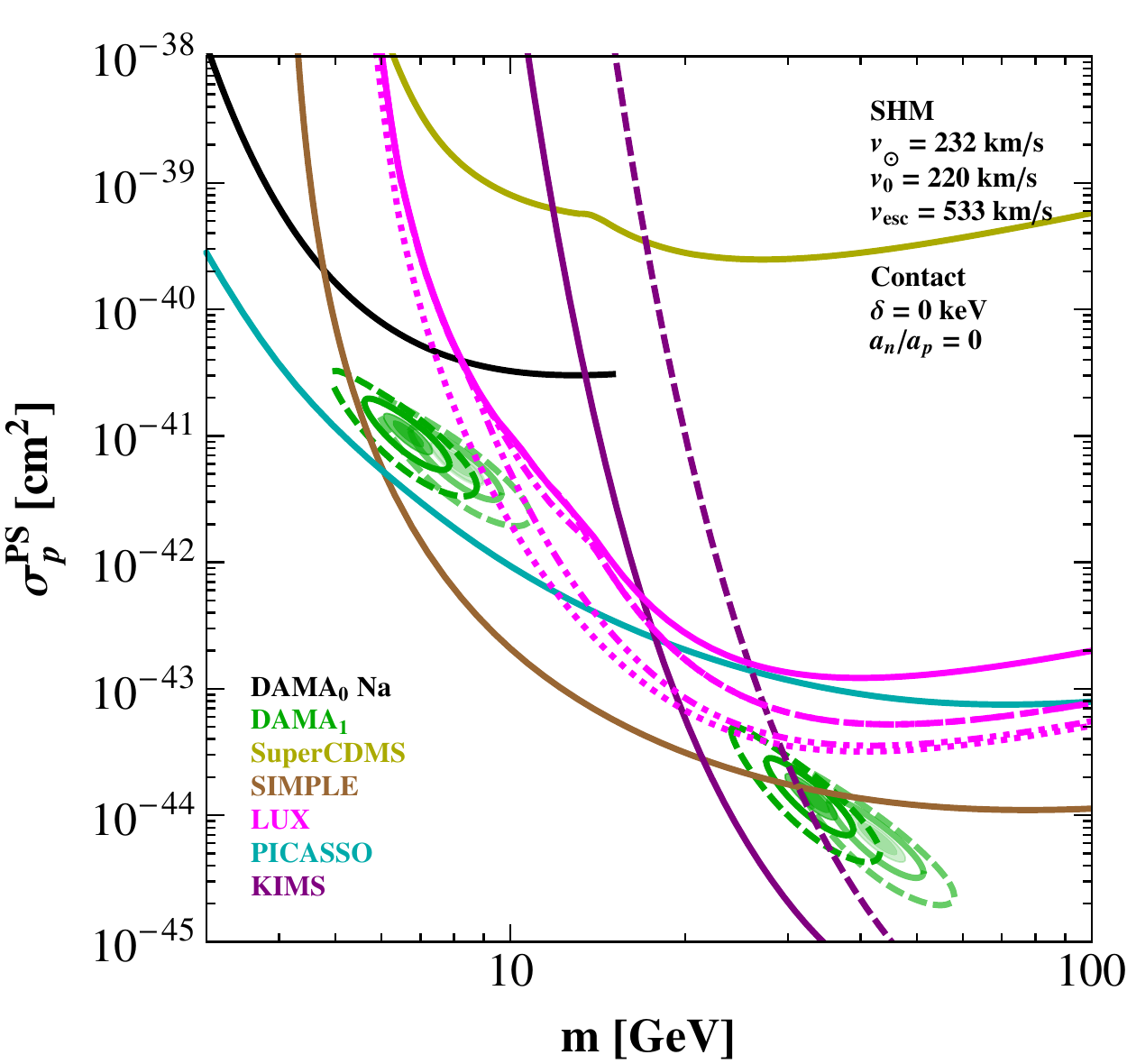}
\caption{\label{fig:PSel}
Same as Fig.~\ref{fig:AVel} but for flavor-universal (left) and proton-only (right) PS interactions.}
\end{figure}

Fig.~\ref{fig:PSel} is the same as Fig.~\ref{fig:AVel}, but for PS interactions with flavor-universal coupling $a_n/a_p=-1/16.4$ (left panel), and with proton-only coupling $a_n=0$ (right panel). As expected, the only limits that change from one case to the other are those of LUX and SuperCDMS, due to their enhanced sensitivity to DM-neutron couplings. The DAMA regions for WIMP scattering off Na are entirely excluded by SIMPLE and PICASSO, and the regions for scattering off I are excluded by KIMS when assuming similar values for the I quenching factor in both experiments. This result is different from what was found in Ref.~\cite{Arina:2014yna}, where some portion of the Na and I DAMA regions are compatible with all null experiments for the PS flavor-universal coupling.

\begin{figure}[t!]
\centering
\includegraphics[width=0.49\textwidth]{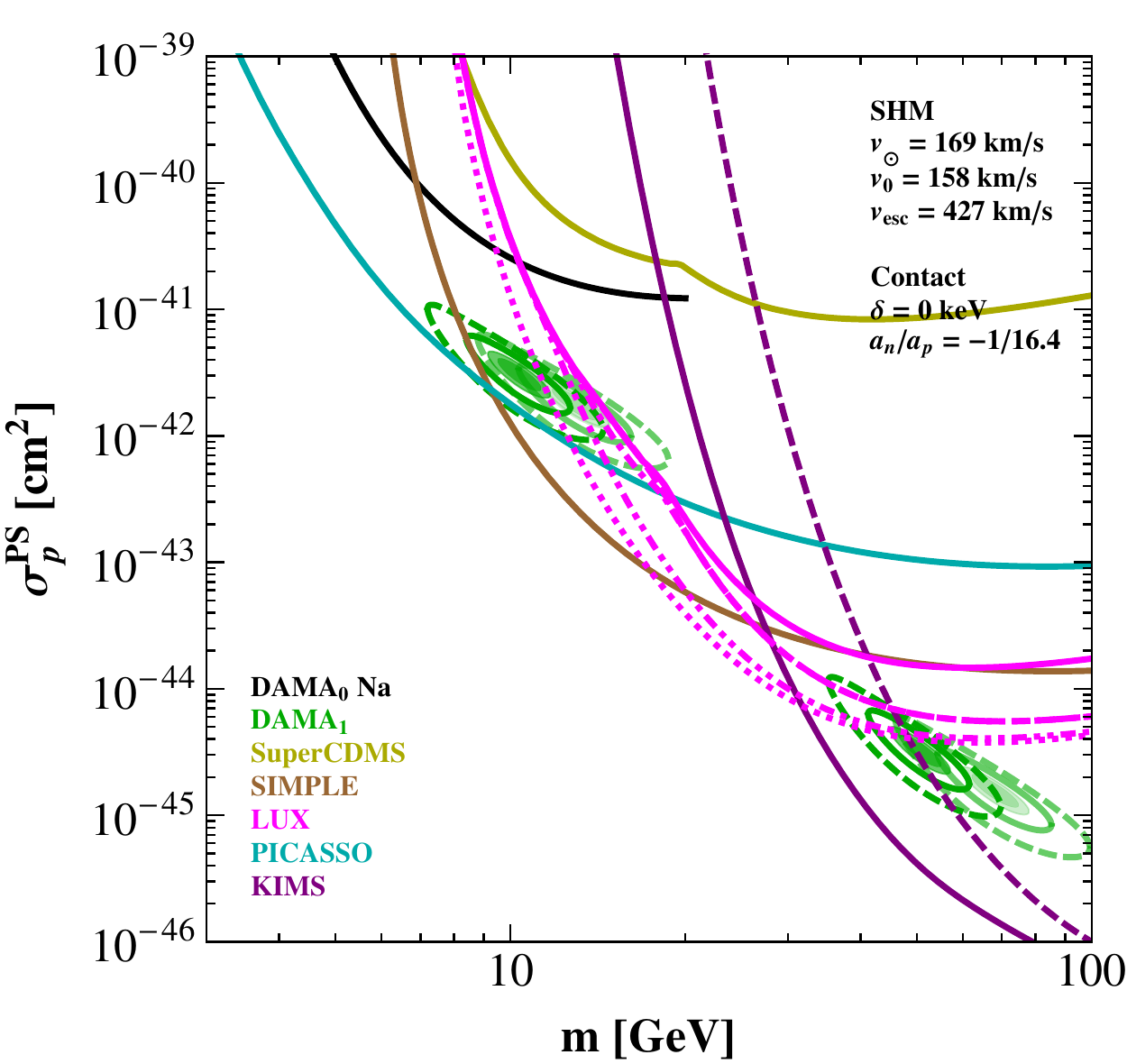}
\includegraphics[width=0.49\textwidth]{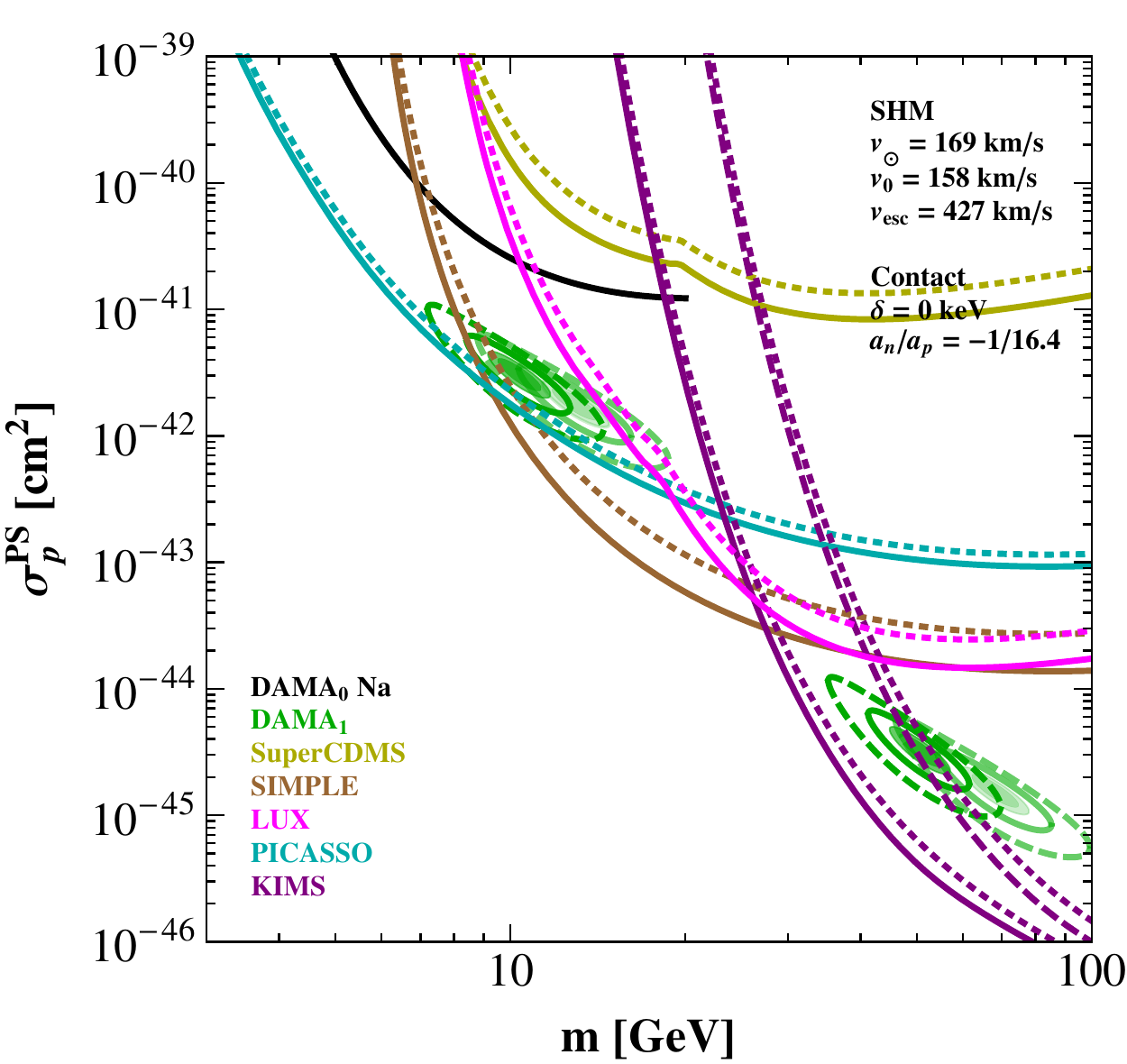}
\caption{\label{fig:PSel_otherSHM}
(left) Same as Fig.~\ref{fig:PSel} but for $v_0$ and $\vesc$ taken to be $3\sigma$ lower than the central values assumed in Ref.~\cite{Arina:2014yna}. (right) Same as the left panel, but showing in addition $99\%$ CL upper bounds (dotted lines) (only the LUX upper bound for $24$ events is presented here).
}
\end{figure}

\begin{figure}[t!]
\centering
\includegraphics[width=0.47\textwidth]{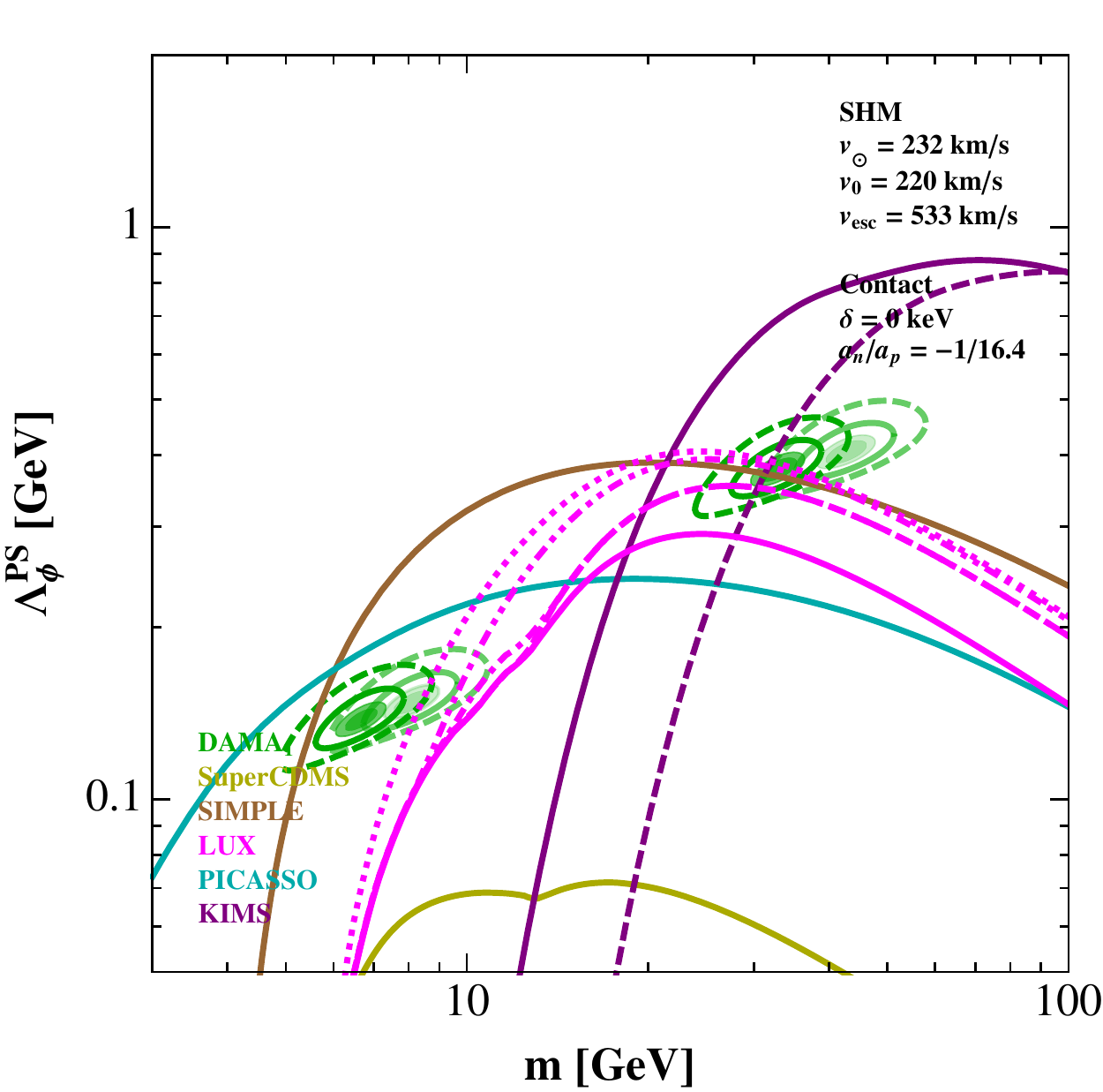}
\includegraphics[width=0.47\textwidth]{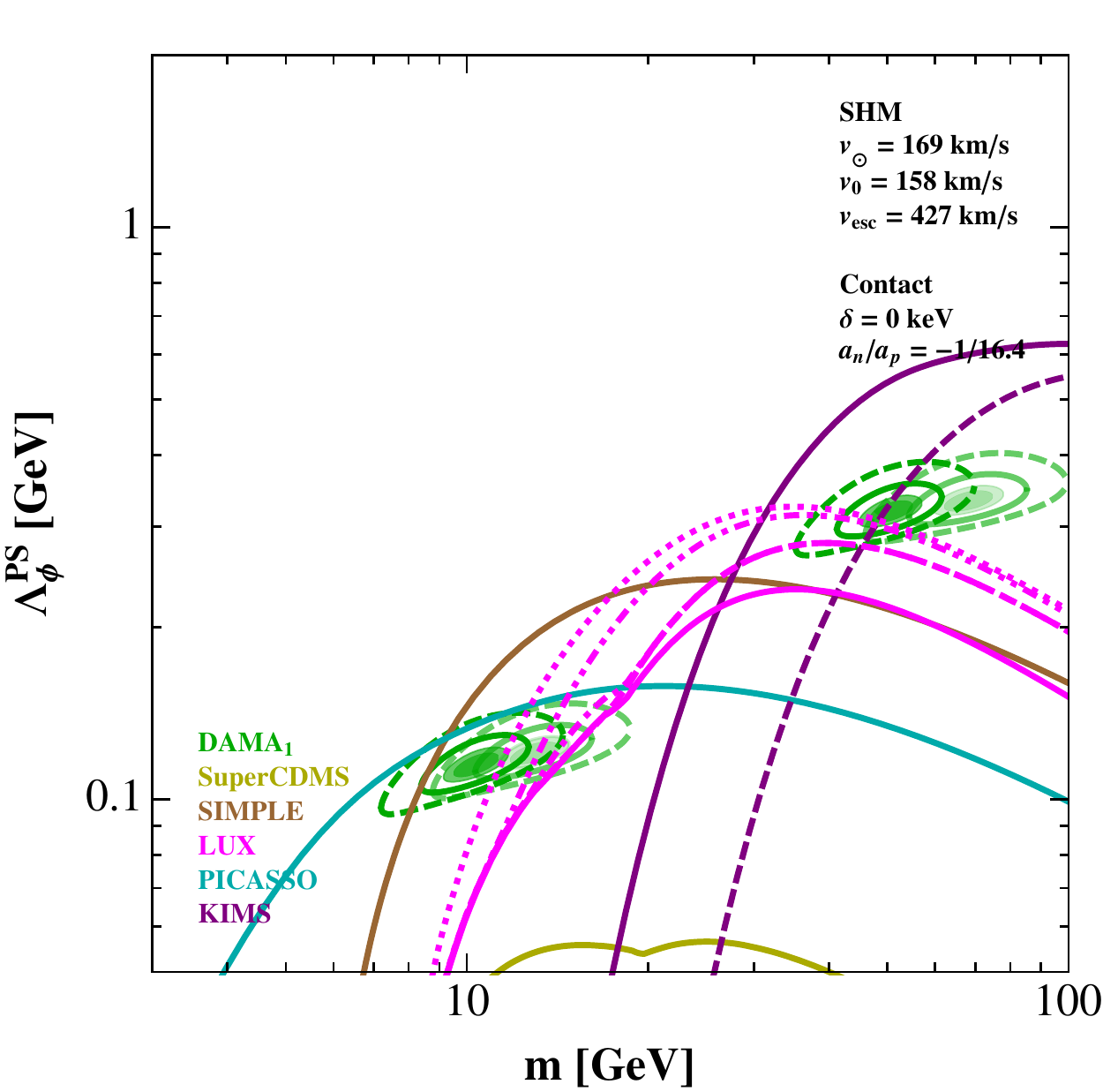}
\caption{\label{fig:PSel_ArinaAxis}
Same as the left panels in Figs.~\ref{fig:PSel} and \ref{fig:PSel_otherSHM}, but plotted in the $\Lambda_\phi \equiv m_\phi / \sqrt{g_\text{DM} g}$ vs WIMP mass $m$ plane as in Fig. 1 of Ref.~\cite{Arina:2014yna}.}
\end{figure}

Ref.~\cite{Arina:2014yna} uses Bayesian statistics to infer $99\%$ credible level exclusion limits, and $90\%$ and $99\%$ credible regions for DAMA, marginalizing over the SHM parameters using Gaussian priors (taking central values for the velocities $\overline{v_0} = 230$ km/s and $\overline{v_\text{esc}}=544$ km/s, and for the local WIMP density $\overline{\rho} = 0.3\text{ GeV/cm}^3$, with standard deviations $\Delta v_0=24.4$ km/s, $\Delta v_\text{esc}=39$ km/s and $\Delta\rho = 0.13\text{ GeV/cm}^3$). As a result, regions and limits at a specific point in parameter space do not necessarily correspond to a fixed set of values for the SHM parameters. In our analysis instead we assumed the same set of SHM parameter values across all experimental results. We found that the regions and limits move approximately in the same manner in the parameter space as we vary the DM velocities, and the DAMA regions fail to escape the upper bounds at the $90\%$ CL. This can be seen in Fig.~\ref{fig:PSel_otherSHM} (left panel), which shows the results for elastic PS interactions with flavor-universal coupling where both $v_\text{esc}$ and $v_0$ are taken $3\sigma$ below their central values in Ref.~\cite{Arina:2014yna} (however, we keep $v^\text{ref}=220$ km/s to plot $\sigma_p$). Our choice for the SHM velocities roughly matches the low mass SIMPLE limit in Fig.~1 of Ref.~\cite{Arina:2014yna}. The right panel of Fig.~\ref{fig:PSel_otherSHM} shows also the $99\%$ CL upper bounds (dotted lines) for the same set of parameters. In this case, the high mass DAMA region corresponding to a quenching factor of $0.09$ escapes the KIMS upper limit for quenching factor $0.05$. The Na component of the DAMA region is still rejected by PICASSO at the $99\%$ CL. In Fig.~\ref{fig:PSel_ArinaAxis} we present our results from the left panels of Figs.~\ref{fig:PSel} and \ref{fig:PSel_otherSHM} in the same plane as Fig.~1 of Ref.~\cite{Arina:2014yna}, namely in the $m$--$\Lambda_\phi$ plane, where $\Lambda_\phi \equiv m_\phi/\sqrt{g_\text{DM}g}$. The allowed DAMA regions shown in Ref.~\cite{Arina:2014yna} are much larger than the regions we found. We believe that this is due to their marginalization over the SHM parameters and experimental parameters including quenching factors.

Ref.~\cite{Dolan:2014ska} found for PS interactions that the LUX bound excludes the DAMA region (see Fig.~9 in Ref.~\cite{Dolan:2014ska}, where the I region in DAMA is completely excluded by LUX in both the contact and long-range limits). This is in disagreement with our conclusions, possibly because of the different analysis of the LUX data.

\subsection{Elastic long-range interactions}

\begin{figure}[t!]
\centering
\includegraphics[width=0.49\textwidth]{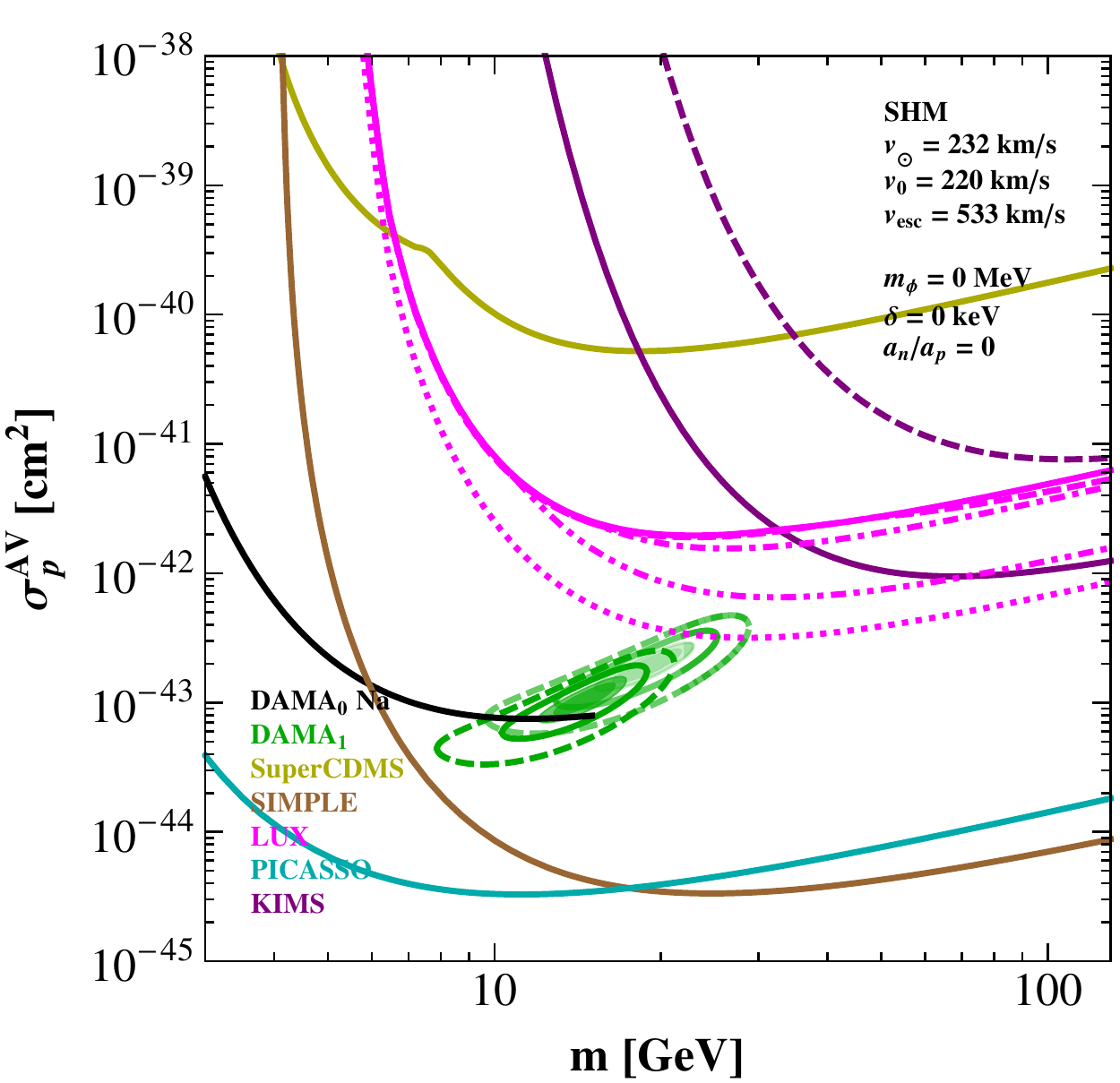}
\includegraphics[width=0.49\textwidth]{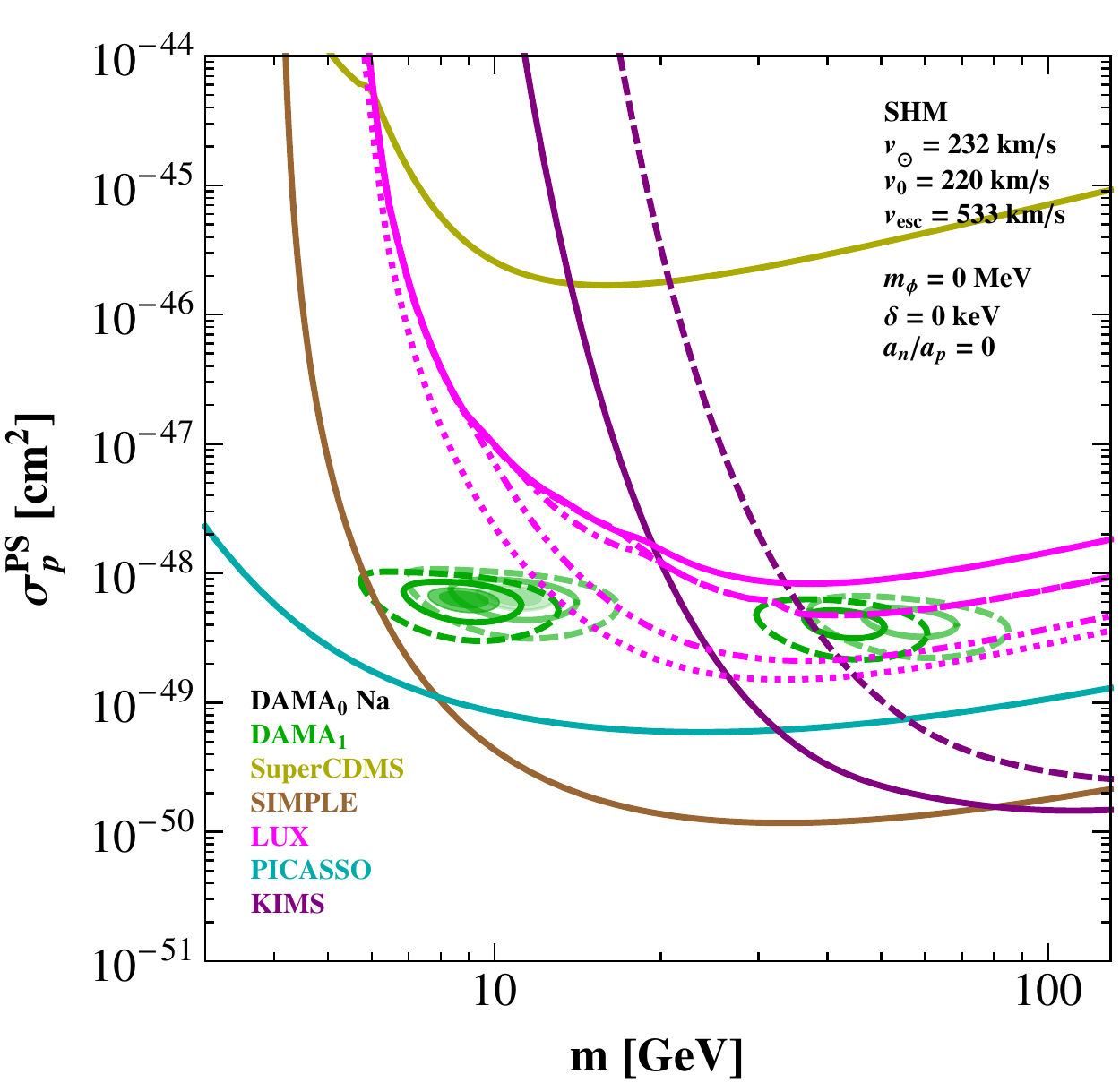}
\caption{\label{fig:LongRange_el}
Same as Fig.~\ref{fig:AVel} but for proton-only elastic AV (left) and PS (right) interactions via a massless mediator.}
\end{figure}

Fig.~\ref{fig:LongRange_el} shows the regions and limits for elastic AV (left panel) and PS (right panel) interactions via a massless mediator. The results are shown in the reference cross section $\sigma_p^\rf$ vs DM mass $m$ plane, where $\sigma_p^\rf=\sigma_p(m_\phi=m_\phi^\rf)$ with $m_\phi^\rf=1$ GeV. Note that the results for the contact AV and long-range PS interactions are very similar up to a shift in the vertical direction (compare Fig.~\ref{fig:AVel} and the right panel of Fig.~\ref{fig:LongRange_el}). This is expected from the $\ER$ dependence of the differential cross sections given in Eqs.~\eqref{eq:AV_diff_cross_section} and \eqref{eq:PS_diff_cross_section_ref}: disregarding the form factors, the differential cross section for the long-range PS and contact AV interactions is proportional to $\ER^0$, for contact PS it is proportional to $\ER^2$, and for long-range AV it is proportional to $\ER^{-2}$. As it can be seen in Fig.~\ref{fig:LongRange_el}, considering long-range elastic interactions does not help to bring compatibility between the DAMA regions and the upper limits from the experiments with null results.

\begin{figure}[h!]
\centering
\includegraphics[width=0.49\textwidth]{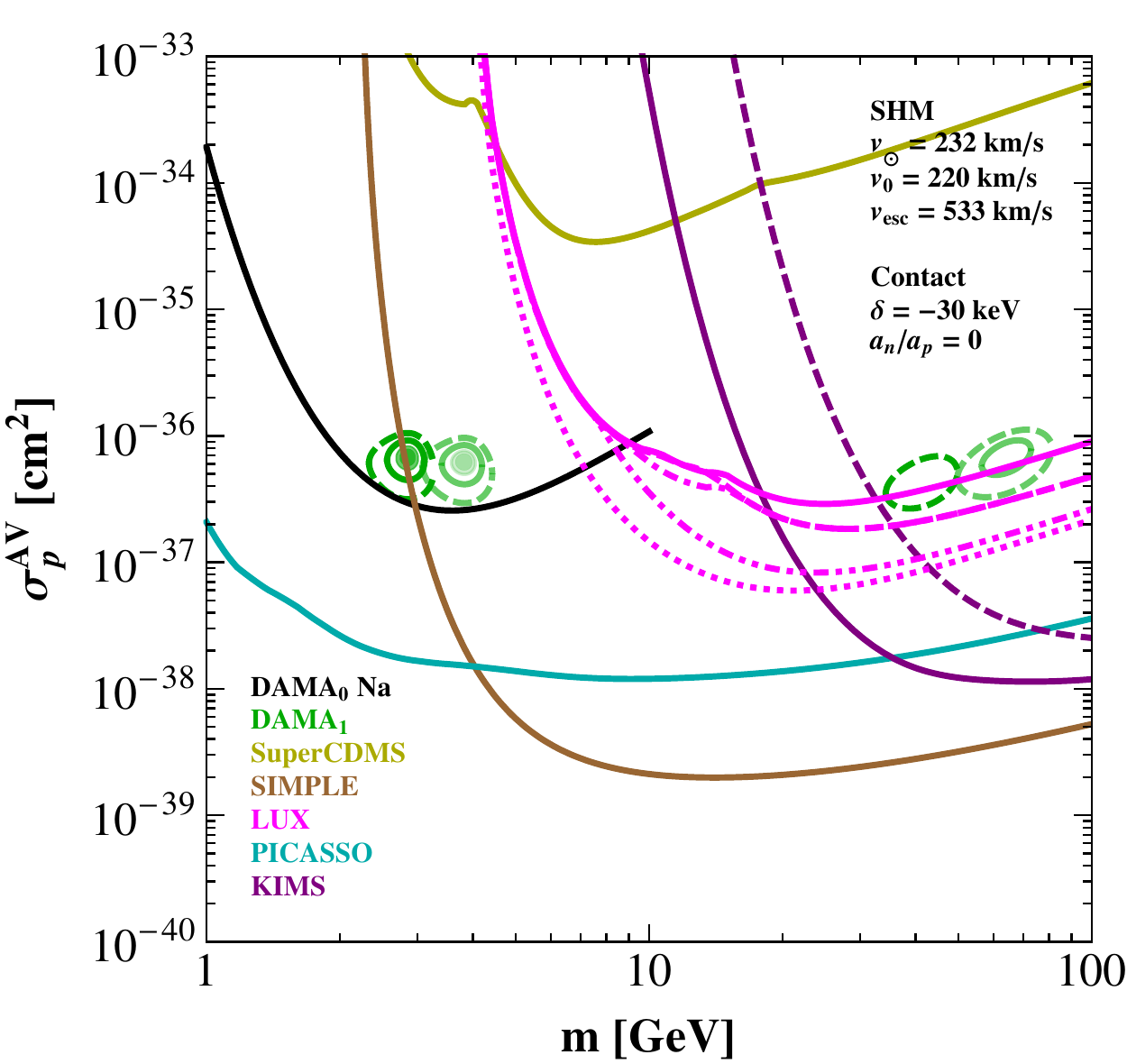}
\includegraphics[width=0.49\textwidth]{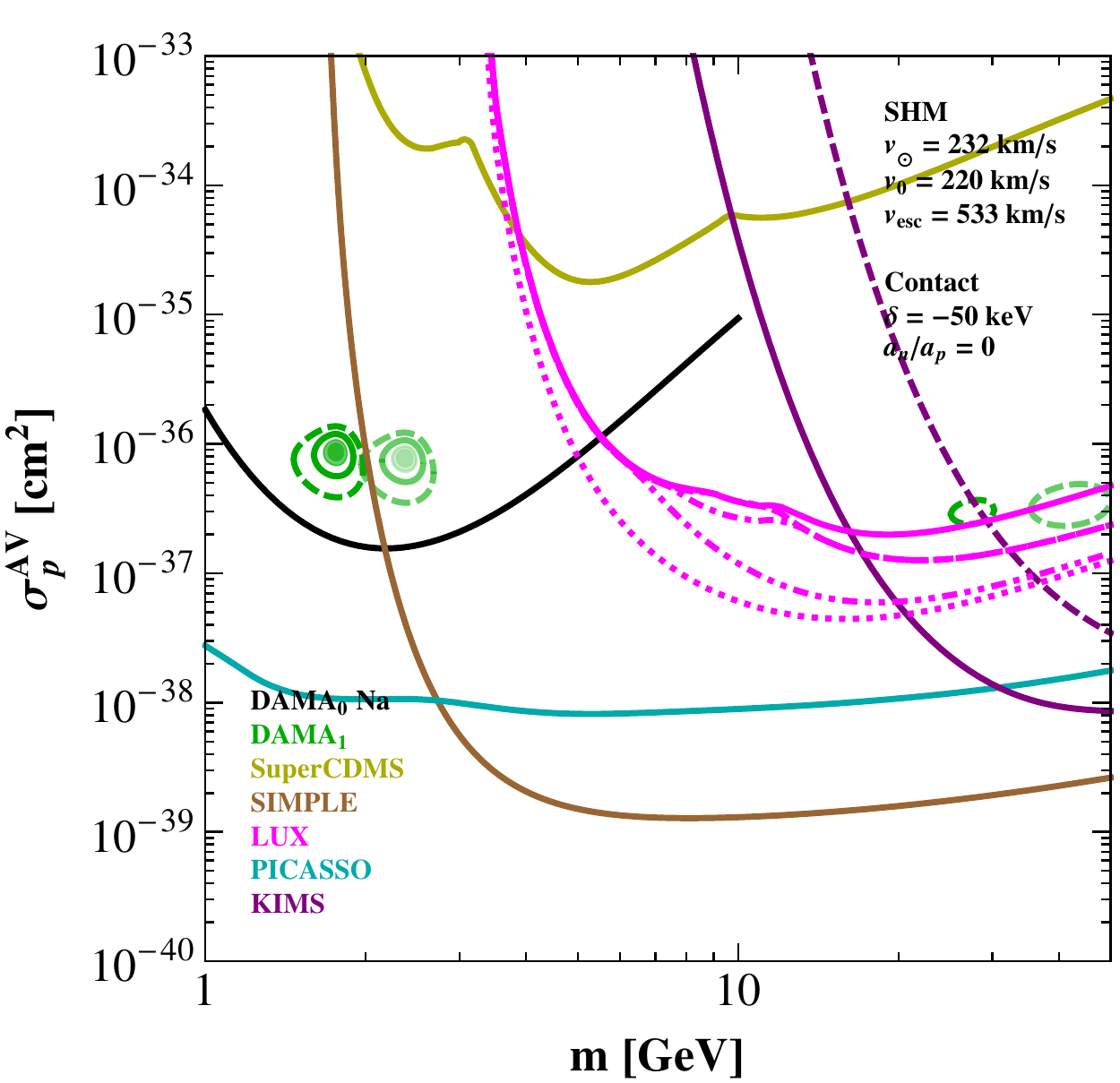}
\caption{\label{fig:AVexo}
Same as Fig.~\ref{fig:AVel} but for exothermic AV interactions with $\delta=-30$ keV (left) and $\delta=-50$ keV (right).}
\end{figure}

\begin{figure}[t!]
\centering
\includegraphics[width=0.49\textwidth]{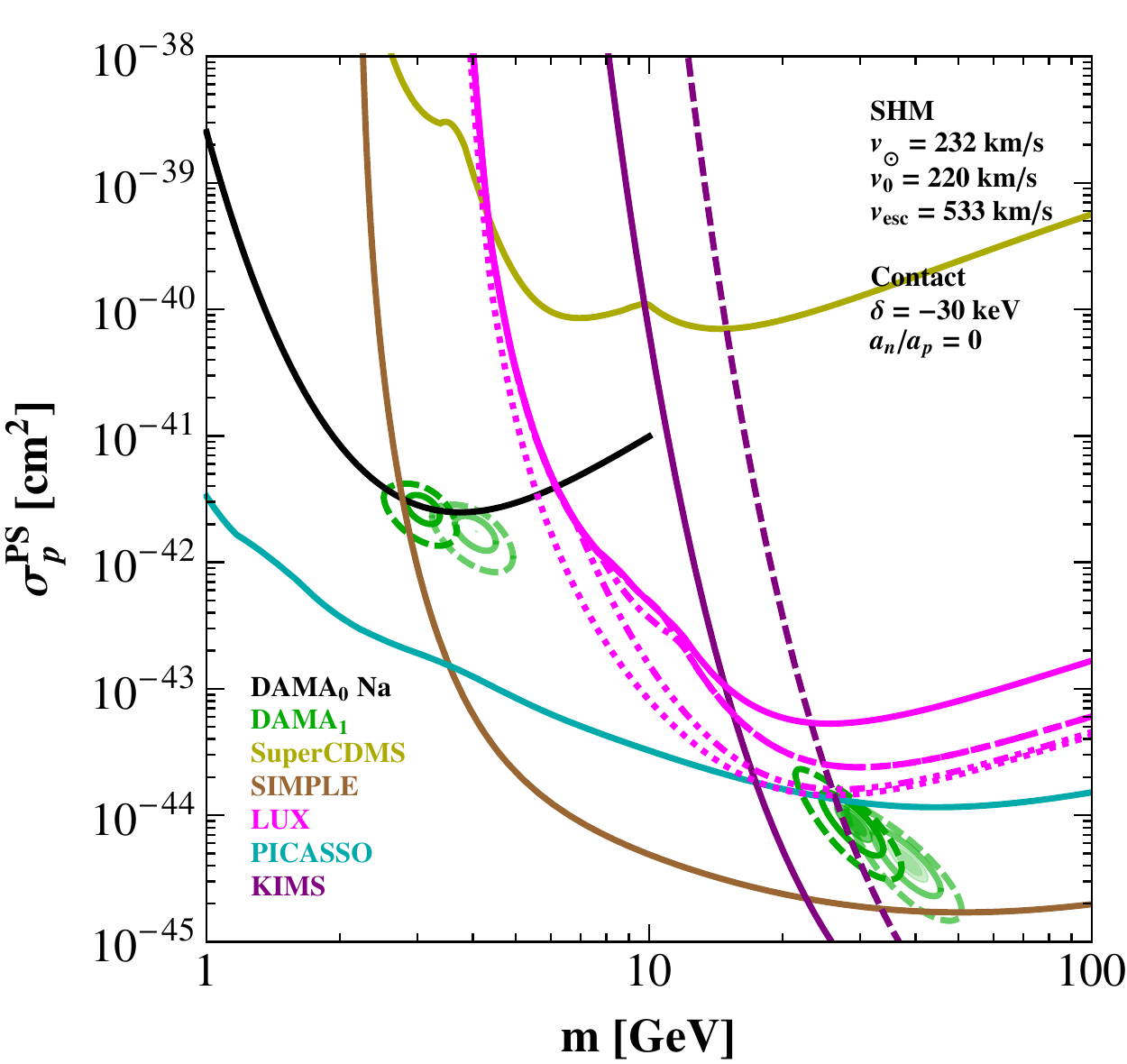}
\includegraphics[width=0.49\textwidth]{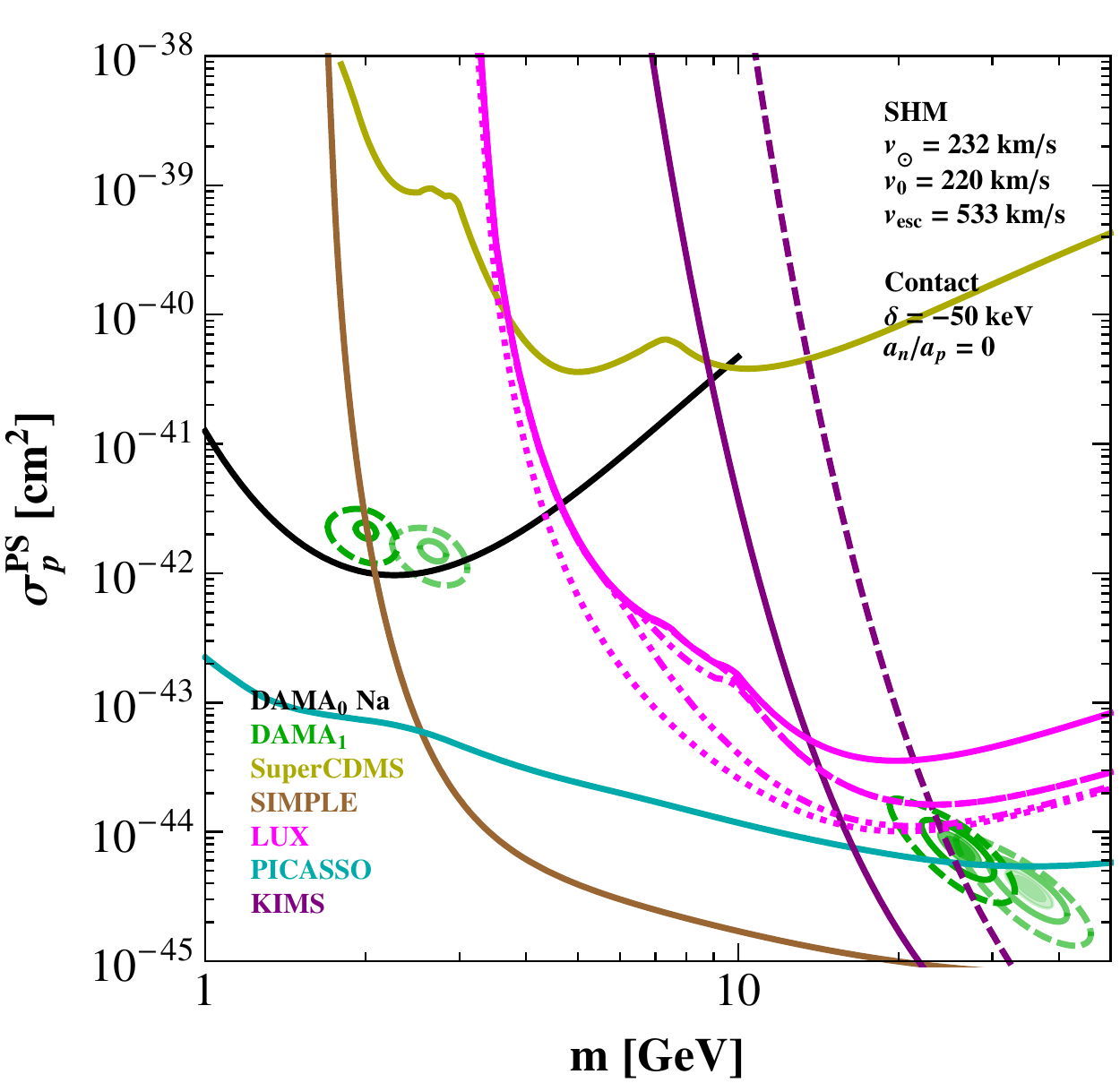}
\caption{\label{fig:PSexo}
Same as Fig.~\ref{fig:AVexo} but for PS interactions.}
\end{figure}

\subsection{Exothermic contact interactions}

Figs.~\ref{fig:AVexo} and \ref{fig:PSexo} show the results for exothermic inelastic proton-only AV and PS interactions, respectively, with $\delta=-30\text{ keV}$ (left panels) and $\delta=-50\text{ keV}$ (right panels). As $|\delta|$ increases, the DAMA regions move to lower masses, for the following reason.

\begin{figure}[t!]
\centering
\includegraphics[width=0.49\textwidth]{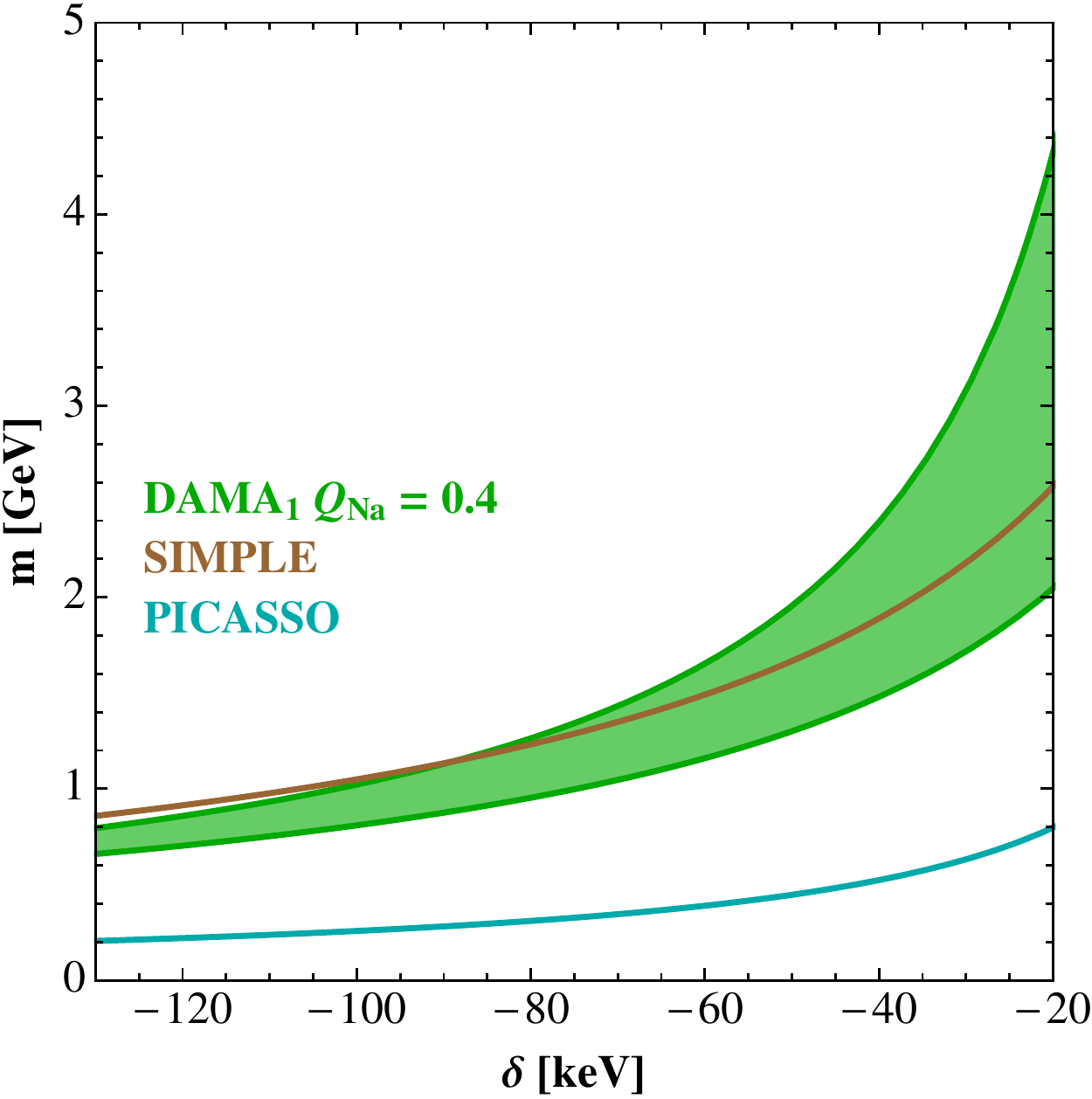}
\caption{\label{fig:PS_mxvsdelta}
Estimated lowest reach in WIMP mass $m$ for SIMPLE and PICASSO, and estimated mass range in which the Na component of the DAMA region with quenching factor $0.40$ is found, as a function of mass splitting $\delta$ for exothermic scattering.
}
\end{figure}

The lowest reach in DM mass for a direct detection experiment is obtained when $\ER^+(v_\text{max})=E_\text{th}$ (see Fig.~\ref{fig:ER_vs_vmin}), where the threshold energy $E_\text{th}$ is the lowest detectable recoil energy. The mass reach can be found by extracting $m$ as a function of $\delta$ and $v$ from Eq.~\eqref{eq:ER_vs_v}, for $\ER^+(v)=E_\text{th}$,
\beq\label{eq:mvsdelta}
\tilde{m}(\delta, v) = \dfrac{E_\text{th} m_T}{v \sqrt{2 E_\text{th} m_T} - E_\text{th} - \delta} \ ,
\eeq
and evaluating it at $v = v_\text{max}$. The DAMA region will be therefore located at WIMP masses higher than $\tilde{m}(\delta, v_\text{max})$, taking Na as the target element. For DAMA, $E_\text{th}=5$ keV for scattering off Na with quenching factor $0.40$. The lowest reach of DAMA is the lower green line plotted in Fig.~\ref{fig:PS_mxvsdelta}. Also shown in the same figure are the mass reaches of PICASSO and SIMPLE, for which we used $E_\text{th}=1.7$ keV and $8$ keV, respectively. We only considered scattering off F in both experiments.

An estimated upper limit on the DM mass for the DAMA region comes from requiring that DM particles with speeds below $200$ km/s always scatter below threshold and are therefore undetectable, because otherwise DAMA should have observed a sign change in the modulation amplitude. In other words, scatterings of DM particles slower than $200$ km/s would yield a different phase for the modulated signal with respect to that measured by DAMA, and therefore an acceptable fit requires these scatterings to occur below threshold. Since a fixed nuclear recoil energy can be imparted by heavier DM particles traveling at lower velocities, the condition $\vmin(E_\text{th}) > 200$ km/s implies an upper limit on the DM mass in DAMA, given by $\tilde{m}(\delta, v = 200 \text{ km/s})$. This upper limit is the higher green line plotted in Fig.~\ref{fig:PS_mxvsdelta}. The other possible condition to avoid scatterings of DM particles slower than $200$ km/s, \ie having a large enough $\ER^-(v = 200 \text{ km/s})$, would imply a very odd spectrum in DAMA, with more events at higher energy instead of the observed spectrum vanishing at high energy.

Since exothermic scattering decreases the value of $v_\text{min}$ for a given recoil energy, the modulation amplitude becomes smaller with respect to the time-average rate. For large enough $|\delta|$ the DAMA modulation signal becomes inconsistent with the DAMA time-average rate.  For values of $\delta$ lower than about $-30\text{ keV}$ (for AV interactions) and $-50\text{ keV}$ (for PS interactions), the DAMA total rate limit rules out the modulation signal in Na, as indicated by the black curve excluding the DAMA Na region in the right panels of Figs.~\ref{fig:AVexo} and \ref{fig:PSexo}. 

For the values of $\delta$ allowed by the DAMA rate, the limit by PICASSO (and also the SIMPLE limit in most instances) rejects the allowed regions. For each value of $\delta$ on the horizontal axis of Fig.~\ref{fig:PS_mxvsdelta}, the DAMA region spans a mass range enclosed within the green belt, while the SIMPLE and PICASSO lines indicate the mass value where the limits in the $m$--$\sigma_p$ plane become vertical. From the plot it becomes clear that exothermic scattering brings compatibility between SIMPLE and DAMA for large enough $|\delta|$, as suggested by Figs.~\ref{fig:AVexo} and \ref{fig:PSexo}, however the region is rejected by the DAMA average rate measurements.  While this is true for $Q_\text{Na} = 0.40$, smaller quenching factors move the DAMA region to larger DM masses, thus potentially compromising this compatibility with SIMPLE. In any case, the DAMA region does not escape the PICASSO limit.

\subsection{Exothermic long-range interactions}

\begin{figure}[h!]
\centering
\includegraphics[width=0.49\textwidth]{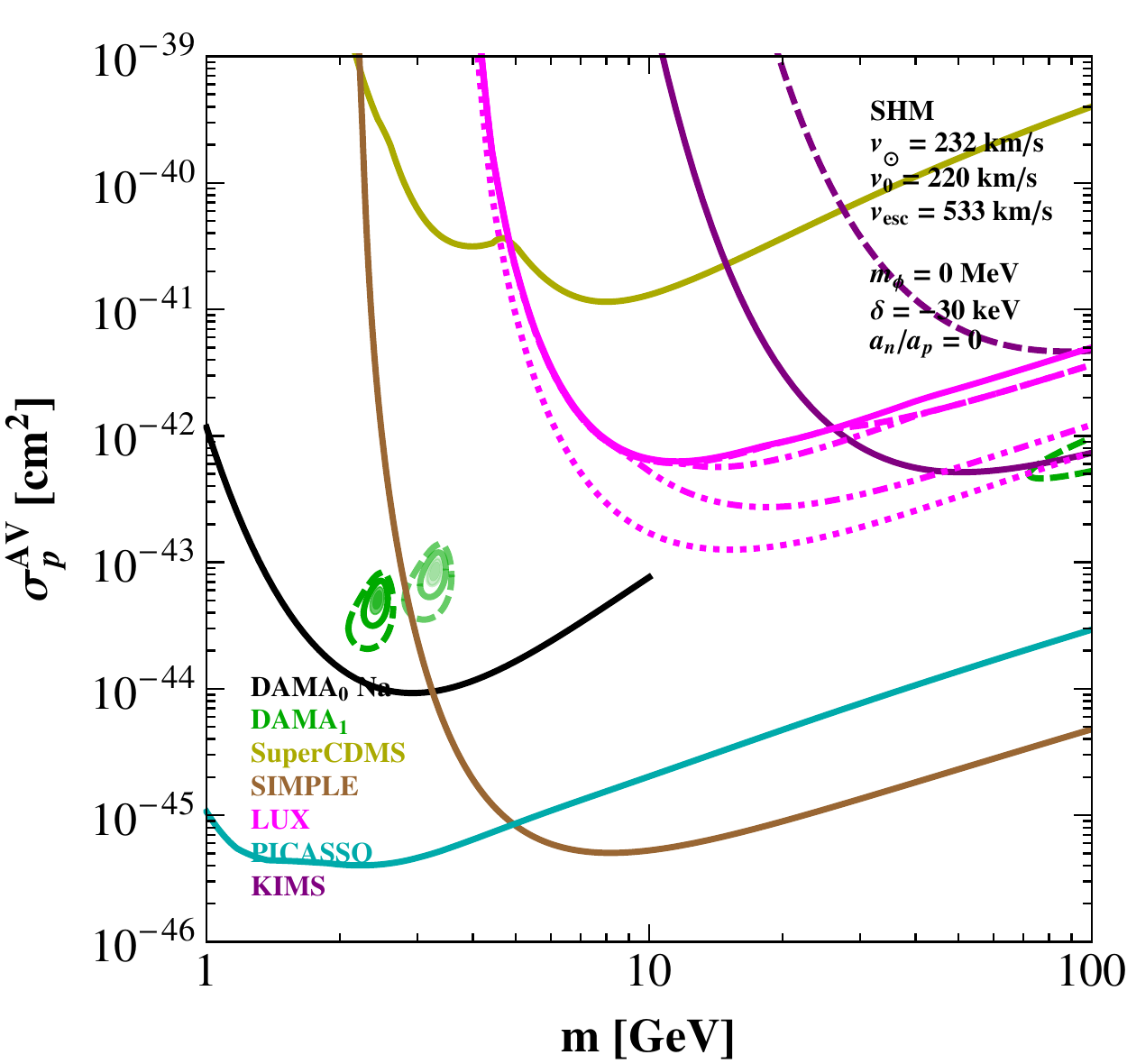}
\caption{\label{fig:LongRange_exo}
Same as Fig.~\ref{fig:AVexo} (left) but for a massless mediator.}
\end{figure}

Fig.~\ref{fig:LongRange_exo} shows the AV interaction via a massless mediator for $\delta=-30\text{ keV}$. Here as well, all the DAMA regions are rejected by the null experiments. We do not plot the results for long-range PS interactions as these are qualitatively similar to those for contact AV interactions, as commented above.

\subsection{Endothermic contact interactions}

\begin{figure}[h!]
\centering
\includegraphics[width=0.49\textwidth]{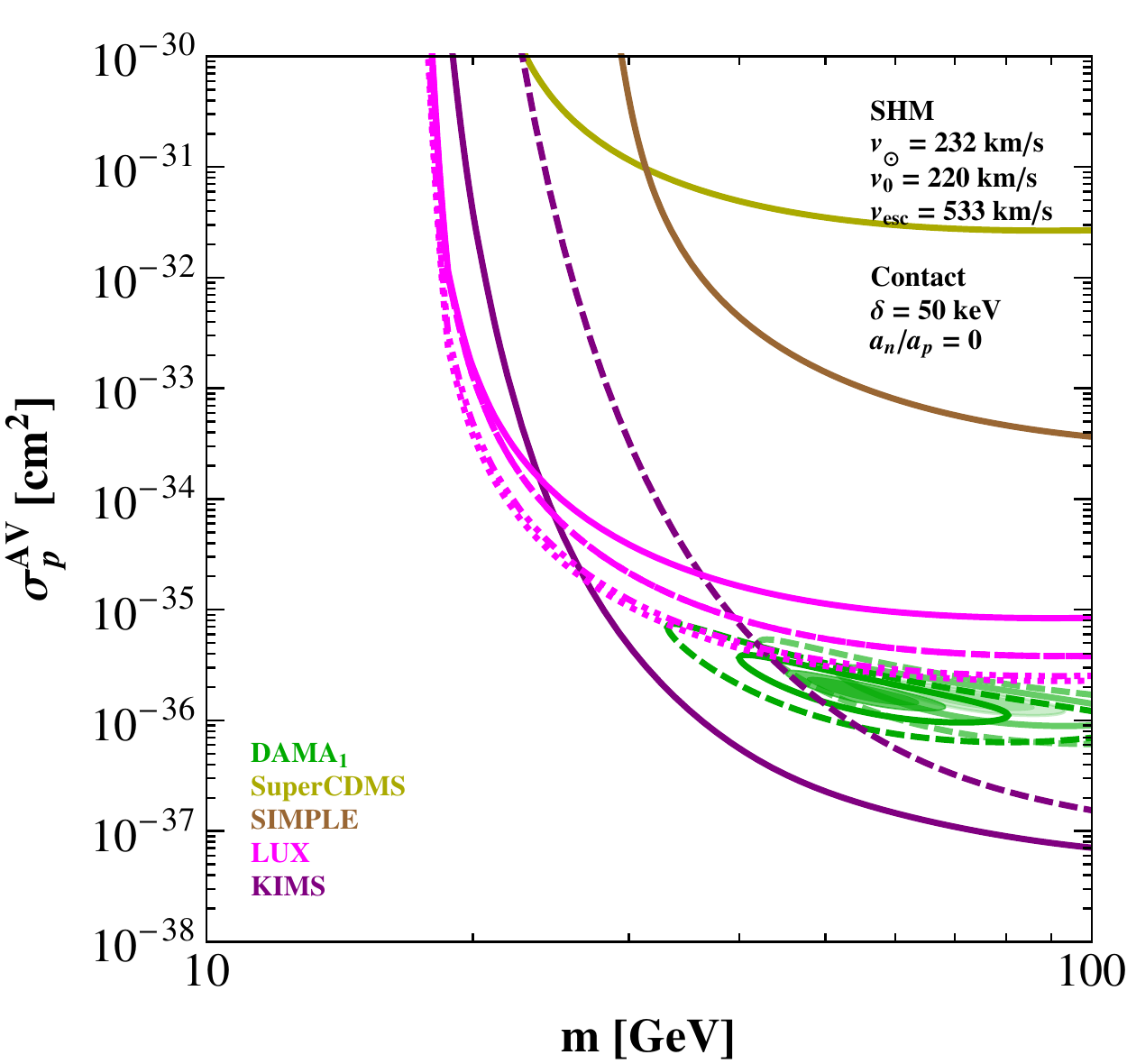}
\includegraphics[width=0.49\textwidth]{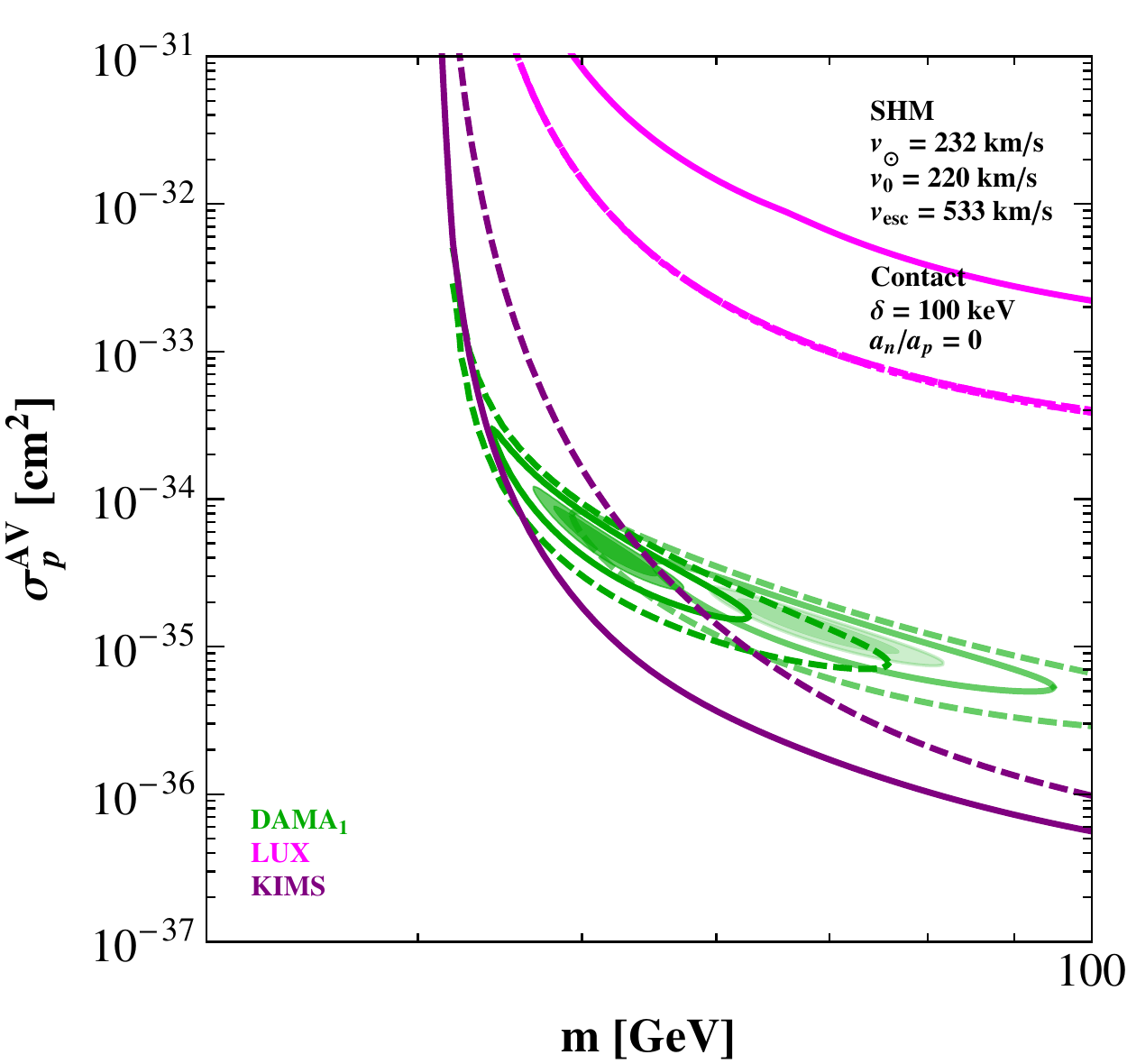}
\caption{\label{fig:AVendo}
Same as Fig.~\ref{fig:AVel} but for endothermic AV interactions with $\delta=50$ keV (left) and $\delta=100$ keV (right).}
\end{figure}

\begin{figure}[h!]
\centering
\includegraphics[width=0.49\textwidth]{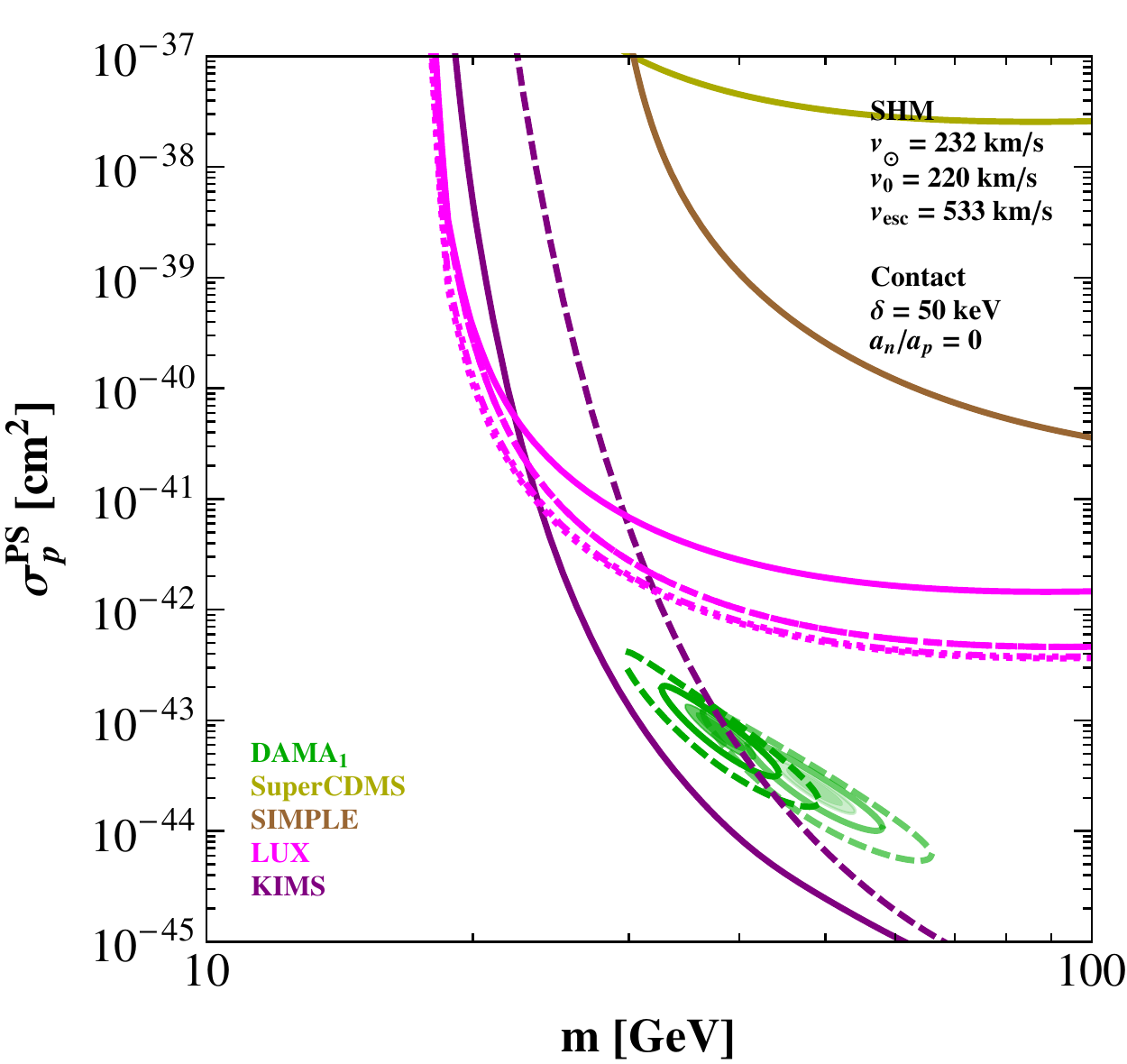}
\includegraphics[width=0.49\textwidth]{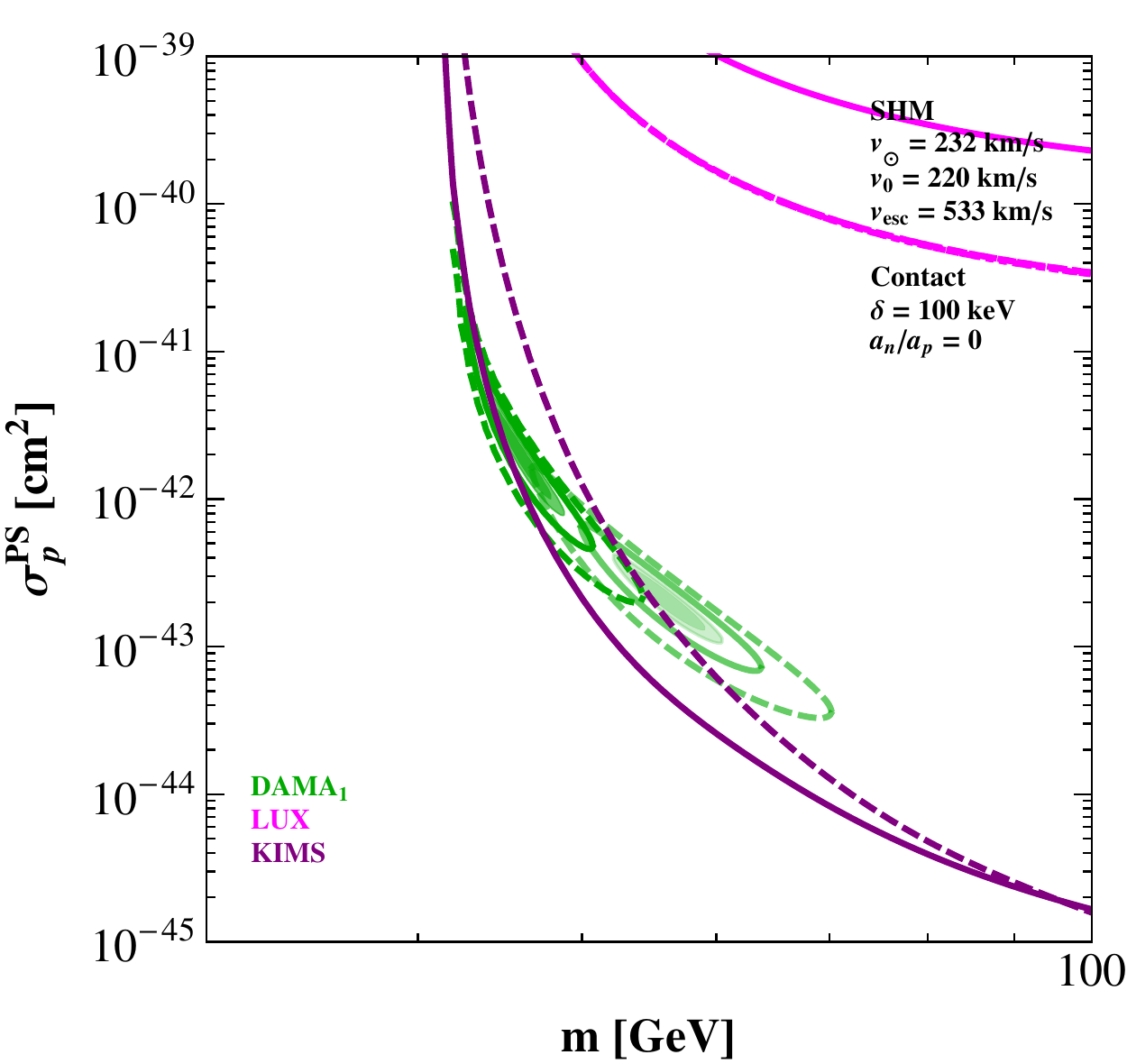}
\caption{\label{fig:PSendo}
Same as Fig.~\ref{fig:AVendo} but for PS interactions.}
\end{figure}

Figs.~\ref{fig:AVendo} and \ref{fig:PSendo} show the result for the proton-only spin dependent endothermic scattering with AV and PS interactions, respectively, in the contact limit, with $\delta=50\text{ keV}$ (left panels) and $\delta=100\text{ keV}$ (right panels). As $\delta$ increases, scattering off light targets becomes kinematically forbidden since $v^T_\delta$ becomes larger than $v_\text{max}$. For $\delta=100\text{ keV}$, the only remaining limits are from KIMS and LUX. We can see that the DAMA region for I scattering moves towards the left compared to the KIMS upper bound as $\delta$ increases. For PS interactions with $\delta=50\text{ keV}$, it is only the combination of larger quenching factor $Q_\text{I}=0.09$ for I in DAMA and smaller quenching factor $Q_\text{I}=0.05$ in KIMS that allows the DAMA signal to be compatible with all upper limits. For $\delta=100\text{ keV}$, the I region corresponding to $Q_\text{I}=0.06$ barely escapes the limit with $Q_\text{I}=0.05$ from KIMS, and the situation remains tense for quenching factors $0.09$ (DAMA) and $0.10$ (KIMS). Raising $\delta$ further makes it progressively more difficult to find a region of the DAMA signal that is kinematically allowed. For AV interactions the DAMA regions are even more severely constrained: only the larger quenching factor $Q_\text{I}=0.09$ DAMA region is allowed by the KIMS upper bound with smaller quenching factor $Q_\text{I}=0.05$ for $\delta=100\text{ keV}$. 

These results are largely consistent with those of Ref.~\cite{Barello:2014uda}, where the framework of non-relativistic operators introduced in Ref.~\cite{Fitzpatrick:2012ix} was generalized to inelastic scattering and a model-independent analysis was performed on a series of effective operators. In Table V of Ref.~\cite{Barello:2014uda} the AV and PS interactions correspond to fermion operators $15$ and $4$, respectively. For those interactions, Ref.~\cite{Barello:2014uda} quotes best fit parameters for DAMA (corresponding to $\delta=106\text{ keV}$ for AV and $\delta=57\text{ keV}$ for PS) that are consistent with the KIMS data only for DAMA quenching factor $0.09$ and KIMS quenching factor $0.05$. In Ref.~\cite{Barello:2014uda}, however, scattering off Cs in KIMS was neglected due to the lack of the form factor in Ref.~\cite{Fitzpatrick:2012ix}. Since the contribution of Cs to the scattering rate is sizable for interaction with protons, we adopt an approximate form factor for Cs as discussed in Sec.~\ref{sec:nuclearFF}, resulting in stronger KIMS bounds.  

At this point it is important to recall the flavor physics bounds we mentioned in the introduction. Fig.~9 of Ref.~\cite{Dolan:2014ska} shows that the quark couplings needed for PS inelastic scattering on I to fit the DAMA data with a one-particle exchange process,
\beq
 g\:g_\text{DM} = 
 \begin{cases}
  \kappa \left(\dfrac{m_\phi}{100\text{ MeV}}\right)^2 &\text{for } m_\phi \gg 100\text{ MeV}  \\
  \kappa' & \text{for } m_\phi \ll 100\text{ MeV},
 \end{cases}
\eeq
where $\kappa\simeq 0.2$ and $\kappa'\simeq0.1$ for $\delta=50$ keV, and $\kappa\simeq 1.6$ and $\kappa'\simeq 0.7$ for $\delta=100$ keV, is rejected \cite{Dolan:2014ska} for any reasonable value of $g_\text{DM}$ (unless $g_\text{DM}>10^5$). Note that the quark coupling used in Ref.~\cite{Dolan:2014ska} is $g/\sqrt{2}$.

\subsection{Endothermic long-range interactions}

\begin{figure}[h!]
\centering
\includegraphics[width=0.49\textwidth]{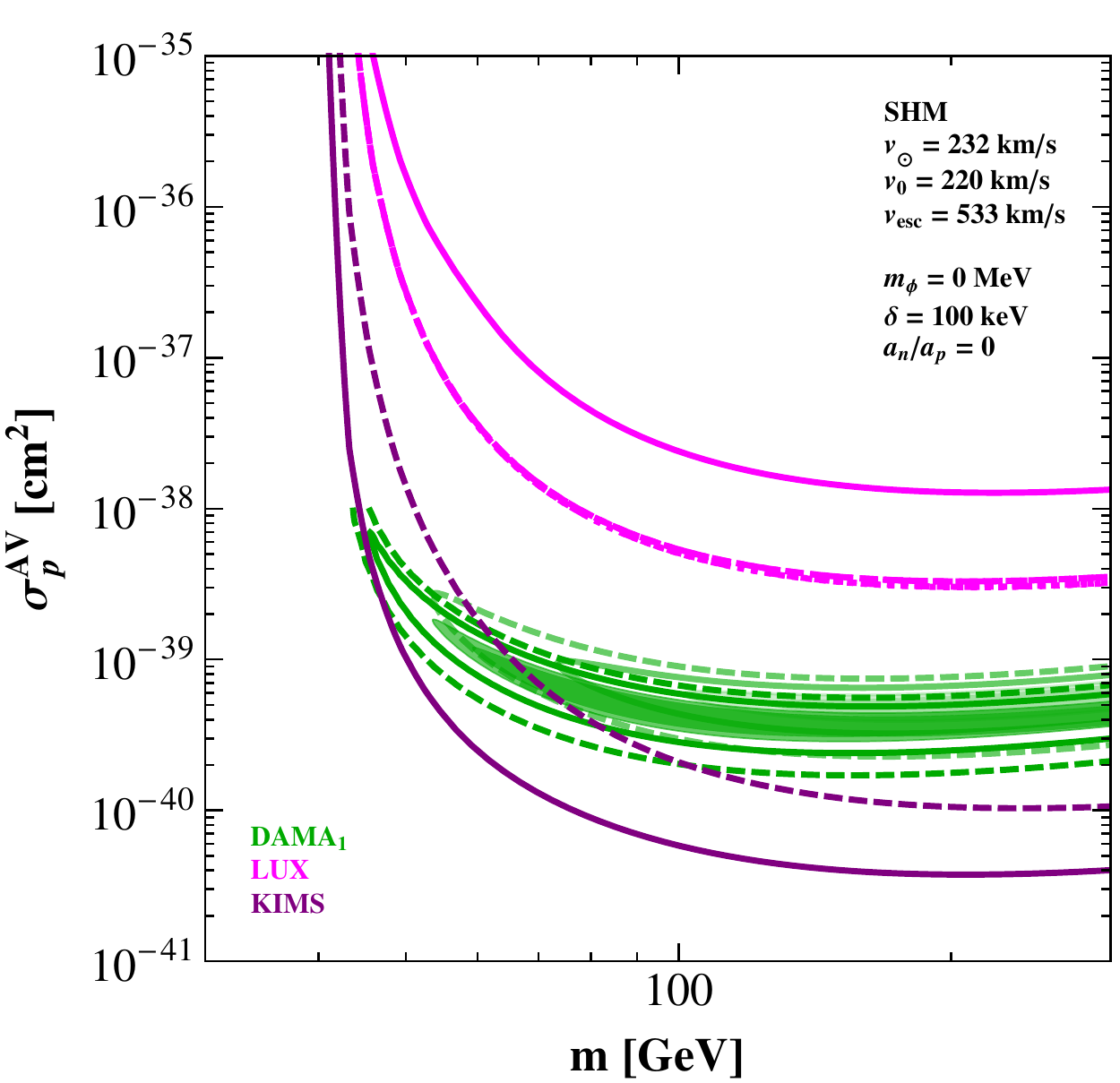}
\caption{\label{fig:LongRange_endo}
Same as Fig.~\ref{fig:AVendo} (right) but for a massless mediator.}
\end{figure}

For the AV interaction via a massless mediator shown in Fig.~\ref{fig:LongRange_endo}, only a very small portion of the I DAMA region with quenching factor $0.09$ escapes the KIMS limit with quenching factor $0.05$. Therefore, quite different quenching factors are needed for I in DAMA and KIMS to have the DAMA region escape the KIMS limit. Although quenching factors of a given element in different crystals can have different values in general, large differences may be questionable. Again, we do not plot the results for long-range PS interactions since these are qualitatively similar to those for contact AV interactions, as commented above.

\section{Halo-independent analysis} \label{sec:halo_indep}

So far we have assumed a particular model for the dark halo of our galaxy. It is however possible to compare direct detection data without making any assumption about the local density or velocity distribution of the dark matter particles \cite{DelNobile:2013cta, DelNobile:2013gba, DelNobile:2013cva, DelNobile:2014sja, Fox:2010bu, Fox:2010bz, McCabe:2011sr, Frandsen:2011gi, HerreroGarcia:2011aa, HerreroGarcia:2012fu, Gondolo:2012rs, Bozorgnia:2013hsa, Frandsen:2013cna, Cherry:2014wia, Fox:2014kua, Scopel:2014kba, Gelmini:2014psa, Feldstein:2014gza, Feldstein:2014ufa, Herrero-Garcia:2015kga, Anderson:2015xaa, Scopel:2015baa, Ferrer:2015bta} (in particular we follow the analysis of Refs.~\cite{DelNobile:2013cta, DelNobile:2013gba, DelNobile:2013cva, DelNobile:2014sja}). This method consists of extracting from the data, instead of just the (reference) WIMP-proton cross section $\sigma_p$, the function $\tilde\eta$ of $\vmin$ defined below in Eq.~\eqref{eta} which encloses all the dependence of the rate on the DM velocity distribution. Since this function is experiment-independent, data from different experiments can be directly compared in the $\vmin$--$\tilde\eta$ plane.

In order to perform the halo-independent analysis we have to assume a value for the DM mass $m$, together with all the other interaction parameters such as the mass splitting $\delta$ and the neutron to proton coupling ratio $a_n / a_p$. We study parameter values that seem promising in our SHM analysis to make a DM interpretation of the DAMA data compatible with all other experiments when relaxing the assumption on the dark halo. Taking into account Figs.~\ref{fig:PSel} and \ref{fig:PSel_otherSHM} we select a WIMP mass below $10$ GeV, one close to $30$ GeV, and one close to $50$ GeV, for WIMPs with PS interactions and elastic scattering. For the couplings we take $a_n = 0$ rather than $a_n / a_p = -1/16.4$; this choice is conservative in the sense that, while results from experiments employing targets with negligible spin-dependent interactions with neutrons like DAMA, SIMPLE, PICASSO, and KIMS are not affected, the bounds from LUX and SuperCDMS are less constraining when the WIMP-neutron coupling is set to $0$ (see \eg Fig.~\ref{fig:PSel}). We do not consider inelastic exothermic scattering, as the DAMA regions are badly excluded in all cases studied in the previous section. For inelastic endothermic scattering, looking at Fig.~\ref{fig:AVendo} we select $m = 40$ GeV for $\delta = 50$ keV and $m = 52$ GeV for $\delta = 100$ keV, for WIMPs with AV interactions with $a_n = 0$. Analogously, from Fig.~\ref{fig:PSendo} we select $m = 38$ GeV for $\delta = 50$ keV and $m = 45$ GeV for $\delta = 100$ keV, for WIMPs with PS interactions with $a_n = 0$. Notice that some of these choices are similar to the best fit parameters of Ref.~\cite{Barello:2014uda}, i.e.~$m=54.3$ GeV and $\delta=106$ keV for AV interactions and $m=40.8$ GeV and $\delta=57$ keV for PS interactions. Finally, from Fig.~\ref{fig:LongRange_endo} we select $m = 80$ GeV for $\delta = 100$ keV, for long-range AV interactions with $a_n = 0$.

\subsection{More on the direct detection rate}

Since the differential cross section $\ud \sigma_T / \ud \ER$ for the AV and PS interactions is proportional to $1 / v^2$ (see section \ref{sec:cross sections}), and thus $v^2 \ud \sigma_T / \ud \ER$ is independent of $v$, we can express Eq.~\eqref{R-E'} as
\begin{equation}
\label{R-E'-eta}
R_{[E'_1,E'_2]}(t) = \sum_T \frac{C_T}{m_T} \int_0^\infty \ud \ER \,
\tilde\eta(\vmin(\ER), t) \left( \frac{v^2}{\sigma_p} \frac{\ud \sigma_T}{\ud \ER} \right)
\int_{E'_1}^{E'_2} \ud E' \, \epsilon(E',\ER) G_T(\ER, E') \ ,
\end{equation}
where $\sigma_p$ is the (reference) total DM-proton scattering cross section and we defined the velocity integral
\begin{equation}\label{eta}
\tilde\eta(\vmin, t) \equiv \frac{\rho \sigma_p}{m} \int_{v \geqslant v_\text{min}} \frac{f(\bsv, t)}{v} \,  \ud^3 v \ .
\end{equation} 
Notice that, while $\tilde\eta$ as a function of $\vmin$ is independent of the target nuclide, $\vmin(\ER)$ is a function of $\ER$ that depends on the target.

By changing the integration variable from $\ER$ to $\vmin$ in Eq.~\eqref{R-E'-eta}, the event rate can be written in the more compact form
\begin{equation}\label{eq:halo-indep}
R_{[E'_1,E'_2]}(t) = \sum_T \int_{v^T_\delta}^\infty \ud \vmin \,
\tilde\eta(\vmin, t) {\cal R}^T_{[E'_1,E'_2]}(\vmin) \ ,
\end{equation}
with $v^T_\delta=\sqrt{2\delta/\mu_T}$. The target-specific response function ${\cal R}^T_{[E'_1,E'_2]}(\vmin)$ is given by
\begin{multline}
{\cal R}^T_{[E'_1z,E'_2]}(\vmin) \equiv \\
\frac{C_T}{m_T}
\left[
\frac{\ud \ER^+(\vmin)}{\ud \vmin}
\left( \frac{v^2}{\sigma_p} \frac{\ud \sigma_T}{\ud \ER}(\ER^+(\vmin), \bsv) \right)
\int_{E'_1}^{E'_2} \ud E' \, \epsilon(E', \ER^+(\vmin)) G_T(\ER^+(\vmin), E')
\right.
\\
\left.
- \frac{\ud \ER^-(\vmin)}{\ud \vmin}
\left( \frac{v^2}{\sigma_p} \frac{\ud \sigma_T}{\ud \ER}(\ER^-(\vmin), \bsv) \right)
\int_{E'_1}^{E'_2} \ud E' \, \epsilon(E', \ER^-(\vmin)) G_T(\ER^-(\vmin), E')
\right] ,
\end{multline}
with $\ER^\pm(\vmin)$ given in Eq.~\eqref{eq:ER_vs_v}. For elastic scattering ($\delta=0$), $\ER^- = 0$ and $\ER^+$ equals
\beq\label{eq:ER}
\ER(\vmin, m_T) = \frac{2 \mu_T^2}{m_T} \vmin^2 \ ,
\eeq
so that ${\cal R}^T_{[E'_1,E'_2]}(\vmin)$ reduces to
\begin{multline}
{\cal R}^T_{[E'_1,E'_2]}(\vmin) =
\frac{C_T}{m_T} \frac{4 \mu_T^2 \vmin}{m_T}
\left( \frac{v^2}{\sigma_p} \frac{\ud \sigma_T}{\ud \ER}(\ER(\vmin, m_T), \bsv) \right) \\
\times \int_{E'_1}^{E'_2} \ud E' \, \epsilon(E', \ER(\vmin, m_T)) G_T(\ER(\vmin, m_T), E') \ .
\end{multline}

Due to Earth's rotation around the Sun, the velocity integral \eqref{eta} is modulated in time with a $1$ year period:
\begin{equation}\label{eta01}
\tilde\eta(\vmin, t) \simeq \tilde\eta^0(\vmin) + \tilde\eta^1(\vmin) \cos\left( \frac{2 \pi}{\text{yr}} (t - t_0) \right) .
\end{equation}
Since all the time dependence of the rate is contained in $\tilde\eta$, we also have that
\begin{equation}
R_{[E'_1,E'_2]}(t) \simeq R^0_{[E'_1,E'_2]} + R^1_{[E'_1,E'_2]} \cos\left( \frac{2 \pi}{\text{yr}} (t - t_0) \right) ,
\end{equation}
where $t_0$ is the time when the rate reaches its maximum value, and $R^0_{[E'_1,E'_2]}$ and $R^1_{[E'_1,E'_2]}$ are the time-average and annual modulation amplitude of the rate, respectively. In general $\tilde\eta$ can be expanded in a Fourier series, and here we assume that higher modes are not important. The DAMA collaboration did not find any hints of higher modes in their data \cite{Bernabei:2013xsa, Bernabei:2013cfa}, thus when considering the DAMA data we adopt Eq.~\eqref{eta01} with the measured phase $t_0 = $ June $2^\text{nd}$. All other experiments considered here give an upper bound on a time-averaged signal, thus on $\tilde\eta^0$. Given that the annual modulation amplitude cannot be larger than the average rate, $\left|\tilde\eta^1(\vmin)\right| \leqslant \tilde\eta^0(\vmin)$, we can interpret upper bounds on $\tilde\eta^0$ from experiments with null results as (conservative) limits on the $\tilde\eta^1$ signal measured by DAMA. We do not consider the direct CDMS-II bound on $\tilde\eta^1$ \cite{Ahmed:2012vq}, since LUX and SuperCDMS set more stringent constraints, see Refs.~\cite{DelNobile:2013gba, DelNobile:2014eta, DelNobile:2014sja} (notice also that SuperCDMS employs the same target material).

\subsection{Data analysis for the halo-independent method}

For LUX, SuperCDMS, and SIMPLE we follow the procedure developed and described in Refs.~\cite{DelNobile:2013cta, DelNobile:2013gba, DelNobile:2013cva, DelNobile:2014sja}. For PICASSO and KIMS we cannot perform a Maximum Gap analysis as done for LUX and SuperCDMS because the data are binned. We therefore produce a limit on $\tilde\eta^0$ at each $\vmin$ value in the following way. We compute the rate \eqref{eq:halo-indep} adopting a step function, $\tilde\eta^0(\vmin) = \tilde\eta^* \theta(\vmin^* - \vmin)$, because it is the function that allows to draw the most conservative bound on the value $\tilde\eta^*$ taken by $\tilde\eta^0(\vmin)$ at a specific $\vmin$ value $\vmin^*$ \cite{Fox:2010bz, Frandsen:2011gi}. For each value of $\vmin^*$ we compare this predicted rate in each single energy bin with the $90\%$ CL limit on the rate. Imposing the computed rate not to surpass the limit in any of the bins thus fixes the maximum allowed $\tilde\eta^*$ at $\vmin^*$. For KIMS we use the black limit lines in Fig.~4 of Ref.~\cite{Kim:2012rza}, while for PICASSO we translate the upper end of the error bars in Fig.~5 of Ref.~\cite{Archambault:2012pm} into $90\%$ CL upper limits assuming the data are Gaussian distributed and the uncertainty is given at the $1 \sigma$ level.

The halo-independent analysis of the DAMA annual modulation data presented in Refs.~\cite{DelNobile:2013cta, DelNobile:2013gba, DelNobile:2013cva, DelNobile:2014sja} and in Ref.~\cite{Gelmini:2014psa} is only applicable when WIMPs can scatter off only one of the target elements, either Na or I. This happens, for instance, if the DM is so light that elastic scattering off I occurs always below threshold, assuming a reasonable maximum speed with respect to Earth, $\vmax$, for WIMPs in the galaxy. It also happens for inelastic endothermic scattering with $v_\text{max} < v^\text{Na}_\delta = \sqrt{2 \delta / \mu_\text{Na}}$, which makes WIMP scattering off Na kinematically forbidden. Therefore, we can straightforwardly apply the analysis of DAMA data presented in Refs.~\cite{DelNobile:2013cta, DelNobile:2013gba, DelNobile:2013cva, DelNobile:2014sja, Gelmini:2014psa} to the light WIMP with $m \lesssim 10$ GeV scattering elastically and to all considered cases of endothermic scattering. The WIMPs with mass close to $30$ GeV and $50$ GeV scattering elastically need a special treatment, described in the following.

\begin{figure}[t!]
\centering
\includegraphics[width=0.49\textwidth]{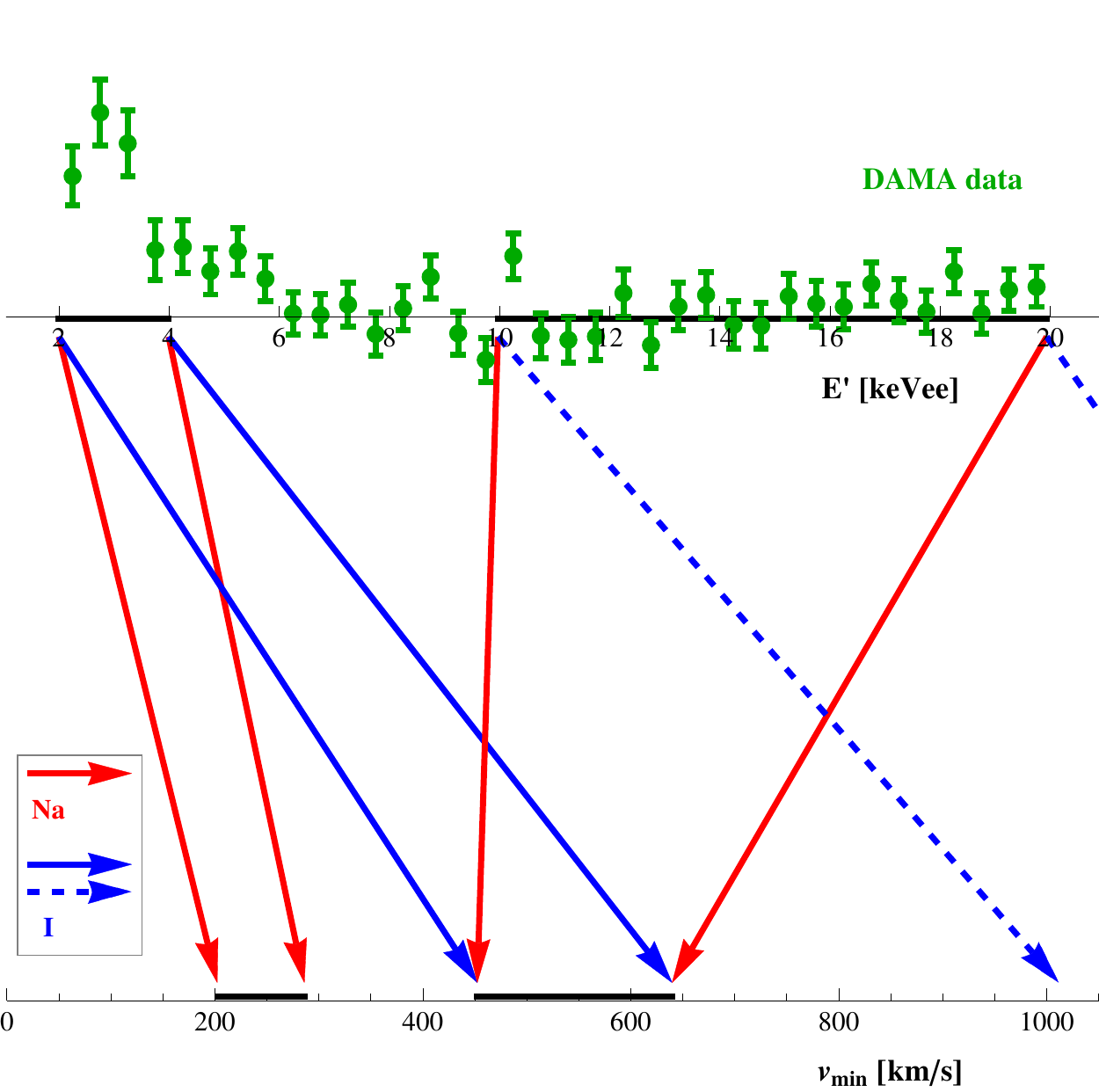}
\caption{\label{fig:PSbinning}
Binning scheme used for $m=30.14$ GeV and elastic scattering. The red and blue arrows show the correspondence between average detected energies $\langle E'\rangle$, shown on the top axis, and minimum speeds $\vmin$, shown on the bottom axis, for Na and I targets, respectively. The dashed arrows show that WIMPs scattering off I could produce events in the highest energy bin only with $\vmin > \vmax = 800$ km/s. The DAMA data are overlaid in green.
}
\end{figure}

When both elements, Na and I, are involved in the scattering, the same value of the detected energy $E'$ is mapped onto different $\vmin$ values because of the different target masses and quenching factors. In order to extract the value of $\tilde\eta$ in a $\vmin$ interval from Eq.~\eqref{eq:halo-indep}, we adapt to the DAMA data the procedure that was developed for CRESST-II in Appendix A.2 of Ref.~\cite{Frandsen:2011gi} and in Ref.~\cite{Gondolo:2012rs}. We start by choosing the highest energy bin at high enough energy so that the heaviest element (I) does not contribute to the scattering rate in it, because the necessary $\vmin$ exceeds the maximum possible speed in the halo. Starting from the highest energy bin we work our way toward the lowest energy bin as depicted in Fig.~\ref{fig:PSbinning}. We compute the $\vmin$ range corresponding to the highest $E'$ bin for Na using the relation $\langle E' \rangle = Q_T \ER(\vmin, m_T)$ for $T = \text{Na}$ (with $\ER(\vmin, m_T)$ given in Eq.~\eqref{eq:ER}). From this $\vmin$ range we derive the second highest energy bin using again the same relation for $T = \text{I}$. The procedure can then be repeated starting from this new bin, until the lightest energy bin above the experimental threshold is built. Since we only wish to consider bins where a significant signal is observed, we require the lowest energy bin to lie within the $2.0$--$6.0$ keVee interval, where the modulation amplitude measured by DAMA is significantly different from zero.

Notice that the DAMA modulation data span $2.0$ to $20.0$ keVee with an original bin size of $0.5$ keVee. In rebinning the DAMA data we want to merge the original bins, and not split existing bins. Therefore, each new bin needs to have boundaries that are multiples of $0.5$ keVee so as to encompass an entire number of original DAMA bins. Notice also that, if $E'$ is the boundary of one of the chosen bins, the corresponding boundary of the next higher energy bin is $r E'$ with $r \equiv Q_\text{Na} \ER(\vmin, m_\text{Na}) / Q_\text{I} \ER(\vmin, m_\text{I}) = Q_\text{Na} \mu_\text{Na}^2 m_\text{I} / Q_\text{I} \mu_\text{I}^2 m_\text{Na}$. Therefore, it is necessary for $r$ to be an integer (or a half-integer) number. For quenching factors that are constant in energy, this can be achieved for particular DM mass values. With $Q_\text{Na} = 0.30$ and $Q_\text{I} = 0.09$, we choose our DM particle masses to be $30.14$ GeV and $47.35$ GeV for which $r = 5.0$ and $3.5$, respectively.

Choosing $\vmax=800$ km/s, for $m = 30.14$ GeV the bin at highest $E'$ must be completely above $Q_\text{I} \ER(v_\text{max}, m_\text{I}) \simeq 6.3$ keVee for $Q_\text{I} = 0.09$. Using $r=5$ we take two bins, $[2.0, 4.0]$ keVee and $[10.0, 20.0]$ keVee. We choose bins well separated in energy and as large as possible to avoid overlapping and to minimize the effect of the tails of the corresponding response functions. The binning scheme for this candidate is shown in Fig.~\ref{fig:PSbinning}. Our choice of $m = 47.35$ GeV comes from a halo model in Fig.~\ref{fig:PSel_otherSHM} with a low $\vmax$ value, close to $\vmax=600$ km/s. Assuming this $\vmax$ value, scattering of I is kinematically forbidden for $E'$ energies above $Q_I \ER(v_\text{max}, m_\text{I}) \simeq 6.97$ keVee for $Q_\text{I} = 0.09$. Using $r=3$ we choose the two bins $[3.0, 6.0]$ keVee and $[10.5, 21.0]$ keVee. Since the highest energy bin surpasses the $2.0$--$20.0$ keVee energy range where the DAMA modulation data is available, for the two additional $0.5$ keVee bins in the $20.0$ to $21.0$ keVee range we assume the same average and mean square error as for the nineteen $0.5$ keVee bins in the $10.5$--$20.0$ keVee range.

\begin{figure}[t!]
\centering
\includegraphics[width=0.49\textwidth]{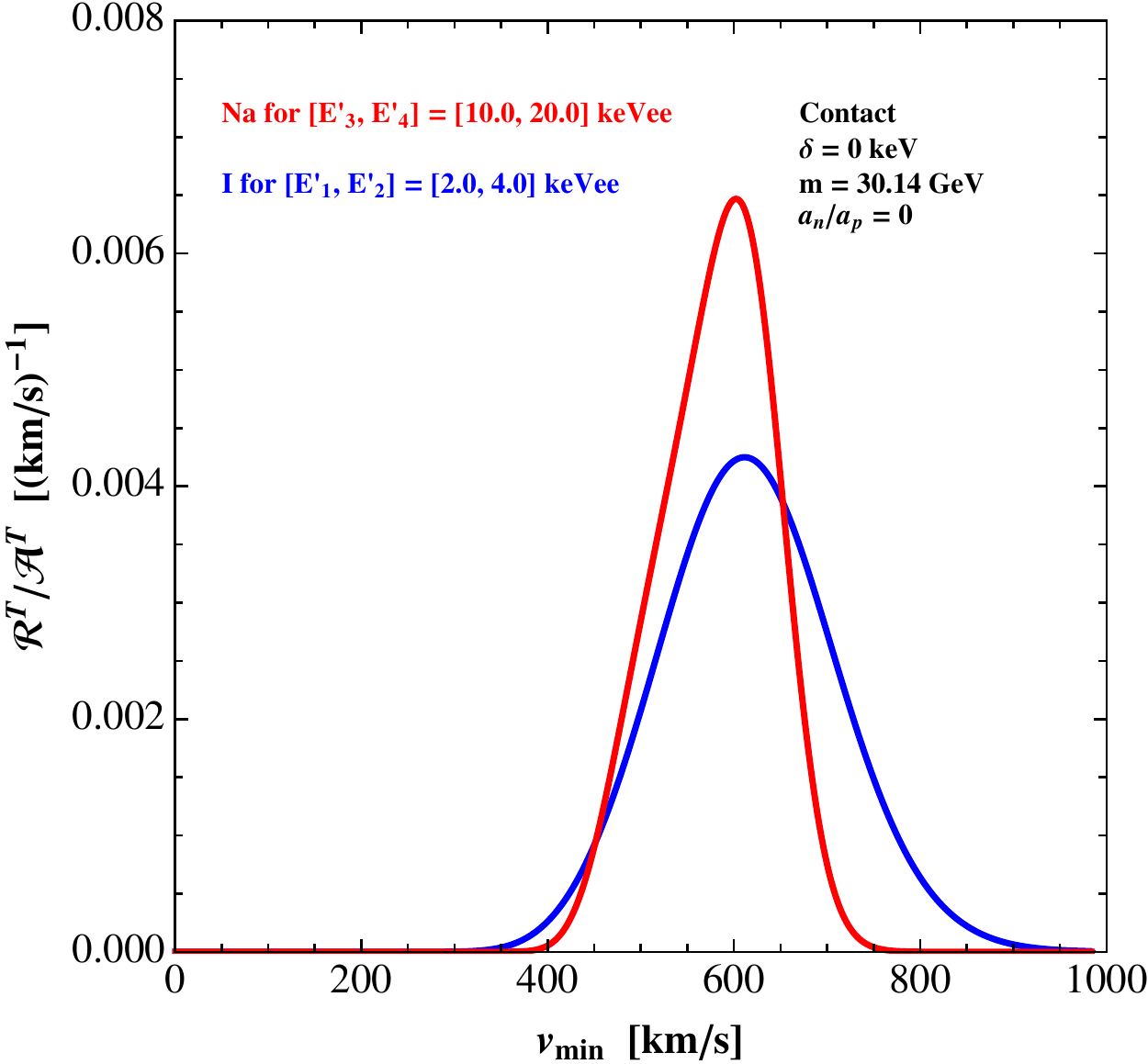}
\includegraphics[width=0.49\textwidth]{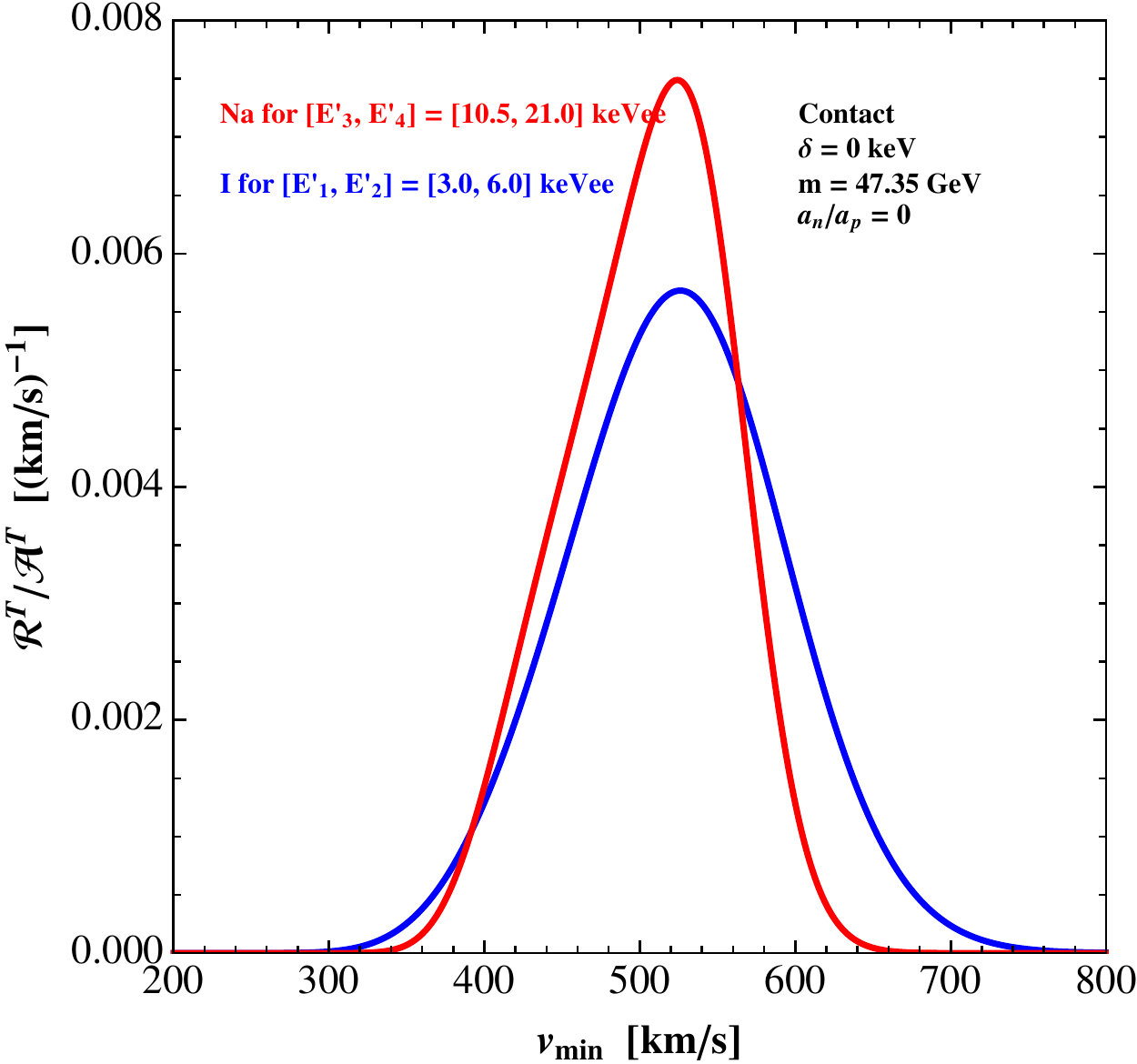}
\caption{\label{fig:PSresponse}
Normalized response functions ${\cal R}^T(\vmin)/{\cal A}^T$ for WIMPs with PS interactions and elastic scattering, for DM mass $m=30.14$ GeV (left) and $m=47.35$ GeV (right). ${\cal R}^{\rm Na}_{[E'_3,E'_4]}(\vmin)/{\cal A}^{\rm Na}_{[E'_3,E'_4]}$ is shown in red and ${\cal R}^{\rm I}_{[E'_1,E'_2]}(\vmin)/{\cal A}^{\rm I}_{[E'_1,E'_2]}$ in blue.
}
\end{figure}

Once the energy bins to be used in the analysis are established, we extract information on the modulated component of the velocity integral, $\tilde\eta^1$, in the following way. For two bins $[E'_1, E'_2]$ and $[E'_3, E'_4]$ (the extension to a larger number of bins is trivial), Eq.~\eqref{eq:halo-indep} reads
\beq\label{eq:rate12}
\begin{split}
R^1_{[E'_1,E'_2]} &= \int_0^\infty \ud \vmin \, \tilde\eta^1(\vmin) \left[ {\cal R}_{[E'_1,E'_2]}^{\rm Na}(\vmin) + {\cal R}_{[E'_1,E'_2]}^{\rm I}(\vmin) \right]
\\
&= {\cal A}^{\rm Na}_{[E'_1,E'_2]} \overline{\tilde\eta^{1, \, \text{Na}}_{[E'_1,E'_2]}} + {\cal A}^{\rm I}_{[E'_1,E'_2]} \overline{\tilde\eta^{1, \, \text{I}}_{[E'_1,E'_2]}}(\vmin)
\end{split}
\eeq
and
\beq\label{eq:rate34}
R^1_{[E'_3,E'_4]} = \int_0^\infty \ud \vmin \, \tilde\eta^1(\vmin) \left[ {\cal R}_{[E'_3,E'_4]}^{\rm Na}(\vmin) + {\cal R}_{[E'_3,E'_4]}^{\rm I}(\vmin) \right] \simeq {\cal A}^{\rm Na}_{[E'_3,E'_4]} \overline{\tilde\eta^{1, \, \text{Na}}_{[E'_3,E'_4]}}
\ ,
\eeq
where scattering off I does not contribute to the rate in the highest energy bin by construction. Here we defined the target-specific average of $\tilde\eta^1_{[E'_1,E'_2]}$ as
\begin{equation}
\overline{\tilde\eta^{1, \, T}_{[E'_1,E'_2]}} \equiv \frac{\int_0^\infty \ud \vmin \, \tilde\eta(\vmin){\cal R}_{[E'_1,E'_2]}^T(\vmin)}{{\cal A}^T_{[E'_1,E'_2]}} \ ,
\end{equation}
with
\beq
{\cal A}^T_{[E'_1,E'_2]} \equiv \int_0^\infty \ud \vmin \, {\cal R}^T_{[E'_1,E'_2]}(\vmin) \ ,
\eeq
and analogous definitions for $[E'_3,E'_4]$. Since the two energy bins are chosen so that scattering off I in $[E'_1,E'_2]$ probes the same $\vmin$ range as scattering off Na in $[E'_3,E'_4]$, we expect that
\beq\label{etasimeq}
\overline{\tilde\eta^{1, \, \text{Na}}_{[E'_3,E'_4]}} \simeq \overline{\tilde\eta^{1, \, \text{I}}_{[E'_1,E'_2]}} \ ,
\eeq
and thus that ${\cal R}_{[E'_3,E'_4]}^{\rm Na}(\vmin) / {\cal A}^{\rm Na}_{[E'_3,E'_4]}\simeq{\cal R}_{[E'_1,E'_2]}^{\rm I}(\vmin) / {\cal A}^{\rm I}_{[E'_1,E'_2]}$. Fig.~\ref{fig:PSresponse} shows the comparison between these quantities for WIMPs with PS interactions and elastic scattering, and gives an idea of the extent to which the assumption in Eq.~\eqref{etasimeq} is correct. In Fig.~\ref{fig:PSresponse} we can see that the $\vmin$ range associated with a certain detected energy bin through the average relation $\langle E' \rangle = Q_T \ER(\vmin, m_T)$ determines only approximately the $\vmin$ range in which the corresponding response functions ${\cal R}^T(\vmin)$ are significantly different from zero. The difference between the response functions of Na and I shown in Fig.~\ref{fig:PSresponse} is due to the width of the energy resolution function, which depends on energy, and the $\vmin$ (or alternatively the $\ER$) dependence of the scattering cross section. 

Finally, using Eqs.~\eqref{eq:rate12}, \eqref{eq:rate34} and \eqref{etasimeq} we get the system of equations
\beq
\begin{cases}
R^1_{[E'_1,E'_2]} = {\cal A}^{\rm Na}_{[E'_1,E'_2]} \overline{\tilde\eta^{1, \, \text{Na}}_{[E'_1,E'_2]}} + {\cal A}^{\rm I}_{[E'_1,E'_2]} \overline{\tilde\eta^{1, \, \text{I}}_{[E'_1,E'_2]}} \ ,
\\
R^1_{[E'_3,E'_4]} \simeq{\cal A}^{\rm Na}_{[E'_3,E'_4]} \overline{\tilde\eta^{1, \, \text{Na}}_{[E'_3,E'_4]}} \simeq {\cal A}^{\rm Na}_{[E'_3,E'_4]} \overline{\tilde\eta^{1, \, \text{I}}_{[E'_1,E'_2]}} \ , \rule{0pt}{8mm}
\end{cases}
\eeq
which can be solved for $\overline{\tilde\eta^{1, \, \text{Na}}_{[E'_1,E'_2]}}$ and $\overline{\tilde\eta^{1, \, \text{I}}_{[E'_1,E'_2]}}$. We compute the $68\%$ CL error on these quantities by propagating the $1 \sigma$ uncertainty of the DAMA data.

\subsection{Results of the halo-independent data comparison}

\begin{figure}[t!]
\centering
\includegraphics[width=0.49\textwidth]{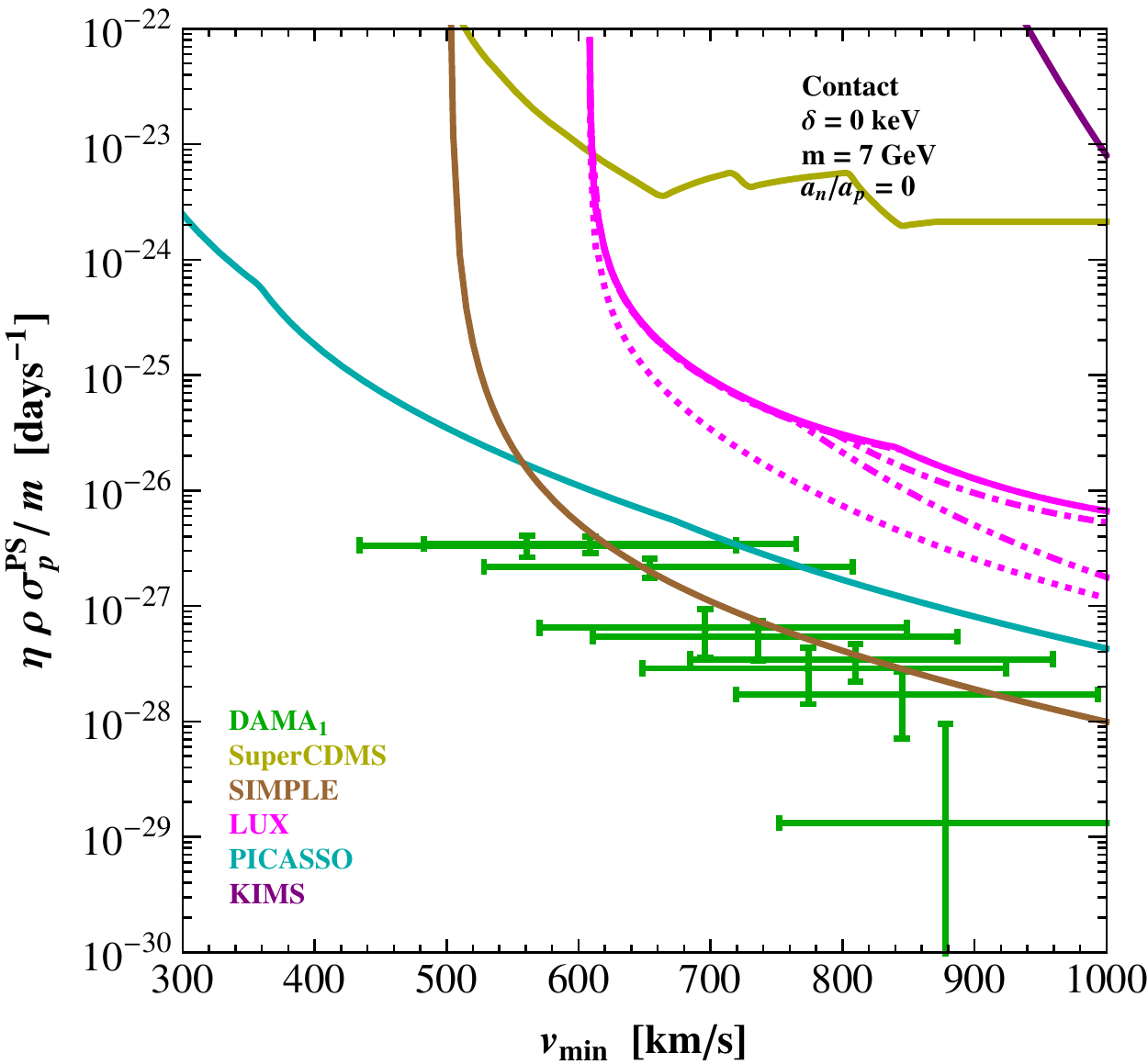}
\\
\includegraphics[width=0.49\textwidth]{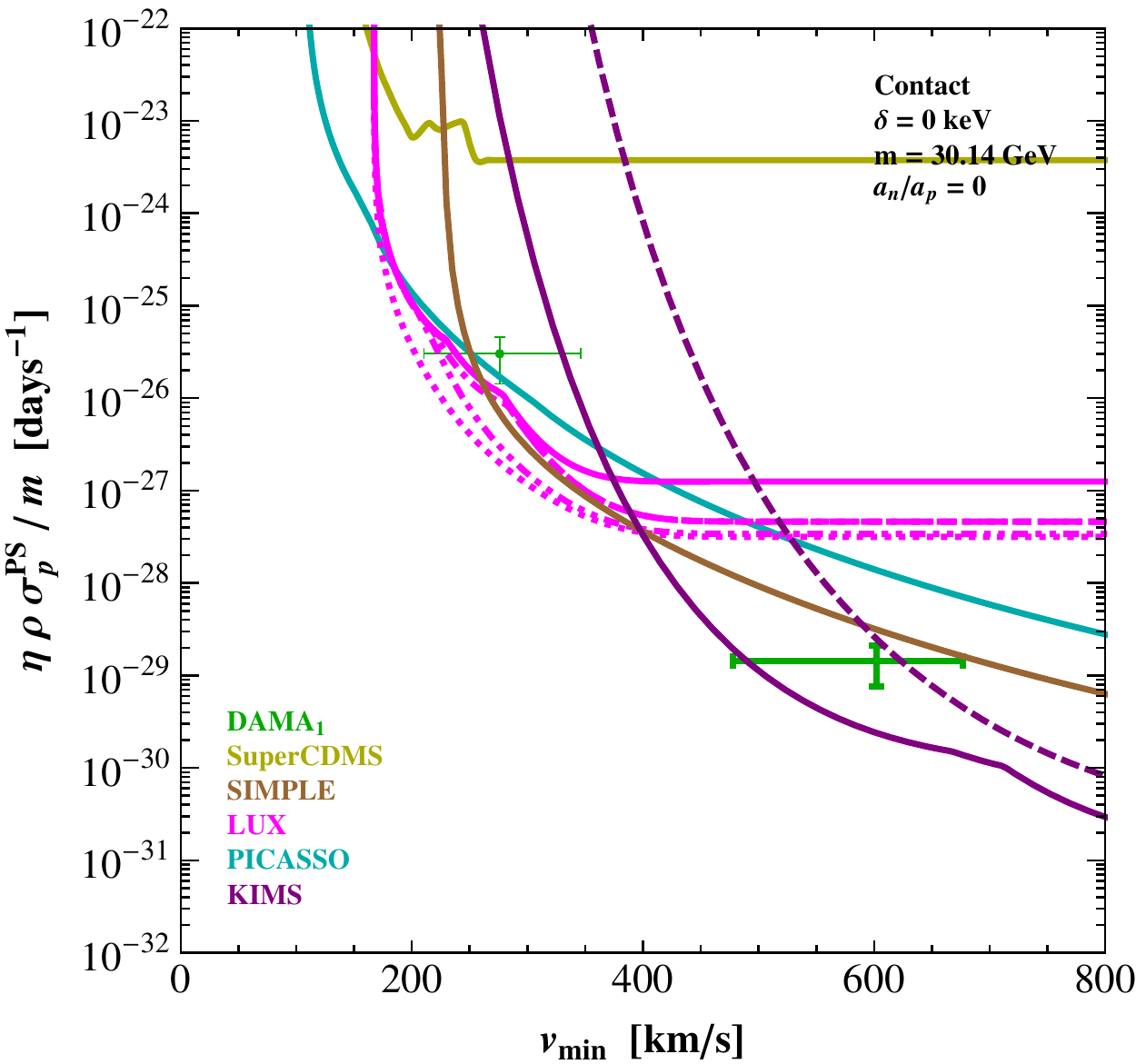}
\includegraphics[width=0.49\textwidth]{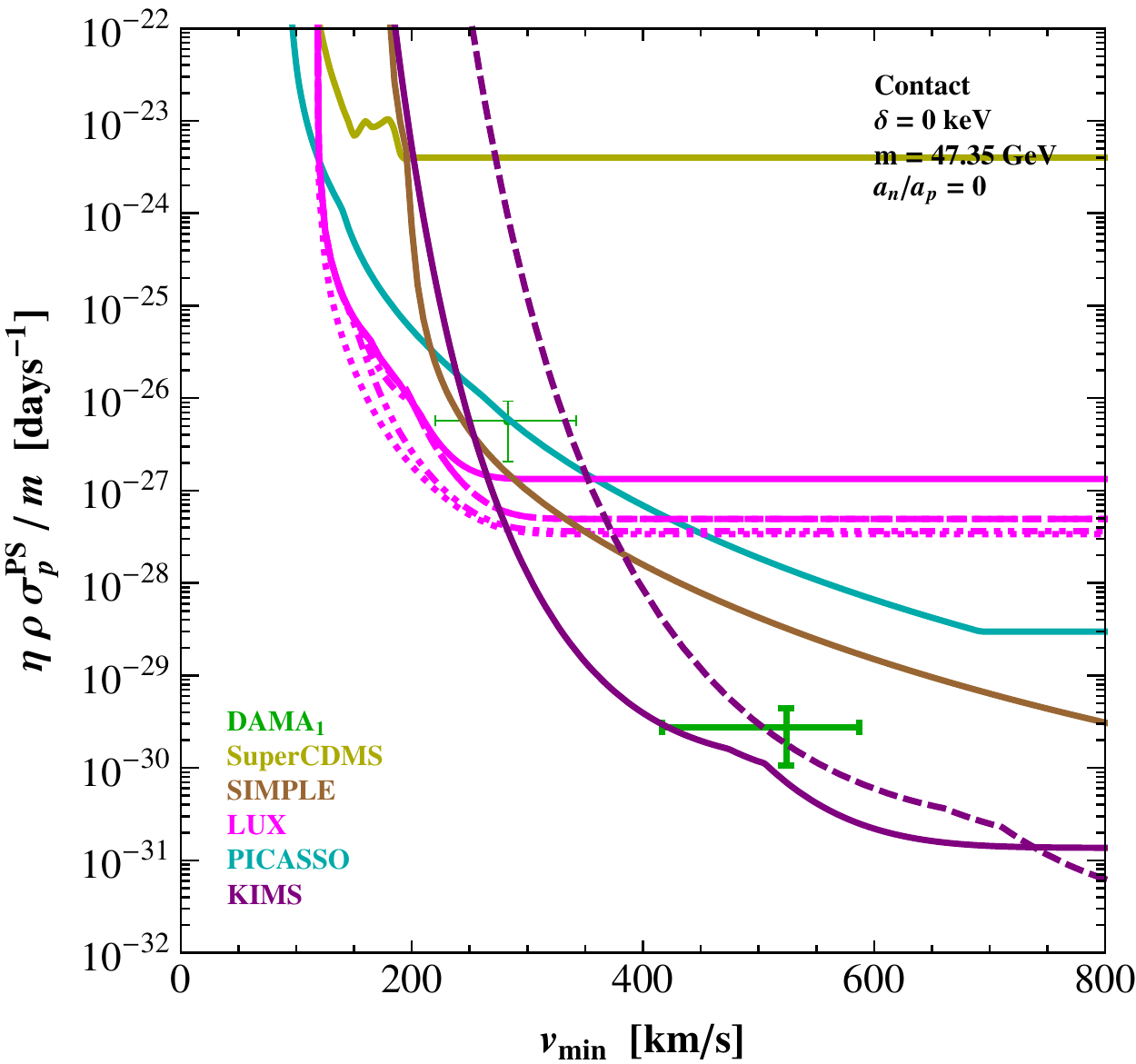}
\caption{\label{fig:halo_ind_PS_elastic}
90\% CL upper bounds on $\tilde\eta^0(\vmin)$ from LUX, SuperCDMS, SIMPLE, PICASSO, and KIMS, and measurements of $\tilde\eta^1(\vmin)$ with $68\%$ CL vertical error bars for DAMA with $Q_\text{Na}=0.30$, $Q_\text{I}=0.09$, for a WIMP with mass $m=7$ GeV (top), $m=30.14$ GeV (bottom left) and $47.35$ GeV (bottom right), contact PS interactions and elastic scattering. Different line styles for the LUX bound correspond, from most constraining to least constraining, to $0$, $1$, $3$, $5$ and $24$ observed events (see Section \ref{sec:data analysis}). The KIMS bound is shown for both $Q_{\rm I} = Q_{\rm Cs} = 0.10$ (solid line) and $0.05$ (dashed line). The thin DAMA crosses show the absolute value of $\overline{\tilde\eta^1}$ when this is negative.
}
\end{figure}

\begin{figure}[t!]
\centering
\includegraphics[width=0.49\textwidth]{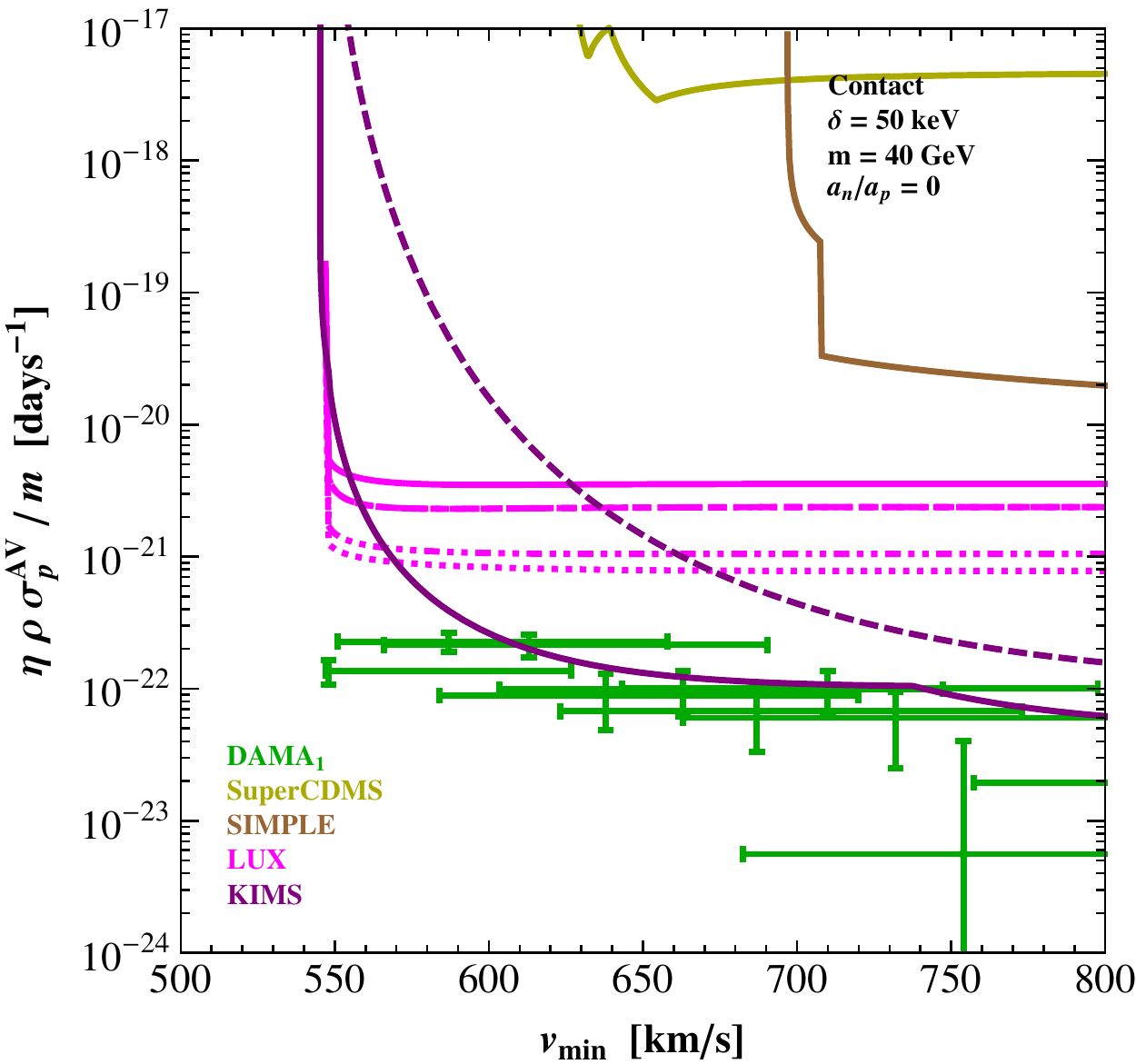}
\includegraphics[width=0.49\textwidth]{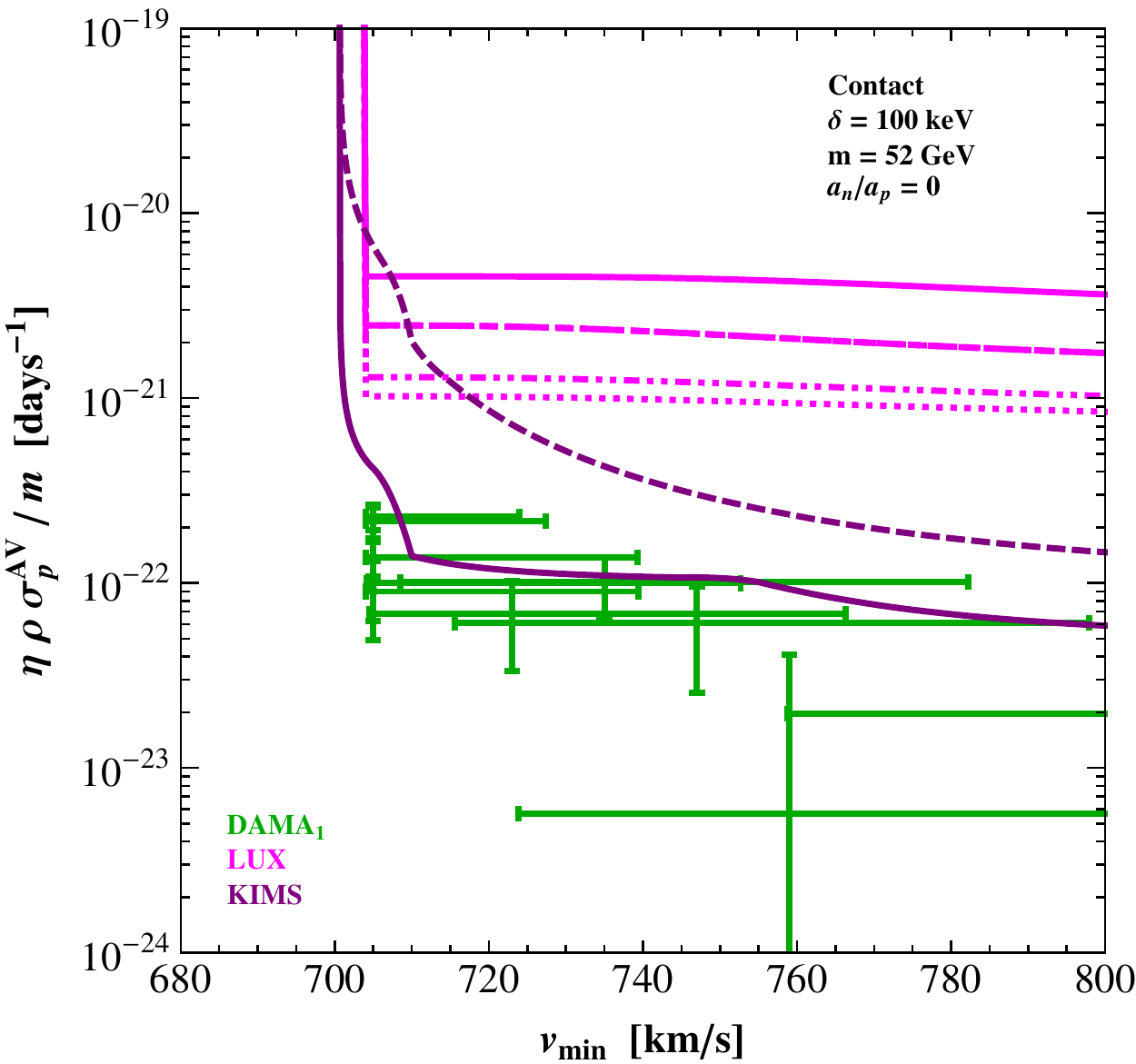}
\caption{\label{fig:halo_ind_AV_endo}
Same as Fig.~\ref{fig:halo_ind_PS_elastic} but for WIMPs with inelastic endothermic contact AV interactions and $m = 58$ GeV, $\delta=50$ keV (left), and $m = 52$ GeV, $\delta=100$ keV (right).
}
\end{figure}

\begin{figure}[t!]
\centering
\includegraphics[width=0.49\textwidth]{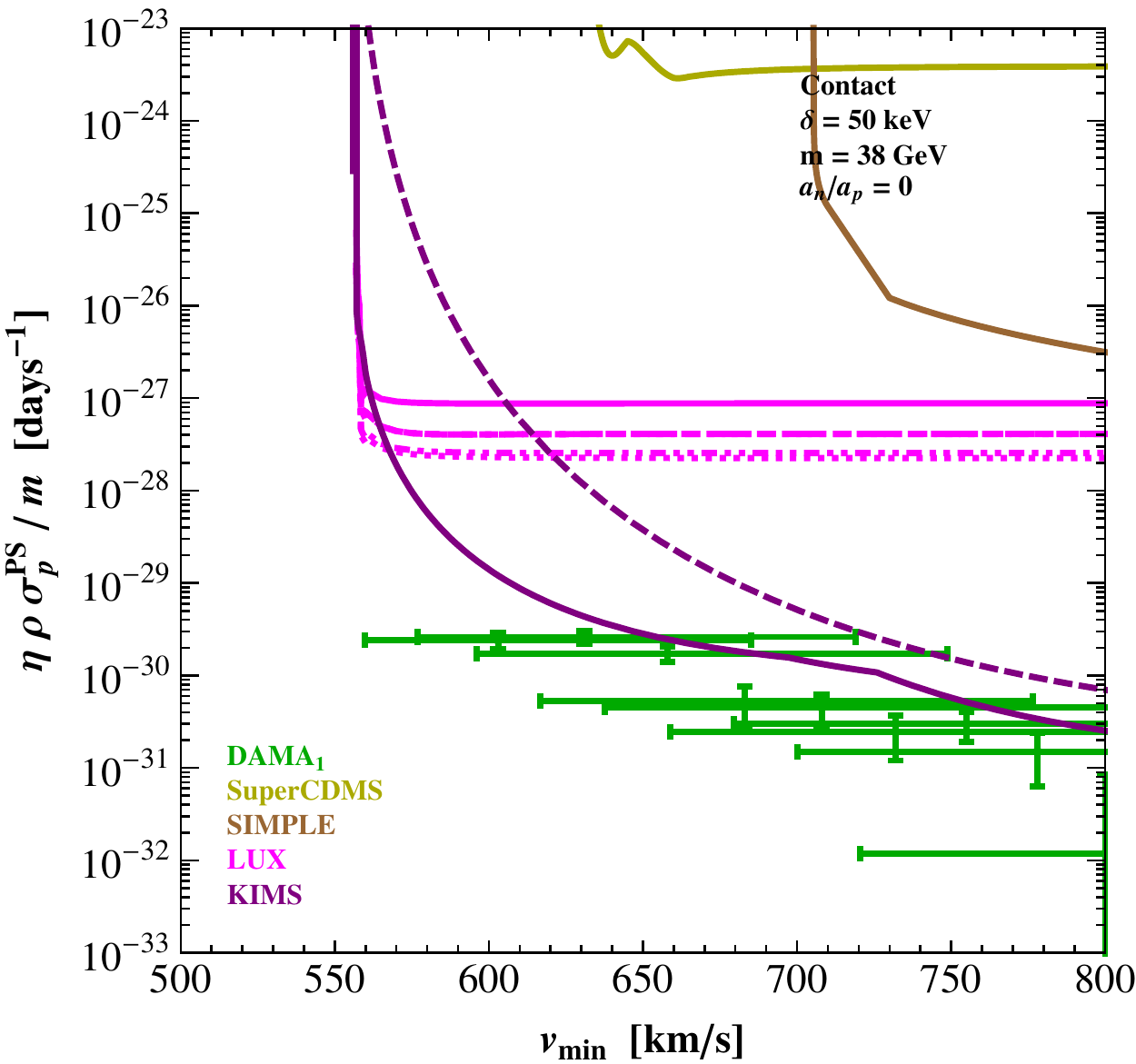}
\includegraphics[width=0.49\textwidth]{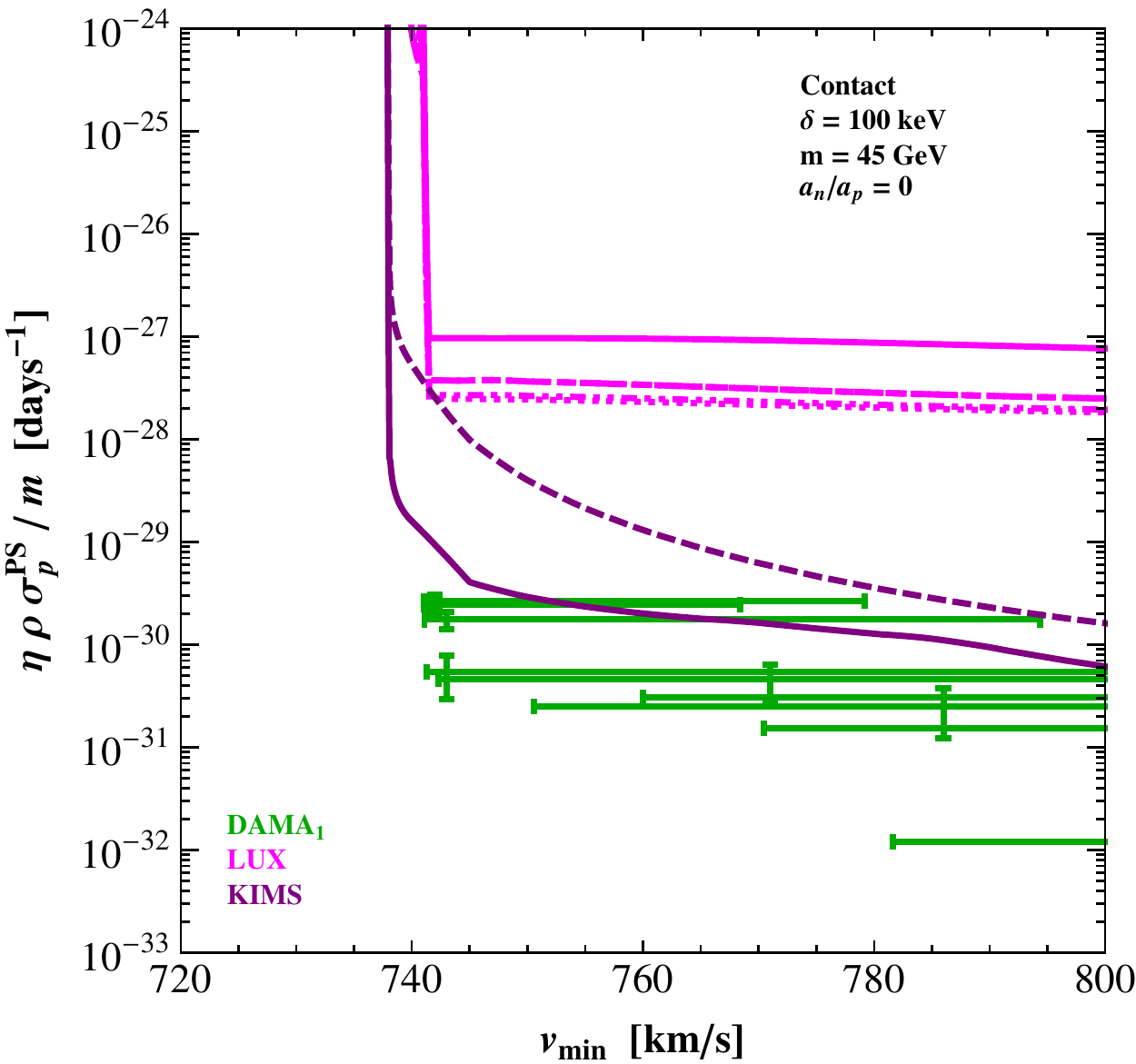}
\caption{\label{fig:halo_ind_PS_endo}
Same as Fig.~\ref{fig:halo_ind_AV_endo} but for WIMPs with inelastic endothermic contact PS interactions and $m = 38$ GeV, $\delta=50$ keV (left), and $m = 45$ GeV, $\delta=100$ keV (right).
}
\end{figure}

\begin{figure}[t!]
\centering
\includegraphics[width=0.49\textwidth]{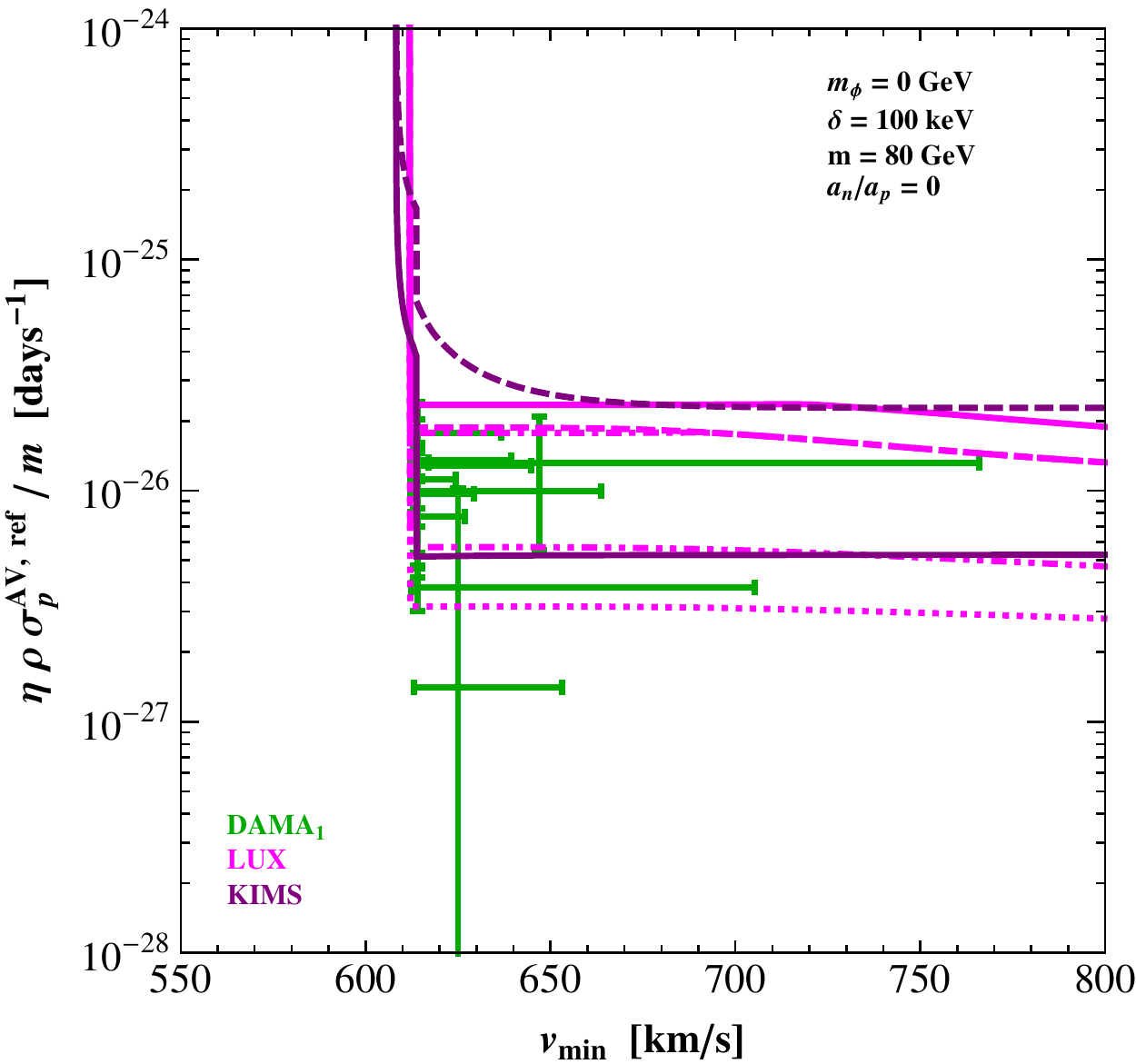}
\caption{\label{fig:halo_ind_AV_endo_longrange}
Same as Fig.~\ref{fig:halo_ind_AV_endo} but for WIMPs with inelastic endothermic AV long-range interactions and $m = 80$ GeV, $\delta=100$ keV.
}
\end{figure}

The plots in Figs.~\ref{fig:halo_ind_PS_elastic}--\ref{fig:halo_ind_AV_endo_longrange} show $90\%$ CL upper bounds on $\tilde\eta^0(\vmin)$ from LUX, SuperCDMS, SIMPLE, PICASSO, and KIMS with $Q_\text{I}=Q_\text{Cs}=0.10$ (solid purple line) and $0.05$ (dashed purple line). The DAMA measurements of the annual modulation amplitude $\tilde\eta^1(\vmin)$ are shown as crosses, where the vertical bars show the $68\%$ CL uncertainty and are located at the position of the maximum of the relevant response functions ${\cal R}^T(\vmin)$. The horizontal bar of each cross indicates the $\vmin$ interval where $90\%$ of the integral of ${\cal R}^T(\vmin)$ about the peak is included. When both Na and I contribute to a cross we choose the response function extended over the larger $\vmin$ interval, which is ${\cal R}^\text{I}(\vmin)$ (see Fig.~\ref{fig:PSresponse}). We assume the most commonly adopted values of the quenching factors, $Q_\text{Na}=0.30$ and $Q_\text{I}=0.09$. 

For WIMPs with PS interactions and elastic scattering (see Fig.~\ref{fig:halo_ind_PS_elastic}), we selected three masses from the DAMA regions shown in Figs.~\ref{fig:PSel} and \ref{fig:PSel_otherSHM}. We show results for $a_n=0$, as explained above. Only scattering off Na is kinematically accessible in DAMA for $m=7$ GeV (top panel of Fig.~\ref{fig:halo_ind_PS_elastic}), since $E'>2$ keVee would require $\vmin>1644$ km/s for I recoils. In this case the limits of PICASSO and SIMPLE cut across the DAMA points (each corresponding to the DAMA bins of width $0.5$ keVee from $2.0$ to $6.5$ keVee), except for the highest energy bin. This shows incompatibility between the DAMA and the PICASSO and SIMPLE data, unless the modulation amplitude $\left|\tilde\eta^1(\vmin)\right|$ can be as large as the time-average $\tilde\eta^0(\vmin)$, which is not possible in the whole $\vmin$ range.

For the bottom panels of Fig.~\ref{fig:halo_ind_PS_elastic}, where $m=30.14$ GeV and $47.35$ GeV, both I and Na contribute (see above). Although these masses were chosen within the DAMA regions in our SHM analysis, the $\overline{\tilde\eta^1}$ crosses shown in the two panels correspond to a halo that significantly differs from the SHM. We obtained negative $\overline{\tilde\eta^1}$ values in the lowest $\vmin$ ranges in both cases, both of them above $\vmin=200$ km/s, while in the SHM $\tilde\eta^1$ is negative only below the $\vmin=200$ km/s value. The absolute value of the crosses with negative $\overline{\tilde\eta^1}$ values are shown in Fig.~\ref{fig:halo_ind_PS_elastic} with thinner lines than those with positive $\overline{\tilde\eta^1}$ values. In both cases the combination of KIMS, PICASSO, SIMPLE and LUX bounds reject the DAMA crosses, showing strong incompatibility between the DAMA result and the just mentioned limits.

Our results are similar to those of Ref.~\cite{Scopel:2015baa}, although the halo-independent analysis done in this reference is different. In Ref.~\cite{Scopel:2015baa} the ``minimal'' $\tilde\eta^1(\vmin)$ compatible with the $1 \sigma$ DAMA error bars is identified with the piecewise continuous function touching the lower end of these error bars. Only the four $0.5$ keVee experimental bins in the $2.0$ to $4.0$ keVee energy interval are considered. $\tilde\eta^1(\vmin)$ is set to zero outside the $\vmin$ range corresponding to this $E'$ range. This amounts to a choice of very low ($m$-dependent) $\vmax$. With this choice, the contribution of I can be neglected for $m \lesssim 60$ GeV. Besides, $\tilde\eta^0(\vmin)$ is assumed to be equal to the ``minimal'' $\tilde\eta^1(\vmin)$ in almost all the considered $\vmin$ bins. This implies the dark halo leading to this velocity integral is quite different from the SHM, as we also find. However, while our results seem similar to those of Ref.~\cite{Scopel:2015baa}, we partially draw different conclusions. The authors of Ref.~\cite{Scopel:2015baa} conservatively conclude that their choice of $\tilde\eta^1$ and $\tilde\eta^0$ for DAMA can be compatible with the limits from other experiments for PS interactions with $m = 7$ and $m = 30$ GeV (while being rejected for AV interactions). This is because they allow $\tilde\eta^1(\vmin)$ to be equal to $\tilde\eta^0(\vmin)$ in almost the whole $2.0$--$4.0$ keVee energy range. Assuming instead that $\left|\tilde\eta^1(\vmin)\right|$ is much smaller than $\tilde\eta^0(\vmin)$, as it happens in most halo models, the DM interpretation of the DAMA data is in strong tension with the limits.

Fig.~\ref{fig:halo_ind_AV_endo} shows the result of our halo-independent analysis for WIMPs with AV interactions and inelastic endothermic scattering. The left panel is for $m=40$ GeV and $\delta=50$ keV, for which $v_\delta^\text{Na}=802.7$ km/s, and the right panel is for $m=52$ GeV and $\delta=100$ keV, for which $v_\delta^\text{Na}=1135.2$ km/s. Therefore, only scattering off I is kinematically allowed for $v_\text{max} = 800$ km/s. The DAMA data seems in disagreement with the KIMS bound for $Q_\text{I}=Q_\text{Cs}=0.10$, but not if $Q_\text{I}=Q_\text{Cs}=0.05$ in KIMS (however, this last value is much smaller than the $Q_\text{I}=0.09$ taken for DAMA).

Fig.~\ref{fig:halo_ind_PS_endo} shows the result of our halo-independent analysis for WIMPs with PS interactions and inelastic endothermic scattering, for $m=38$ GeV and $\delta=50$ keV in the left panel, resulting in $v_\delta^\text{Na}=810.1$ km/s, and $m=45$ GeV and $\delta=100$ keV in the right panel, resulting in $v_\delta^\text{Na}=1113.0$ km/s. WIMPs scattering off Na are again kinematically forbidden for $v_\text{max}=800$ km/s. The DAMA data are in disagreement with the KIMS bound, unless $Q_\text{I}=Q_\text{Cs}=0.05$ and $\left|\tilde\eta^1(\vmin)\right| \simeq \tilde\eta^0(\vmin)$.

Fig.~\ref{fig:halo_ind_AV_endo_longrange} shows our results for a WIMP with AV interactions and long-range inelastic endothermic scattering. Here $m=80$ GeV and $\delta=100$ keV, thus $v_\delta^\text{Na}=1031.5$ km/s. The DAMA data are excluded by our most stringent KIMS ($Q_\text{I}=Q_\text{Cs}=0.10$) and LUX bounds, but they remain in strong tension with the null results even when the less stringent KIMS and LUX bounds are considered.

\section{Conclusions}\label{sec:conclusion}

We investigated the possibility of interpreting the annual modulation signal observed in the DAMA experiment as due to WIMPs with spin-dependent coupling mostly to protons. We considered both an axial-vector (AV) interaction, which is what is usually referred to as `spin-dependent interaction', and a pseudo-scalar (PS) interaction, proposed in Ref.~\cite{Arina:2014yna} to reconcile DAMA with the null experiments. We also extended our analysis to inelastic scattering, and considered both contact and long-range interactions. Due to the similar $\ER$ dependence of the differential cross sections, we find for the long-range PS interaction the same results as for contact AV interactions, up to a shift in $\sigma_p$. We analyzed the data both assuming the Standard Halo Model (SHM) and in a halo-independent manner.

Spin-dependent WIMP couplings mostly to protons effectively weaken the bounds from experiments using Xe and Ge as target elements, whose spin is due mostly to neutrons. However, the bounds from experiments with F and I targets, such as PICASSO, SIMPLE and KIMS, remain relevant since their spin is due mostly to protons. 

Assuming the SHM, for elastic scattering (see Figs.~\ref{fig:AVel} to \ref{fig:LongRange_el}) we found that, in all the cases investigated here, the DAMA regions for Na are entirely excluded by SIMPLE and PICASSO, while the regions for I are excluded by KIMS.

For exothermic scattering (see Figs.~\ref{fig:AVexo}, \ref{fig:PSexo} and \ref{fig:LongRange_exo}), the DAMA regions move progressively to smaller WIMP masses with respect to the upper limits as $|\delta|$ increases, because the modulation phase observed by DAMA forces $v_\text{min}>200$ km/s, and this is possible only for progressively lighter WIMPs (see Fig.~\ref{fig:PS_mxvsdelta}). Thus, exothermic scattering brings compatibility between the DAMA region for Na and the upper bounds from SIMPLE. However, it does not suppress the PICASSO limit which continues to rule out the DAMA region. Furthermore, exothermic scattering reduces the modulation amplitude with respect to the time-average rate. Thus, the upper limit derived from the DAMA average rate measurement rejects the interpretation of the signal as due to scattering off Na for values of $\delta < -30\text{ keV}$. The DAMA region for scattering off I is excluded by the SIMPLE and KIMS upper bounds.

For endothermic scattering (see Figs.~\ref{fig:AVendo} to \ref{fig:LongRange_endo}), only KIMS provides relevant bounds. Scattering in all detectors besides KIMS and LUX becomes kinematically forbidden for large enough $\delta$. We showed results for $\delta=50$ and $100$ keV, because as $\delta$ increases further, scattering off I becomes kinematically forbidden as well. For $\delta=50\text{ keV}$, only assuming a larger quenching factor $Q_\text{I}=0.09$ for I in DAMA, and a smaller quenching factor $Q_\text{I}=Q_\text{Cs}=0.05$ in KIMS, the allowed DAMA region is compatible with all present limits for PS couplings. However, the possibility that the same nuclide has such different quenching factors in different crystals may be questionable. The same holds for contact and long-range AV and long-range PS interactions for $\delta=100$ keV. For contact PS interactions and $\delta=100\text{ keV}$, a small sleeve of the $90\%$ CL DAMA region for scattering off I escapes the $90\%$ CL KIMS limit with similar $Q_\text{I}$ for both experiments. These results are largely consistent with the results of Ref.~\cite{Barello:2014uda}. However, for PS interactions the DAMA regions are rejected by flavor physics bounds on the PS coupling to quarks \cite{Dolan:2014ska} (unless $g_\text{DM}$ can be very large $g_\text{DM}>10^5$). In our analysis we assumed that the scattering process can be approximated by one-particle exchange. The inclusion of multi-particle exchange processes may change the form of the WIMP-nucleus scattering cross section, and therefore all bounds shall be reconsidered.

For WIMP mass values within the DAMA regions derived assuming the SHM and close to the upper limits rejecting them, we performed also a halo-independent analysis. This is presented in Fig.~\ref{fig:halo_ind_PS_elastic} for elastic scattering with contact PS interactions and in Figs.~\ref{fig:halo_ind_AV_endo}--\ref{fig:halo_ind_AV_endo_longrange} for inelastic endothermic scattering with contact AV, contact PS, and long-range AV interactions. We again find strong tension between the DAMA data and upper bounds, except for contact AV and PS interactions with inelastic endothermic scattering if $Q_\text{I}$ in KIMS is much smaller ($Q_\text{I}=0.05$) than the $Q_\text{I}=0.09$ assumed for DAMA, although this choice of different $Q_\text{I}$ values for both experiments may be questionable.

\section*{Acknowledgements}
The authors were supported in part by the Department of Energy under Award Number DE-SC0009937. This research was also supported in part by the National Science Foundation under Grant No.~PHY11-25915 (through the Kavli Institute for Theoretical Physics, KITP, at the University of California, Santa Barbara, where G.G.~carried out part of the work).

\end{document}